\documentclass[twocolumn,english,aps,prl,showpacs,superscriptaddress]{revtex4-2}
\usepackage[utf8]{inputenc}
\usepackage{geometry}
\usepackage{array}
\usepackage{booktabs}
\usepackage{amsmath}
\usepackage{amssymb}
\usepackage{graphicx}
\usepackage{wasysym}
\usepackage{esint}

\makeatletter

\providecommand{\tabularnewline}{\\}

\usepackage{babel}

\makeatother

\usepackage{babel}
\begin{document}
\title{High-fidelity and robust controlled-Z gates for superconducting transmon
qubits. %
\mbox{%
I%
}. Intrinsic errors and decoherence}
\author{Chia-Hsien Huang}
\affiliation{Department of Physics and Center for Theoretical Physics, National
Taiwan University, Taipei 10617, Taiwan}
\affiliation{Center for Quantum Science and Engineering, National Taiwan University,
Taipei 10617, Taiwan}
\author{Hsi-Sheng Goan}
\email{goan@phys.ntu.edu.tw}
\affiliation{Department of Physics and Center for Theoretical Physics, National
Taiwan University, Taipei 10617, Taiwan}
\affiliation{Center for Quantum Science and Engineering, National Taiwan University,
Taipei 10617, Taiwan}
\affiliation{Physics Division, National Center for Theoretical Sciences, Taipei 106319, Taiwan}
\date{\today}
\begin{abstract}
Fault-tolerant quantum computing requires quantum gates that combine
short duration, high fidelity, and robustness to noise. As quantum
processors scale, robustness becomes increasingly important because
exhaustive characterization and frequent recalibration of individual
gates become impractical. We present a comprehensive optimal-control
framework for controlled-Z (CZ) gates between directly coupled, frequency-tunable
transmons that systematically accounts for intrinsic Hamiltonian errors
and realistic decoherence mechanisms, including energy relaxation
and dephasing. Using smooth, single-shot flux pulses, we obtain $12\text{-ns}$
CZ gates---close to the theoretical speed limit of $11.6\,\mathrm{ns}$---with
intrinsic errors suppressed to the $2\times10^{-5}$ level. We construct
dephasing noise models, calibrated against published experimental
measurements, that capture fast dephasing from room-temperature control
electronics and slow dephasing from $1/f$ flux noise. Using these
models, we identify dephasing sweet spots that suppress sensitivity
to dephasing noise. We further exploit transmon asymmetry as a design
parameter to access distinct noise-resilient operating regimes. For
transmons with the optimal Josephson energy ratio $a_{E_{J}}=E_{J1}/E_{J2}=1.31$,
operation near the joint fast- and slow-dephasing sweet spot yields
a total gate infidelity of $4.66\times10^{-4}$, predominantly limited
by energy relaxation ($4\times10^{-4}$ for $T_{1}=30\,\mu\mathrm{s}$).
The optimized gate maintains an infidelity below $10^{-3}$ despite
up to a sevenfold increase in fast-dephasing noise strength or a fourfold
increase in slow-dephasing noise strength relative to the corresponding
experimentally characterized levels. Strongly ($a_{E_{J}}=11.0$)
and weakly ($a_{E_{J}}=1.1$) asymmetric transmons provide tolerance
to slow- and fast-dephasing noise up to $90\times$ and $10\times$
their experimentally characterized levels, respectively, while maintaining
an infidelity below $10^{-3}$. These results establish an experimentally
applicable framework for balancing gate speed, fidelity, and noise
robustness while reducing characterization and recalibration overhead.

\global\long\def\figurename{FIG.}%
\makeatletter 
\global\long\def\thefigure{\@arabic\c@figure}%
\makeatother 
\end{abstract}
\maketitle

\section{Introduction}

Fault-tolerant universal quantum computing has the potential to transform
fields ranging from cryptography and materials science to drug discovery
and artificial intelligence \citep{Shor1994,AspuruGuzik2005,Babbush2018,Santagati2024,Biamonte2017}.
Realizing this potential requires quantum gates that combine high
fidelity with robustness to parameter uncertainties and environmental
noise. High-fidelity gates reduce the physical-qubit overhead associated
with quantum error correction \citep{Acharya2023}, while robust gates
can maintain high-fidelity operation despite device variations and
noise fluctuations \citep{HuangGoan2017,Propson2022,Wright2026}.
In current quantum processors, achieving and maintaining high-fidelity
operation relies on extensive characterization and periodic recalibration
of individual qubits and gates. Although such procedures are effective
for processors containing tens of qubits, their time and resource
requirements increase substantially with system size \citep{Wittler2021,Werninghaus2021,Xu2022},
making characterization and calibration a growing operational bottleneck
for large-scale quantum hardware. 
Robust control offers a route to mitigate this scaling challenge by engineering gate operations to be less sensitive to parameter variations and noise fluctuations while maintaining high fidelity, therefore reducing the need for precise characterization and frequent recalibration.

Superconducting transmon qubits have emerged as a leading platform
for quantum computation \citep{KochYuGambettaEtAl2007,Kjaergaard2020},
with experimentally demonstrated two-qubit controlled-Z (CZ) gate
fidelities above $99\%$, reaching up to $99.93\%$ \citep{BarendsKellyMegrantEtAl2014,Barends2019,Rol2019,Negirneac2021,Foxen2020,Sung2021,Li2025a,Marxer2026,Li2024,Li2025}.
Further improving the gate fidelity requires a comprehensive understanding
of the underlying error mechanisms and their interplay. Errors in
CZ gates can be broadly categorized into three classes: (i) intrinsic
errors arising from coherent Hamiltonian dynamics in the absence of
decoherence and parameter uncertainties; (ii) decoherence-induced
errors due to energy relaxation and dephasing; and (iii) systematic
errors arising from parameter uncertainties and calibration drift.
Significant progress has been made in understanding individual error
mechanisms \citep{MartinisGeller2014,Barends2019} and developing
a range of control strategies, including diabatic and fast-adiabatic
protocols \citep{MartinisGeller2014,Barends2019}, flux-pulse engineering
\citep{Rol2019,Negirneac2021}, optimization-based and reinforcement-learning
approaches \citep{KellyBarendsCampbellEtAl2014,Zong2021,Glaser2025,Egger2013,GarciaRipoll2020,Li2024,Sarma2025},
dynamical decoupling \citep{Zuk2024}, parametric coupling \citep{Jin2025},
and fully microwave-driven implementations \citep{shirai2025highfidelityallmicrowaveczgate}.
Despite these advances, it remains challenging to develop a control
framework that simultaneously accounts for multiple error sources
under realistic experimental constraints while maintaining robustness
to parameter variations and noise fluctuations.

In this work, we develop a unified framework for designing high-fidelity,
robust CZ gates in frequency-tunable transmons. We consider the direct-coupling
architecture \citep{BarendsKellyMegrantEtAl2014}, which provides
a simple setting for systematically studying the fundamental mechanisms
underlying intrinsic errors, decoherence, parameter uncertainties,
and their interplay. The experimentally characterized device and noise
parameters and the detailed CZ-gate error budget reported in Ref.\ \citep{BarendsKellyMegrantEtAl2014}
enable the construction of realistic noise models for quantitative
analysis and the identification of optimal control strategies. The
physical insights and control strategies developed in this work for
this simple architecture provide a foundation for extending the framework
to more complex coupling architectures---including resonator-mediated
coupling \citep{Rol2019,Negirneac2021}, tunable couplers \citep{Foxen2020,Sung2021,Li2025a,Marxer2026},
and double-transmon couplers \citep{Li2024,Li2025}---by incorporating
the corresponding coupling dynamics and associated decoherence mechanisms.

The framework addresses several key questions that are critical for
the practical implementation of high-fidelity, robust CZ gates: What
fundamental trade-offs arise among gate speed, fidelity, and robustness?
How do different noise channels---white noise from room-temperature
control electronics and $1/f$ flux noise---affect optimal control
strategies and robustness requirements? What role does transmon asymmetry
play in accessing noise-resilient operating regimes? To what extent
can optimized gates tolerate parameter variations and increased noise
while maintaining infidelities below $10^{-3}$? How do realistic
experimental constraints, such as arbitrary waveform generator (AWG)
bandwidth limitations, affect achievable gate fidelities? Through
detailed error decomposition and systematic optimization, we address
these questions and establish design principles for high-performance
CZ gates that can reduce calibration overhead in large-scale quantum
processors.

Our main contributions are as follows:

(1) Unified optimal-control framework: We develop a unified optimal-control
framework that integrates intrinsic Hamiltonian dynamics and decoherence
models calibrated against published experimental measurements into
the gate-design process, enabling the simultaneous optimization of
gate fidelity and robustness. The framework adopts a modular error
decomposition strategy, in which the total gate infidelity is expressed
as the sum of contributions from individual error sources. This modular
representation allows the cost function to be tailored to the dominant
error mechanisms in a given device and to be readily extended to incorporate
additional error sources.

(2) Physics-informed pulse design: We propose a physics-informed flux-pulse
design that decomposes the control waveform into a Pull-Back-Pulse
component and a Fine-Tune-Pulse component, $\Phi(t)=\Phi^{\mathrm{PBP}}(t)+\Phi^{\mathrm{FTP}}(t)$,
with the control parameters of both components optimized to minimize
the cost function. Specifically, the Pull-Back-Pulse $\Phi^{\mathrm{PBP}}(t)$
rapidly drives the system to a favorable operating point identified
from the system Hamiltonian and the noise model, and maintains the
flux bias there for a prescribed duration. During this interval, the
Fine-Tune-Pulse $\Phi^{\mathrm{FTP}}(t)$ is superimposed to introduce
local modulation around the operating point, providing additional
control flexibility for further cost-function minimization. By establishing
a favorable starting point for local pulse refinement, the proposed
pulse design enables the optimization to begin its search from a lower-cost
region of the control landscape, thereby facilitating efficient convergence
to high-fidelity and robust gate solutions.

(3) Detailed intrinsic error decomposition: We develop a detailed
numerical decomposition of the intrinsic gate error into phase, swap,
and leakage contributions. Phase error is further resolved into relative
local phase error and error arising from a shift in the target coupling
phase, while swap error is separated into two distinct swap-rotation
errors within the computational subspace. Leakage error is separated
into odd- and even-quanta leakage paths in the absence of X/Y microwave
control, with each leakage path further resolved according to the
destination non-computational states within each quanta sector. This
hierarchical error decomposition identifies the dominant error mechanisms,
thereby guiding the selection of effective control strategies and
optimization objectives. Furthermore, it enables simulation resources
to be allocated according to the dominant error contributions, avoiding
unnecessary computational overhead during optimization. For example,
when targeting extremely low gate infidelity, analysis of higher-quanta
leakage determines the number of transmon energy levels that must
be included in the numerical simulation.

(4) AWG-filtering-aware fine-tuning procedure: We develop a post-optimization
fine-tuning procedure that incorporates an experimentally relevant
AWG filtering model \citep{MotzoiGambettaMerkelEtAl2011} to account
for waveform distortions caused by the finite bandwidth of AWGs. The
resulting fine-tuned control pulses recover the original gate performance
despite filtering-induced waveform distortions in practical implementations.
For a representative commercial AWG with a cutoff frequency of $\omega_{c}/2\pi=$
$424.7\,\mathrm{MHz}$, the filtering effect increases the intrinsic
gate infidelity from $1.80\times10^{-5}$ to $1.14\times10^{-2}$.
Applying the proposed fine-tuning procedure reduces the intrinsic
gate infidelity to $2.05\times10^{-5}$, effectively restoring the
original gate performance.

(5) Realistic noise modeling: Using published Ramsey coherence measurements
and CZ-gate error budgets \citep{BarendsKellyMegrantEtAl2014}, we
construct realistic noise models that capture fast dephasing associated
with room-temperature control electronics and slow dephasing from
$1/f$ flux noise, respectively. Noise realizations generated from
the constructed models are incorporated into numerical simulations
of the full system dynamics, reproducing the experimentally measured
qubit-frequency dependence of the dephasing times $T_{\phi,1}$ and
$T_{\phi,2}$ associated with fast and slow dephasing, respectively.
This agreement with experimental measurements supports the applicability
of the noise models for simulating realistic device behavior. The
models further enable quantitative prediction of gate performance
under realistic experimental conditions and separation of the contributions
from fast and slow dephasing to the total gate infidelity, providing
a quantitative basis for gate optimization across different noise
regimes.

(6) Noise-informed sweet-spot identification and exploitation: Building
on the constructed noise models, we systematically identify the slow-dephasing
sweet spot, where the derivative of the qubit frequency $\Delta$
with respect to the magnetic flux $\Phi$ satisfies $d\Delta/d\Phi=0$,
and the fast-dephasing sweet spot in the lower-frequency regime. We
further demonstrate that operating near these sweet spots suppresses
the corresponding dephasing mechanisms. In particular, for asymmetric
transmons, the two sweet spots coincide at $\Phi/\Phi_{0}=0.5$, where
$\Phi_{0}$ is the superconducting flux quantum, thereby offering
the potential for simultaneous suppression of both fast and slow dephasing.
This provides a principled approach to selecting favorable operating
points that balance competing error sources while enhancing robustness
against dephasing.

(7) Asymmetry-assisted gate design: We reveal that transmon asymmetry---quantified
by the Josephson energy ratio $a_{E_{J}}=E_{J1}/E_{J2}$---provides
tunable access to distinct noise-resilient operating regimes. Strongly
asymmetric transmons, which exhibit lower overall $\left|d\Delta/d\Phi\right|$
across the tunable frequency range, provide enhanced robustness against
slow dephasing. This is consistent with experimental observations
of reduced sensitivity to $1/f$ flux noise in transmons with lower
tunability \citep{Hutchings2017}. Conversely, weakly asymmetric transmons,
which provide access to lower qubit frequencies through their larger
tunable frequency range, extend the robustness window against fast
dephasing. Under realistic noise conditions, the optimal asymmetry
is determined by the relative strengths of fast and slow dephasing:
stronger slow dephasing favors larger asymmetry, whereas stronger
fast dephasing favors smaller asymmetry. The resulting optimal asymmetry
balances the contributions from fast and slow dephasing, thereby minimizing
the total gate infidelity and providing an optimal trade-off between
robustness against the two dephasing mechanisms.

(8) Exceptional robustness with quantified tolerance margins: The
optimized gates implemented on strongly or weakly asymmetric transmons
exhibit substantial robustness to increased noise strength, thereby
reducing characterization and calibration demands. Specifically, strongly
asymmetric transmons ($a_{E_{J}}=11.0$) provide flux-noise tolerance
up to $90\times$ the experimentally characterized level while maintaining
a total gate infidelity below $10^{-3}$, and potentially offer a
viable route to operation in highly inhomogeneous flux-noise environments.
Conversely, weakly asymmetric transmons ($a_{E_{J}}=1.1$) tolerate
fast-dephasing noise up to $10\times$ the experimentally characterized
level under the same gate-infidelity criterion. These quantified tolerance
margins translate directly into relaxed characterization and calibration
requirements, facilitating the scalable deployment of high-fidelity
gates in large-scale quantum processors.

(9) Near-speed-limit, high-fidelity, robust gates: We achieve a $12\text{-ns}$
CZ gate using transmons with the optimal asymmetry $a_{E_{J}}=1.31$
under the realistic noise conditions reported in Ref.\ \citep{BarendsKellyMegrantEtAl2014}.
The gate time is close to the theoretical speed limit of $11.6\,\mathrm{ns}$.
The optimized total infidelity is $4.66\times10^{-4}$, dominated
by the energy-relaxation contribution of $4\times10^{-4}$ for $T_{1}=30\,\mu\mathrm{s}$,
while the intrinsic error is $2.46\times10^{-5}$ and the total dephasing
contribution is $4.11\times10^{-5}$. With state-of-the-art coherence
times ($T_{1}=360\,\mu\mathrm{s}$, achievable with tantalum or similar
materials \citep{Place2021}), the projected total infidelity decreases
to $9.90\times10^{-5}$, below commonly cited fault-tolerance thresholds.
This optimized robust gate also maintains a total gate infidelity
below $10^{-3}$ when fast- and slow-dephasing noise strengths are
independently increased to $7\times$ and $4\times$ their experimentally
characterized levels, respectively, with the other noise component
held at its experimental level. Thus, in a large-scale processor,
individual gates can retain high fidelity despite substantial spatial
variations in local noise, demonstrating the potential of the proposed
design for robust gate operation in inhomogeneous device environments.

This paper is organized as follows. Section\ \ref{sec:High_fidelity_CZ_gate}
presents the system Hamiltonian, analyzes intrinsic error mechanisms
(leakage, swap, and phase errors), and demonstrates optimization of
smooth flux pulses to suppress these errors while respecting AWG bandwidth
constraints. Sections\ \ref{subsec:Noise_characterization_dephasing}
and \ref{subsec:Noise_characterization_from_CZ_infid_budget} develop
detailed decoherence models calibrated against experimental data,
decomposing dephasing into fast (white noise) and slow ($1/f$ noise)
components and deriving their contributions to gate infidelity. Section\ \ref{subsec:HF_robust_CZ_against_noise}
combines these elements to identify optimal operating regimes, demonstrating
high-fidelity, robust CZ gates across symmetric and asymmetric transmon
configurations, with quantitative analysis of robustness margins.
Section\ \ref{sec:Conclusion} concludes with a discussion of broader
implications and a preview of the companion paper (Part $\mathrm{II}$)
\citep{Huang2026a}, which extends this framework to address robustness
against systematic parameter uncertainties---the remaining critical
challenge for large-scale fault-tolerant quantum processors, where
uncertainties in system parameters arising from imperfect parameter
knowledge, device-to-device variations, and temporal drift must be
managed without exhaustive individual characterization.

\section{Intrinsic errors}

\label{sec:High_fidelity_CZ_gate}

Currently, the infidelity of CZ gates is predominantly limited by
decoherence. However, as advances in technology reduce decoherence,
intrinsic gate errors are expected to become the dominant source of
CZ gate infidelity. Therefore, in the initial stage of this work,
we focus on an ideal transmon system, free from noise and uncertainties,
to investigate the intrinsic errors of CZ gates and subsequently design
high-fidelity gates using optimized control flux pulses. In Sec.\ \ref{subsec:Error_analysis_ideal_CZ_gate},
we introduce the intrinsic errors of CZ gates in an ideal transmon
system and analyze them through simulations of the drift Hamiltonian
evolution. In Sec.\ \ref{subsec:Control_scheme}, we describe our
control scheme and optimization method for constructing high-fidelity
CZ gates. The optimization results are presented in Sec.\ \ref{subsec:Optimization_results},
followed by a detailed analysis of the contributions to gate infidelity
in Sec.\ \ref{subsec:Detailed_infidelity_contribution}. In Sec.\ \ref{subsec:Filtering_effects},
we incorporate experimental constraints arising from the finite bandwidth
of arbitrary waveform generators (AWGs) and assess their impact on
the performance of the optimized CZ gates. Sections \ref{subsec:Error_analysis_ideal_CZ_gate}--\ref{subsec:Filtering_effects}
focus on symmetric transmon qubits. Finally, in Sec.\ \ref{subsec:Optimization_asymmetric_transmon},
we extend the construction of optimized CZ gates to asymmetric transmons
and compare their performance with that of symmetric transmons.

\subsection{Intrinsic errors of CZ gates in an ideal transmon system}

\label{subsec:Error_analysis_ideal_CZ_gate}

\subsubsection{Ideal Hamiltonian}

\label{subsec:Ideal_Hamiltonian}

The ideal Hamiltonian for two directly coupled superconducting transmon
qubits, in the absence of microwave XY controls and with flux Z controls
only, can be expressed as

\begin{eqnarray}
\mathcal{H}_{I}(t)/\hbar & = & \Delta_{A}(t)\cdot a_{A}^{\dagger}a_{A}+\dfrac{\alpha_{A}}{2}a_{A}^{\dagger}a_{A}(a_{A}^{\dagger}a_{A}-1)\nonumber \\
 &  & +\Delta_{B}(t)\cdot a_{B}^{\dagger}a_{B}+\dfrac{\alpha_{B}}{2}a_{B}^{\dagger}a_{B}(a_{B}^{\dagger}a_{B}-1)\nonumber \\
 &  & +g_{AB}(-i)(a_{A}-a_{A}^{\dagger})(-i)(a_{B}-a_{B}^{\dagger}).\label{eq:full_ideal_H}
\end{eqnarray}
Here, the subscripts A and B denote qubit-A ($Q_{A}$) and qubit-B
($Q_{B}$), respectively; $a_{A/B}^{\dagger}$ and $a_{A/B}$ are
the corresponding creation and annihilation operators; $g_{AB}$ is
the coupling strength between the two qubits; and $\alpha_{A/B}$
and $\Delta_{A/B}$ are the anharmonicities and qubit frequencies.
The ideal qubit frequencies

\begin{align}
 & \Delta_{A/B}(t)=\alpha_{A/B}+\nonumber \\
 & (\omega_{0,A/B}-\alpha_{A/B})\left[(1-d^{2})\cos^{2}\left(\frac{\pi\Phi_{A/B}(t)}{\Phi_{0}}\right)+d^{2}\right]^{\dfrac{1}{4}}\label{eq:freq_definition}
\end{align}
are controlled via the magnetic flux $\Phi_{A/B}(t)$ threading the
SQUID loop of the frequency-tunable transmons. Here, $\omega_{0,A/B}$
are the zero-flux qubit frequencies, $\Phi_{0}$ is the superconducting
flux quantum, and $d=(a_{E_{J}}-1)/(a_{E_{J}}+1)$ with the Josephson
energy ratio $a_{E_{J}}=E_{J1}/E_{J2}$ ($E_{J1}\geq E_{J2}$), where
$\hbar E_{J1}$ and $\hbar E_{J2}$ are the Josephson energies of
the two junctions forming the transmon \citep{KochYuGambettaEtAl2007,Hutchings2017}.
For symmetric transmons, $E_{J1}=E_{J2}$, so that $a_{E_{J}}=1$
and $d=0$; for asymmetric transmons, $E_{J1}>E_{J2}$, yielding $a_{E_{J}}>1$
and $0<d<1$.

\begin{figure*}
\includegraphics[scale=0.4]{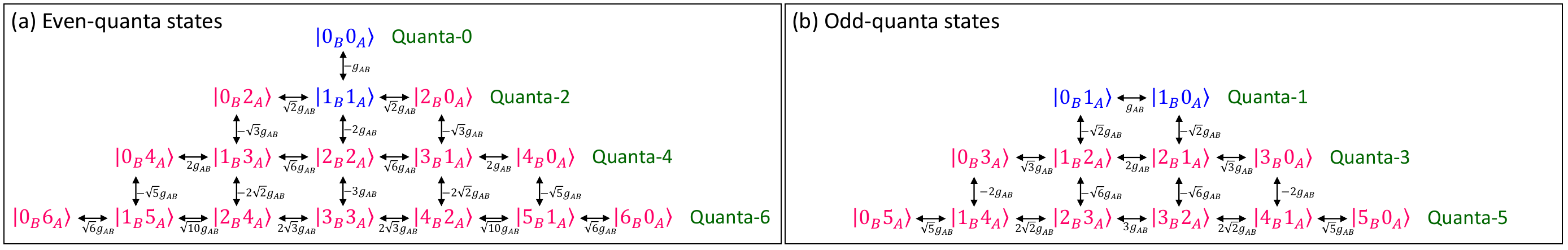}

\caption{Decoupled (a) even-quanta and (b) odd-quanta states. Leakage pathways
from computational basis states (blue) to non-computational states
(pink) are shown. States $\left|n_{B}m_{A}\right\rangle $ within
each row share the same total number of quanta $q=n_{B}+m_{A}$, labeled
as Quanta-$q$ (green) at the end of the row. The coupling strength
between two states is indicated in black near the double-arrow connecting
them.}

\label{fig:Even_Odd_Quanta_Paths} 
\end{figure*}

Next, we define the state $\left|n_{B}m_{A}\right\rangle $ with quanta
$n_{B}+m_{A}$, where $n_{B}$ and $m_{A}$ are the quantum numbers
of $Q_{B}$ and $Q_{A}$, respectively. In the target system governed
by the Hamiltonian $\mathcal{H}_{I}(t)$ in Eq.\ (\ref{eq:full_ideal_H}),
the absence of microwave XY control terms proportional to $-i(a_{A/B}-a_{A/B}^{\dagger})$
limits the allowed state transfers $\left|n_{B}m_{A}\right\rangle \leftrightarrow\left|n_{B}'m_{A}'\right\rangle $
to two distinct types. The first type is the same-quanta transfer,
for which $n_{B}+m_{A}=n_{B}'+m_{A}'$, associated with the coupling
Hamiltonian $\hbar g_{AB}(a_{A}a_{B}^{\dagger}+a_{A}^{\dagger}a_{B})$.
The second type is the jump-2-quanta transfer, for which $n_{B}+m_{A}=n_{B}'+m_{A}'\pm2$,
corresponding to the Hamiltonian $-\hbar g_{AB}(a_{A}a_{B}+a_{A}^{\dagger}a_{B}^{\dagger})$.
Both coupling Hamiltonians originate from the $Q_{A}$--$Q_{B}$
interaction term $+\hbar g_{AB}(-i)(a_{A}-a_{A}^{\dagger})(-i)(a_{B}-a_{B}^{\dagger})$
in $\mathcal{H}_{I}(t)$. As a result, even-quanta states $\left|n_{B}m_{A}\right\rangle $
(with $n_{B}+m_{A}$ even) cannot couple to odd-quanta states $\left|n_{B}'m_{A}'\right\rangle $
(with $n_{B}'+m_{A}'$ odd), and vice versa. In other words, even-
and odd-quanta states are decoupled in this system. The corresponding
coupling strengths and leakage pathways from the computational even-quanta
basis states $\left\{ \left|0_{B}0_{A}\right\rangle ,\left|1_{B}1_{A}\right\rangle \right\} $
and odd-quanta basis states $\left\{ \left|0_{B}1_{A}\right\rangle ,\left|1_{B}0_{A}\right\rangle \right\} $
are illustrated in Figs.\ \ref{fig:Even_Odd_Quanta_Paths}(a) and
\ref{fig:Even_Odd_Quanta_Paths}(b), respectively.

The matrix representation of the ideal Hamiltonian $\mathcal{H}_{I}(t)$
from Eq.\ (\ref{eq:full_ideal_H}) in the dominant basis states $\left\{ \left|0_{B}0_{A}\right\rangle ,\left|0_{B}1_{A}\right\rangle ,\left|1_{B}0_{A}\right\rangle ,\left|1_{B}1_{A}\right\rangle ,\left|0_{B}2_{A}\right\rangle ,\left|2_{B}0_{A}\right\rangle \right\} $,
used to implement a CZ gate, can be expressed as

\begin{align}
 & \mathcal{H}_{I}/\hbar=\nonumber \\
 & \left(\begin{array}{cccccc}
0 & 0 & 0 & -g_{AB} & 0 & 0\\
0 & \Delta_{A} & g_{AB} & 0 & 0 & 0\\
0 & g_{AB} & \Delta_{B} & 0 & 0 & 0\\
-g_{AB} & 0 & 0 & \Delta_{A}+\Delta_{B} & \sqrt{2}g_{AB} & \sqrt{2}g_{AB}\\
0 & 0 & 0 & \sqrt{2}g_{AB} & 2\Delta_{A}+\alpha_{A} & 0\\
0 & 0 & 0 & \sqrt{2}g_{AB} & 0 & 2\Delta_{B}+\alpha_{B}
\end{array}\right).\label{eq:ideal_H_in_6_basis_states}
\end{align}
The component of the Hamiltonian in the subspace $\left\{ \left|0_{B}1_{A}\right\rangle ,\left|1_{B}0_{A}\right\rangle \right\} $,
denoted $\mathcal{H}_{I_{1}}$, commutes with the component in the
subspace $\left\{ \left|0_{B}0_{A}\right\rangle ,\left|1_{B}1_{A}\right\rangle ,\left|0_{B}2_{A}\right\rangle ,\left|2_{B}0_{A}\right\rangle \right\} $,
denoted $\mathcal{H}_{I_{2}}$. The Hamiltonian $\mathcal{H}_{I_{1}}$
can be further decomposed into two commuting components,

\begin{equation}
\mathcal{H}_{I_{11}}=\hbar\left(\begin{array}{cc}
\dfrac{\Delta_{A}+\Delta_{B}}{2} & 0\\
0 & \dfrac{\Delta_{A}+\Delta_{B}}{2}
\end{array}\right),\label{eq:ideal_H_1_1}
\end{equation}
and 
\begin{equation}
\mathcal{H}_{I_{12}}=\hbar\left(\begin{array}{cc}
\dfrac{1}{2}\delta_{01-10} & g_{AB}\\
g_{AB} & -\dfrac{1}{2}\delta_{01-10}
\end{array}\right),\label{eq:ideal_H_1_2}
\end{equation}
where we define the detuning $\delta_{01-10}\equiv\Delta_{A}-\Delta_{B}$.
Furthermore, owing to the large detuning $\delta_{00-11}\equiv\Delta_{A}+\Delta_{B}$
between $\left|0_{B}0_{A}\right\rangle $ and $\left|1_{B}1_{A}\right\rangle $,
we neglect the basis state $\left|0_{B}0_{A}\right\rangle $ in $\mathcal{H}_{I_{2}}$
and decompose the remaining $\mathcal{H}_{I_{2}}$ into two commuting
components, 
\begin{equation}
\mathcal{H}_{I_{21}}=\hbar\left(\begin{array}{ccc}
\Delta_{A}+\Delta_{B} & 0 & 0\\
0 & \Delta_{A}+\Delta_{B} & 0\\
0 & 0 & \Delta_{A}+\Delta_{B}
\end{array}\right)\label{eq:ideal_H_2_1}
\end{equation}
and 
\begin{equation}
\mathcal{H}_{I_{22}}=\hbar\left(\begin{array}{ccc}
0 & \sqrt{2}g_{AB} & \sqrt{2}g_{AB}\\
\sqrt{2}g_{AB} & \delta_{A} & 0\\
\sqrt{2}g_{AB} & 0 & \delta_{B}
\end{array}\right).\label{eq:ideal_H_2_2}
\end{equation}
Here, the detunings are defined as $\delta_{A}\equiv\Delta_{A}-\Delta_{B}+\alpha_{A}$
and $\delta_{B}\equiv\Delta_{B}-\Delta_{A}+\alpha_{B}$, satisfying
$\delta_{A}+\delta_{B}=\alpha_{A}+\alpha_{B}$.

\subsubsection{CZ gate implementation}

\label{subsec:CZ_gate_implementation}

To realize a two-qubit CZ gate, 
\begin{equation}
\mathrm{CZ}=\left(\begin{array}{cccc}
1 & 0 & 0 & 0\\
0 & 1 & 0 & 0\\
0 & 0 & 1 & 0\\
0 & 0 & 0 & -1
\end{array}\right),\label{eq:CZ_gate_target}
\end{equation}
in the computational basis states $\left\{ \left|0_{B}0_{A}\right\rangle ,\left|0_{B}1_{A}\right\rangle ,\left|1_{B}0_{A}\right\rangle ,\left|1_{B}1_{A}\right\rangle \right\} $,
all computational basis states remain unchanged, except that the $\left|1_{B}1_{A}\right\rangle $
state acquires an additional phase factor $e^{i\phi_{AB}}$ with coupling
phase $\phi_{AB}=\pi$, after the CZ gate operation. This coupling
phase is generated via the Quanta-2-state couplings $\left|1_{B}1_{A}\right\rangle $$\leftrightarrow$$\left|0_{B}2_{A}\right\rangle $
and $\left|1_{B}1_{A}\right\rangle $$\leftrightarrow$$\left|2_{B}0_{A}\right\rangle $,
governed by the Hamiltonian $\mathcal{H}_{I_{22}}$ in Eq.\ (\ref{eq:ideal_H_2_2}).
The strongest coupling occurs at $\delta_{A}=0$ ($\delta_{B}=\alpha_{A}+\alpha_{B}$)
and $\delta_{B}=0$ ($\delta_{A}=\alpha_{A}+\alpha_{B}$). In the
regime $\delta_{A}=0$ with $\left|\delta_{B}\right|\gg\left|\delta_{A}\right|$,
the $\left|1_{B}1_{A}\right\rangle $$\leftrightarrow$$\left|0_{B}2_{A}\right\rangle $
coupling dominates the dynamics. In this case, $\mathcal{H}_{I_{22}}$
can be approximated as $\mathcal{H}_{I_{22}}=\sqrt{2}g_{AB}\cdot\sigma_{X}$
in the basis states $\left\{ \left|1_{B}1_{A}\right\rangle ,\left|0_{B}2_{A}\right\rangle \right\} $,
where $\sigma_{X}$ is the Pauli X matrix. This allows the CZ gate
time to be estimated as $t_{f}=n\pi/(\sqrt{2}g_{AB})$ for positive
odd integers $n$, ensuring that $\left|1_{B}1_{A}\right\rangle $
returns to its initial state with the correct coupling phase $\phi_{AB}=\pi$.
Consequently, the speed limit of the CZ gate occurs at $n=1$, yielding
the shortest gate time $t_{f}=\pi/(\sqrt{2}g_{AB})$.

As $\delta_{A}$ moves away from the strongest coupling point ($\delta_{A}=0$)
in the positive direction ($\delta_{A}>0$), the condition $\left|\delta_{B}\right|\gg\left|\delta_{A}\right|$
remains valid. Consequently, the Hamiltonian $\mathcal{H}_{I_{22}}$can
be approximated in the basis states $\left\{ \left|1_{B}1_{A}\right\rangle ,\left|0_{B}2_{A}\right\rangle \right\} $
as $\mathcal{H}_{I_{22}}=\sqrt{2}g_{AB}\cdot\sigma_{X}-(\delta_{A}/2)\cdot\sigma_{Z}+(\delta_{A}/2)\cdot I$,
where $\sigma_{Z}$ is the Pauli-Z matrix and $I$ is the $2\times2$
identity matrix. The corresponding diagonal propagator element in
the $\left|1_{B}1_{A}\right\rangle $ state is given by $\left[\cos\theta+i\frac{1}{2}\left(\delta_{A}/\sqrt{2g_{AB}^{2}+\delta_{A}^{2}/4}\right)\sin\theta\right]\cdot\exp(-i\delta_{A}t_{f}/2)$
with $\theta=\sqrt{2g_{AB}^{2}+\delta_{A}^{2}/4}\cdot t_{f}$. When
$\theta=n\pi$ for a positive integer $n$, the magnitude of this
element returns to unity, and the associated phase factor becomes
$e^{i\phi_{AB}}=\exp\left(+in\pi-i\delta_{A}t_{f}/2\right)$, yielding
a coupling phase $\phi_{AB}=\left(\sqrt{2g_{AB}^{2}+\delta_{A}^{2}/4}-\delta_{A}/2\right)\cdot t_{f}$.
As $\delta_{A}$ increases, the coupling phase $\phi_{AB}$ approaches
$(2g_{AB}^{2}/\delta_{A})\cdot t_{f}$, so achieving the target $\phi_{AB}=\pi$
requires a longer gate time $t_{f}$. Physically, a larger $\delta_{A}$
weakens the $\left|1_{B}1_{A}\right\rangle $$\leftrightarrow$$\left|0_{B}2_{A}\right\rangle $
coupling, necessitating extended evolution to accumulate the desired
phase. Notably, the conditions of unit probability and correct coupling
phase $\phi_{AB}=\pi$ do not generally coincide, leading to enhanced
errors. Specifically, deviations from unity probability correspond
to leakage out of the computational state $\left|1_{B}1_{A}\right\rangle $,
while deviations from the target phase correspond to phase errors.

Conversely, as $\delta_{A}$ decreases from the strongest coupling
point ($\delta_{A}=0$) in the negative direction ($\delta_{A}<0$),
$\left|\delta_{A}\right|$ still increases while $\left|\delta_{B}\right|$
decreases, since $\delta_{B}=\alpha_{A}+\alpha_{B}-\delta_{A}$. As
a result, the $\left|1_{B}1_{A}\right\rangle $$\leftrightarrow$$\left|0_{B}2_{A}\right\rangle $
coupling weakens, whereas the $\left|1_{B}1_{A}\right\rangle $ $\leftrightarrow$$\left|2_{B}0_{A}\right\rangle $
coupling is enhanced. When $\delta_{A}=\alpha_{A}+\alpha_{B}$, the
system reaches another peak coupling point ($\delta_{B}=0$), dominated
by the $\left|1_{B}1_{A}\right\rangle $ $\leftrightarrow$$\left|2_{B}0_{A}\right\rangle $
interaction, which can also yield the shortest gate time, $t_{f}=\pi/(\sqrt{2}g_{AB})$,
similar to the case at $\delta_{A}=0$.

When the state $\left|1_{B}1_{A}\right\rangle $ accumulates the coupling
phase $\phi_{AB}$, the states $\left|0_{B}1_{A}\right\rangle $ and
$\left|1_{B}0_{A}\right\rangle $ evolve due primarily to the $\left|0_{B}1_{A}\right\rangle \leftrightarrow\left|1_{B}0_{A}\right\rangle $
swap rotation, governed by the Hamiltonian $\mathcal{H}_{I_{12}}$
in Eq.\ (\ref{eq:ideal_H_1_2}), thereby giving rise to a swap error.
Simultaneously, another swap error arises from the $\left|0_{B}0_{A}\right\rangle \leftrightarrow\left|1_{B}1_{A}\right\rangle $
swap rotation, described by the Hamiltonian in Eq.\ (\ref{eq:ideal_H_in_6_basis_states}).
However, this latter swap error is significantly smaller than the
former, owing to the larger detuning $\delta_{00-11}$ in the jump-two-quanta
$\left|0_{B}0_{A}\right\rangle \leftrightarrow\left|1_{B}1_{A}\right\rangle $
coupling, compared with the smaller detuning $\delta_{01-10}$ in
the same-quanta $\left|0_{B}1_{A}\right\rangle \leftrightarrow\left|1_{B}0_{A}\right\rangle $
coupling. Additionally, local phase accumulations, dominantly governed
by the Hamiltonians $\mathcal{H}_{I_{11}}$ and $\mathcal{H}_{I_{21}}$
in Eqs.\ (\ref{eq:ideal_H_1_1}) and (\ref{eq:ideal_H_2_1}), also
contribute to phase errors. These accumulated relative local phases
can, however, be compensated by applying single-qubit Z rotations
after the two-qubit operation, or simultaneously corrected during
the CZ gate using optimized pulse shaping techniques.

\subsubsection{Intrinsic CZ gate errors}

\label{subsec:Intrinsic_CZ_gate_errors}

In the preceding section, we discussed the intrinsic errors induced
during CZ gate implementation, such as leakage errors, swap errors,
and phase errors. These errors can be explicitly derived from the
ideal two-qubit gate infidelity,

\begin{equation}
J_{1}\equiv1-\dfrac{1}{16}\left|\mathrm{Tr}\left(U_{T}^{\dagger}U_{I,4\times4}(t_{f})\right)\right|^{2},\label{eq:J1_definition}
\end{equation}
with detailed derivations provided in Appendix\ \ref{appendix:error_contributions}.
Here, $U_{I,4\times4}(t_{f})$ denotes the projection of the ideal
propagator $U_{I}(t_{f})=\mathcal{T}_{+}\exp\left[-i\int_{0}^{t_{f}}\mathcal{H}_{I}(t')dt'\right]$
onto the subspace spanned by the computational basis states of the
two qubits, $\left\{ \left|0_{B}0_{A}\right\rangle ,\left|0_{B}1_{A}\right\rangle ,\left|1_{B}0_{A}\right\rangle ,\left|1_{B}1_{A}\right\rangle \right\} $,
evaluated at the gate time $t_{f}$. $U_{T}$ denotes the target gate,
and $\mathrm{Tr}$ represents the matrix trace. In this work, the
target gate $U_{T}$ is the CZ gate, as specified in Eq.\ (\ref{eq:CZ_gate_target}).

The leakage error can be evaluated as $1-\left|\mathrm{Tr}\left(U_{I,4\times4}^{\dagger}(t_{f})U_{I,4\times4}(t_{f})\right)\right|^{2}/4.$
This error arises from two distinct leakage pathways, as illustrated
in Figs.\ \ref{fig:Even_Odd_Quanta_Paths}(a) and \ref{fig:Even_Odd_Quanta_Paths}(b),
corresponding to the decoupled even- and odd-quanta states, as discussed
in Sec.\ \ref{subsec:Ideal_Hamiltonian}. To analyze the leakage
in greater detail, we define the Quanta-$m$ (Qt-$m$) leakage error
as the contribution to the gate infidelity arising from probability
leakage from the computational states to the Qt-$m$ non-computational
states. For example, the Qt-$2$ leakage error quantifies the infidelity
due to probability leakage from the $\left|1_{B}1_{A}\right\rangle $
state to the Qt-$2$ non-computational states $\left\{ \left|0_{B}2_{A}\right\rangle ,\left|2_{B}0_{A}\right\rangle \right\} $.
Similarly, the Qt-3 leakage error accounts for the contribution from
the $\left|0_{B}1_{A}\right\rangle $ and $\left|1_{B}0_{A}\right\rangle $
states to the Qt-$3$ non-computational states $\left\{ \left|0_{B}3_{A}\right\rangle ,\left|1_{B}2_{A}\right\rangle ,\left|2_{B}1_{A}\right\rangle ,\left|3_{B}0_{A}\right\rangle \right\} $.
In the evaluation of Qt-$2$, Qt-$4$, and Qt-$6$ leakage errors
reported in this paper, we neglect contributions from the $\left|0_{B}0_{A}\right\rangle $
state due to the large detuning $\delta_{00-11}$, which renders its
contribution negligible compared to that from the $\left|1_{B}1_{A}\right\rangle $
state.

The swap error comprises the $\left|0_{B}1_{A}\right\rangle \leftrightarrow\left|1_{B}0_{A}\right\rangle $
swap error and the $\left|0_{B}0_{A}\right\rangle \leftrightarrow\left|1_{B}1_{A}\right\rangle $
swap error, as discussed in the preceding section and derived in Appendix\ \ref{appendix:error_contributions}.

The phase error originates from incorrect local phases $\phi_{00}$,
$\phi_{01}$, $\phi_{10}$, and $\phi_{11}$ appearing in the diagonal
elements of $U_{I,4\times4}(t_{f})$, expressed as phase factors $e^{i\phi_{00}}$,
$e^{i\phi_{01}}$, $e^{i\phi_{10}}$, and $e^{i\phi_{11}}$, in the
basis sequence $\left|0_{B}0_{A}\right\rangle $, $\left|0_{B}1_{A}\right\rangle $,
$\left|1_{B}0_{A}\right\rangle $, and $\left|1_{B}1_{A}\right\rangle $.
For an ideal CZ gate, all relative local phases should vanish except
for the required coupling phase $\phi_{AB}=\pi$ associated with the
$\left|1_{B}1_{A}\right\rangle $ state. Accordingly, the phase error
can be decomposed into two contributions, as detailed in Appendix\ \ref{appendix:error_contributions}.
The first contribution is the relative local phase error, arising
from differences among local phases $\phi_{00}$, $\phi_{01}$, and
$\phi_{10}$, and is quantified as $\left[(\phi_{01}-\phi_{00})^{2}+(\phi_{10}-\phi_{00})^{2}+(\phi_{01}-\phi_{10})^{2}\right]/16$.
The second contribution is the $\phi_{11}$ phase error, associated
with the deviation of $\phi_{11}$ from the target coupling phase
$\phi_{AB}=\pi$ relative to the other local phases, and is expressed
as $\left[(\phi_{11}-\phi_{00}-\pi)^{2}+(\phi_{11}-\phi_{01}-\pi)^{2}+(\phi_{11}-\phi_{10}-\pi)^{2}\right]/16$.
The relative local phase error can be effectively suppressed by compensating
for the relative local phases using a single-qubit-Z gate

\begin{equation}
U_{Z}=\left(\begin{array}{cccc}
1 & 0 & 0 & 0\\
0 & e^{-i\phi_{A}} & 0 & 0\\
0 & 0 & e^{-i\phi_{B}} & 0\\
0 & 0 & 0 & e^{-i(\phi_{A}+\phi_{B})}
\end{array}\right),\label{eq:local_phase_compensated}
\end{equation}
applied in the basis states $\left\{ \left|0_{B}0_{A}\right\rangle ,\left|0_{B}1_{A}\right\rangle ,\left|1_{B}0_{A}\right\rangle ,\left|1_{B}1_{A}\right\rangle \right\} $,
immediately following the CZ gate operation. Here, $\phi_{A}=(\phi_{01}-\phi_{00})$
and $\phi_{B}=(\phi_{10}-\phi_{00})$. After this compensation, the
phase error is dominated by the $\phi_{11}$ phase error. The single-qubit-Z
gate $U_{Z}$ is implemented as the combination of a Z-rotation on
$Q_{A}$ and a Z-rotation on $Q_{B}$.

\subsubsection{Simulation of intrinsic errors in drift Hamiltonian evolution}

\label{subsec:Drift_H_intrinsic_error}

\begin{figure}
\includegraphics[scale=0.52]{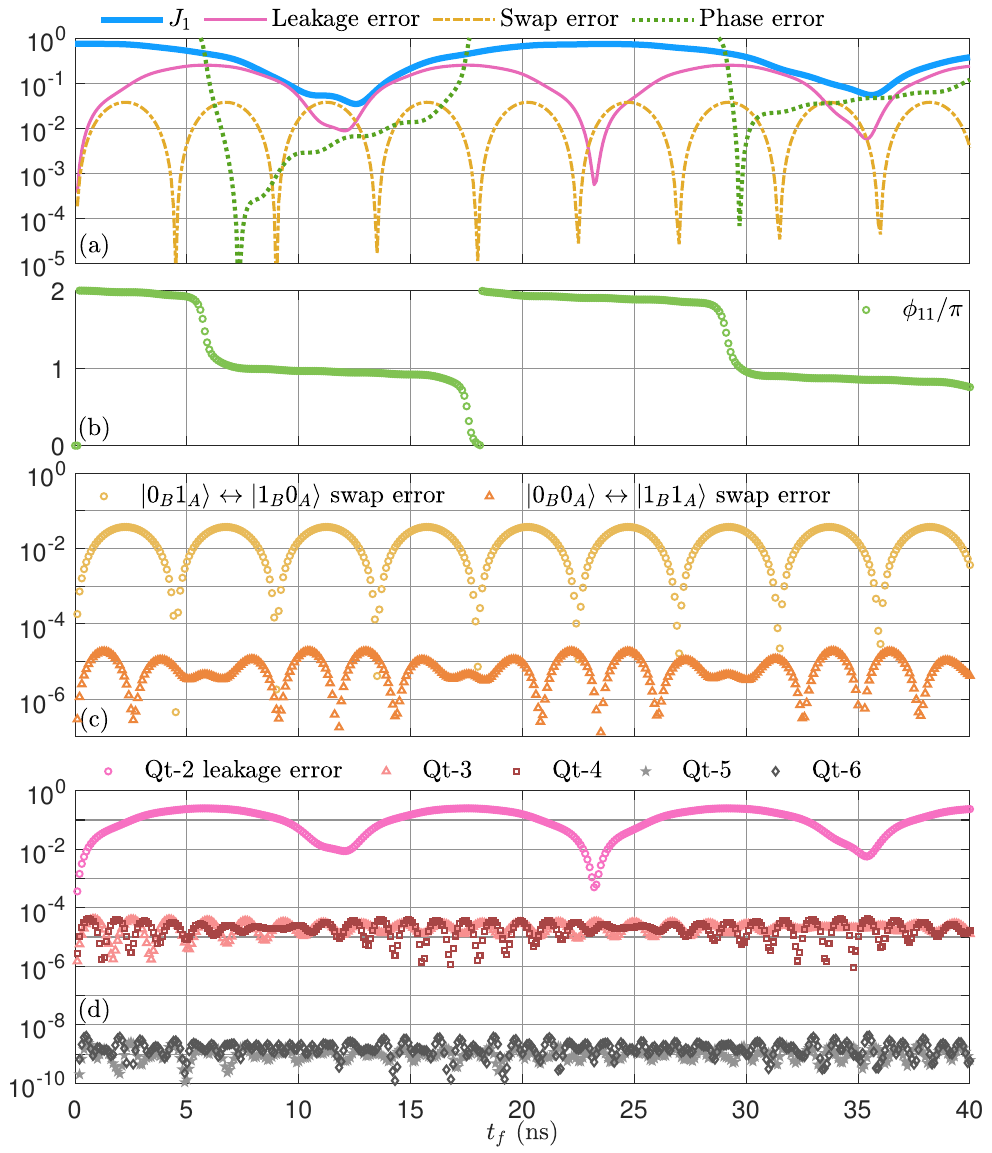}
\caption{Intrinsic CZ gate errors for the propagator evolved freely under the
drift Hamiltonian $\mathcal{H}_{I}$ with fixed $\Delta_{A/B}$ at
$\delta_{A}=0$. (a) The ideal CZ gate infidelity $J_{1}$ (bold blue
solid line) and its constituent intrinsic errors, including the leakage
error (thin pink solid line), the swap error (yellow dash-dotted line),
and the phase error (green dotted line), as functions of the gate
time $t_{f}$ from $0$ to $40\,\mathrm{ns}$. (b) The phase error
is attributed to the deviation of the local phase $\phi_{11}$ (green
circles) from the target coupling $\phi_{AB}=\pi$. (c) The swap error
comprises the $\left|0_{B}1_{A}\right\rangle \leftrightarrow\left|1_{B}0_{A}\right\rangle $
swap error (yellow circles) and the $\left|0_{B}0_{A}\right\rangle \leftrightarrow\left|1_{B}1_{A}\right\rangle $
swap error (orange triangles). (d) The leakage error includes the
Qt-2 to Qt-6 leakage errors, shown as pink circles, light-pink triangles,
brown squares, light-gray pentagrams, and dark-gray diamonds, respectively.}
\label{fig:Drift_H_evolution} 
\end{figure}

In this section, we simulate and analyze the intrinsic errors of a
CZ gate realized via evolution under the drift Hamiltonian $\mathcal{H}_{I}$
{[}Eq.\ (\ref{eq:full_ideal_H}){]}, considering seven energy levels
for each qubit and fixed $\Delta_{A}$ and $\Delta_{B}$ at $\delta_{A}=0$
under the condition of the strongest coupling. In these simulations
(as well as in subsequent simulations presented in this work and in
the companion work \citep{Huang2026a}), we employ experimental system
parameters reported in Ref.\ \citep{BarendsKellyMegrantEtAl2014},
namely the coupling constant $g_{AB}=2\pi\times30.4\,\mathrm{MHz}$,
the anharmonicities $\alpha_{A/B}=2\pi\times214/212\,\mathrm{MHz}$,
and the zero-flux qubit frequencies $\omega_{0,A/B}=2\pi\times5.78\mathrm{/5.06\,GHz}$.
We fix $\Delta_{B}=2\pi\times4.7\,\mathrm{GHz}$, matching the idle
frequency of $Q_{B}$ in the experiment \citep{BarendsKellyMegrantEtAl2014};
consequently, for $\delta_{A}=0$, the fixed $\Delta_{A}$ is determined
as $\Delta_{A}=\delta_{A}+\Delta_{B}-\alpha_{A}=2\pi\times4.914\,\mathrm{GHz}$.
Then, following the experimental procedure for compensating for relative
local phases, we extract the local phases from the propagator of the
drift Hamiltonian to construct the single-qubit-Z gate $U_{Z}$ defined
in Eq.\ (\ref{eq:local_phase_compensated}), which is subsequently
applied after the propagator. The resulting compensated propagator
is subsequently used to evaluate the ideal CZ gate infidelity $J_{1}$
{[}Eq.\ (\ref{eq:J1_definition}){]}, as well as its constituent
intrinsic errors, as detailed in Sec.\ \ref{subsec:Intrinsic_CZ_gate_errors}
and Appendix\ \ref{appendix:error_contributions}. Finally, these
errors are presented as functions of the evolution time $t_{f}$ for
the drift Hamiltonian in Fig.\ \ref{fig:Drift_H_evolution}.

As shown in Fig.\ \ref{fig:Drift_H_evolution}(a), the local minima
of $J_{1}$ occur at $t_{f}\approx12\,\mathrm{ns}$ and $36\,\mathrm{ns}$,
in good agreement with the estimated permissible CZ gate times under
the strong $\left|1_{B}1_{A}\right\rangle \leftrightarrow\left|0_{B}2_{A}\right\rangle $
coupling at $\delta_{A}=0$, as introduced in Sec.\ \ref{subsec:CZ_gate_implementation}.
Specifically, these times satisfy $t_{f}=n\pi/(\sqrt{2}g_{AB})\approx n\times11.6\,\mathrm{ns}$
for the odd integers $n=1$ and $n=3$, respectively.

Because the calibrated single-qubit-Z gate $U_{Z}$ is inserted to
compensate for relative local phases in the simulations, the relative
local phase error is effectively eliminated. Consequently, the phase
error shown in Fig.\ \ref{fig:Drift_H_evolution}(a) is predominantly
determined by the $\phi_{11}$ phase error, which arises from the
deviation of the compensated local phase $\phi_{11}$ from the target
coupling $\phi_{AB}=\pi$, as illustrated in Fig.\ \ref{fig:Drift_H_evolution}(b).
The step-like behavior of $\phi_{11}$ is primarily due to the strong
$\left|1_{B}1_{A}\right\rangle \leftrightarrow\left|0_{B}2_{A}\right\rangle $
coupling at $\delta_{A}=0$, governed by the Hamiltonian $\mathcal{H}_{I_{22}}=\sqrt{2}g_{AB}\cdot\sigma_{X}$
in the basis $\left\{ \left|1_{B}1_{A}\right\rangle ,\left|0_{B}2_{A}\right\rangle \right\} $.
For the $\left|1_{B}1_{A}\right\rangle $ basis state, the diagonal
element of the corresponding propagator is $\mathrm{cos}\theta$,
where $\theta=\sqrt{2}g_{AB}\cdot t_{f}$. For $\theta$ in the intervals
$\left[0.5\pi+2n\pi,1.5\pi+2n\pi\right]$, with $n$ an integer, $\mathrm{cos}\theta$
is negative and can be expressed as $e^{i\pi}\left|\mathrm{cos}\theta\right|$,
thereby contributing a $\pi$ phase to $\phi_{11}$. In the remaining
intervals, $\mathrm{cos}\theta$ is positive and can be written as
$e^{i0\pi}\left|\mathrm{cos}\theta\right|$ or $e^{i2\pi}\left|\mathrm{cos}\theta\right|$,
contributing $0$ (or $2\pi$) phase. Consequently, $\phi_{11}=\pi$
for $t_{f}$ satisfying $(0.5\pi+2n\pi)/(\sqrt{2}g_{AB})<t_{f}<(1.5\pi+2n\pi)/(\sqrt{2}g_{AB})$,
i.e., $(5.8\,\mathrm{ns}+n\cdot23.3\,\mathrm{ns})<t_{f}<(17.4\,\mathrm{ns}+n\cdot23.3\,\mathrm{ns})$,
and $\phi_{11}=0$ (or $2\pi$) otherwise, resulting in the characteristic
step-like structure. A weaker coupling via $\left|1_{B}1_{A}\right\rangle \leftrightarrow\left|2_{B}0_{A}\right\rangle $
interaction, with large detuning $\left|\delta_{B}\right|=\left|\alpha_{A}+\alpha_{B}\right|$,
contributes a small continuous phase to $\phi_{11}$, smoothing the
transition between $\pi$ and $0$ (or $2\pi$) as shown in Fig.\ \ref{fig:Drift_H_evolution}(b).
This contribution also slightly distorts the $\phi_{11}$ step, causing
the minima of the phase error to shift relative to the minima of $J_{1}$
observed in Fig.\ \ref{fig:Drift_H_evolution}(a).

The swap error illustrated in Fig.\ \ref{fig:Drift_H_evolution}(a)
is primarily dominated by the $\left|0_{B}1_{A}\right\rangle \leftrightarrow\left|1_{B}0_{A}\right\rangle $
swap error, with only a minor contribution arising from the $\left|0_{B}0_{A}\right\rangle \leftrightarrow\left|1_{B}1_{A}\right\rangle $
swap error, as shown in Fig.\ \ref{fig:Drift_H_evolution}(c). This
behavior can be attributed to the smaller detuning $\delta_{01-10}\equiv\Delta_{A}-\Delta_{B}=2\pi\times0.214\mathrm{\,GHz}$
in the $\left|0_{B}1_{A}\right\rangle \leftrightarrow\left|1_{B}0_{A}\right\rangle $
coupling, governed by the Hamiltonian $\mathcal{H}_{I_{12}}=\frac{1}{2}\delta_{01-10}\cdot\sigma_{Z}+g_{AB}\cdot\sigma_{X}$
{[}Eq.\ (\ref{eq:ideal_H_1_2}){]}, compared with the much larger
detuning $\delta_{00-11}\equiv\Delta_{A}+\Delta_{B}=2\pi\times9.614\,\mathrm{GHz}$
in the $\left|0_{B}0_{A}\right\rangle \leftrightarrow\left|1_{B}1_{A}\right\rangle $
coupling, described by the Hamiltonian $-\frac{1}{2}\delta_{00-11}\cdot\sigma_{Z}-g_{AB}\cdot\sigma_{X}$
in the basis $\left\{ \left|0_{B}0_{A}\right\rangle ,\left|1_{B}1_{A}\right\rangle \right\} $
from $\mathcal{H}_{I}$ {[}Eq.\ (\ref{eq:ideal_H_in_6_basis_states}){]}.
Referring to Appendix\ \ref{appendix:error_contributions}, the maximum
$\left|0_{B}1_{A}\right\rangle \leftrightarrow\left|1_{B}0_{A}\right\rangle $
swap error can be estimated as $\frac{1}{2}\cdot g_{AB}^{2}/(g_{AB}^{2}+(\delta_{01-10}/2)^{2})\approx3.7\times10^{-2}$
with a corresponding swap period $\pi/\sqrt{g_{AB}^{2}+(\delta_{01-10}/2)^{2}}\approx4.5\,\mathrm{ns}$.
By contrast, for the $\left|0_{B}0_{A}\right\rangle \leftrightarrow\left|1_{B}1_{A}\right\rangle $
swap, the estimated error $\frac{1}{2}\cdot g_{AB}^{2}/(g_{AB}^{2}+(\delta_{00-11}/2)^{2})\approx2.0\times10^{-5}$
is significantly smaller, and the corresponding swap period $\pi/\sqrt{g_{AB}^{2}+(\delta_{00-11}/2)^{2}}\approx0.1\,\mathrm{ns}$
is much shorter. These estimates are in good agreement with the simulations
presented in Fig.\ \ref{fig:Drift_H_evolution}(c). The $\left|0_{B}0_{A}\right\rangle \leftrightarrow\left|1_{B}1_{A}\right\rangle $
swap period is not directly resolved in Fig.\ \ref{fig:Drift_H_evolution}(c)
because it is comparable to the simulation sampling interval of $0.1\,\mathrm{ns}$.

The leakage error depicted in Fig.\ \ref{fig:Drift_H_evolution}(a)
is primarily dominated by the Qt-2 leakage error, as illustrated in
Fig.\ \ref{fig:Drift_H_evolution}(d). This same-quanta Qt-2 leakage
error originates from the $\left|1_{B}1_{A}\right\rangle \leftrightarrow\left|0_{B}2_{A}\right\rangle $
and $\left|1_{B}1_{A}\right\rangle \leftrightarrow\left|2_{B}0_{A}\right\rangle $
couplings, which are governed by the Hamiltonian $\mathcal{H}_{I_{22}}$
in Eq.\ (\ref{eq:ideal_H_2_2}). For the $\left|1_{B}1_{A}\right\rangle \leftrightarrow\left|0_{B}2_{A}\right\rangle $
rotation at zero detuning ($\delta_{A}=0$), the rotation period is
$\pi/(\sqrt{2}g_{AB})\approx11.6\,\mathrm{ns}$, leading to a maximum
$\left|1_{B}1_{A}\right\rangle \rightarrow\left|0_{B}2_{A}\right\rangle $
leakage error of $0.25$, as estimated using the method introduced
in Appendix\ \ref{appendix:error_contributions}. In contrast, the
$\left|1_{B}1_{A}\right\rangle \leftrightarrow\left|2_{B}0_{A}\right\rangle $
rotation experiences a larger detuning $\left|\delta_{B}\right|=\left|\alpha_{A}+\alpha_{B}\right|=2\pi\times426\,\mathrm{MHz}$,
resulting in a shorter rotation period $\pi/\sqrt{(\sqrt{2}g_{AB})^{2}+(\delta_{B}/2)^{2}}\approx2.3\,\mathrm{ns}$
and a smaller maximum $\left|1_{B}1_{A}\right\rangle \rightarrow\left|2_{B}0_{A}\right\rangle $
leakage error $\frac{1}{4}\cdot(\sqrt{2}g_{AB})^{2}/[(\sqrt{2}g_{AB})^{2}+(\delta_{B}/2)^{2}]\approx1.0\times10^{-2}$.
Thus, the Qt-2 leakage error is dominated by the $\left|1_{B}1_{A}\right\rangle \rightarrow\left|0_{B}2_{A}\right\rangle $
leakage error, while the $\left|1_{B}1_{A}\right\rangle \rightarrow\left|2_{B}0_{A}\right\rangle $
leakage error becomes appreciable only when the former reaches its
minima at $t_{f}=n\pi/(\sqrt{2}g_{AB})\approx n\times11.6\,\mathrm{ns}$
for positive integers $n$. These analytical estimates are in excellent
agreement with the simulated Qt-2 leakage shown in Fig.\ \ref{fig:Drift_H_evolution}(d).
Additionally, we also demonstrate the minor leakage contributions
from the Qt-3 to Qt-6 leakage errors in Fig.\ \ref{fig:Drift_H_evolution}(d).
The jump-2-quanta Qt-3 and Qt-4 leakage errors, involving a single
jump-2-quanta coupling described by the Hamiltonian $\mathcal{H}_{I}$
in Eq.\ (\ref{eq:full_ideal_H}), correspond to the state probability
leakages from Qt-1 states $\left\{ \left|0_{B}1_{A}\right\rangle ,\left|1_{B}0_{A}\right\rangle \right\} $
to Qt-3 states and from the Qt-2 state $\left|1_{B}1_{A}\right\rangle $
to Qt-4 states, as shown in Figs.\ \ref{fig:Even_Odd_Quanta_Paths}(a)
and \ref{fig:Even_Odd_Quanta_Paths}(b), respectively. These jump-2-quanta
leakage errors scale as $g_{AB}^{2}/\Delta_{A}^{2}$ and are on the
order of $10^{-5}$. The jump-4-quanta Qt-5 and Qt-6 leakage errors
involve two jump-2-quanta rotations via the leakage pathways from
Qt-1 states $\left\{ \left|0_{B}1_{A}\right\rangle ,\left|1_{B}0_{A}\right\rangle \right\} $
to Qt-5 states and from the Qt-2 state $\left|1_{B}1_{A}\right\rangle $
to Qt-6 states, as shown in Figs.\ \ref{fig:Even_Odd_Quanta_Paths}(a)
and \ref{fig:Even_Odd_Quanta_Paths}(b), respectively. These jump-4-quanta
leakage errors scale as $g_{AB}^{4}/\Delta_{A}^{4}$ and are on the
order of $10^{-9}$.

In summary, the analysis of Fig.\ \ref{fig:Drift_H_evolution} shows
that the dominant contributions to the ideal CZ gate infidelity $J_{1}$
are the phase error, the Qt-2 leakage error, and the $\left|0_{B}1_{A}\right\rangle \leftrightarrow\left|1_{B}0_{A}\right\rangle $
swap error. Therefore, theoretically, by minimizing the cost function
$J_{1}$---simulated using only the Qt-0, Qt-1, and Qt-2 states---through
optimization of the qubit frequencies $\Delta_{A}$ and $\Delta_{B}$,
these dominant errors can be synchronized to their minima at the gate
time $t_{f}$, yielding an ideal CZ gate infidelity $J_{1}$, limited
by the Qt-3 and Qt-4 leakage errors, on the order of $10^{-5}$. This
procedure is referred to as the Qt-2 $J_{1}$ optimization. To achieve
$J_{1}<10^{-5}$, the optimization must include the additional Qt-3
and Qt-4 leakage errors in the cost function $J_{1}$ by expanding
the simulation subspace up to Qt-4 states---this procedure is referred
to as the Qt-4 $J_{1}$ optimization. Similarly, to reach even lower
infidelity, $J_{1}<10^{-9}$, the procedure should be further extended
to the Qt-6 $J_{1}$ optimization, which expands the simulation subspace
up to Qt-6 states to incorporate the Qt-5 and Qt-6 leakage errors
into the cost function.

\subsection{Control method}

\label{subsec:Control_scheme}

In this section, we present a control method developed to suppress
contributions to the CZ gate infidelity arising from intrinsic errors
(Sec.\ \ref{sec:High_fidelity_CZ_gate}), decoherence (Sec.\ \ref{sec:robust_against_noises}),
and system parameter uncertainties \citep{Huang2026a}. This method
consists of: designing a single-shot smooth flux pulse $\Phi(t)$,
formulating a cost function tailored to the target error sources,
and optimizing the control parameters to minimize this cost function,
thereby achieving high-fidelity and robust CZ gates.

The designed flux pulse $\Phi(t)$ consists of a Pull-Back-Pulse (PBP),
$\Phi^{\mathrm{PBP}}(t)$, and a Fine-Tune-Pulse (FTP), $\Phi^{\mathrm{FTP}}(t)$,
such that $\Phi(t)=\Phi^{\mathrm{PBP}}(t)+\Phi^{\mathrm{FTP}}(t)$.
The role of $\Phi^{\mathrm{PBP}}(t)$ is to adjust the detunings $\delta_{A/B}$
to their target values, thereby bringing the system to a configuration
favorable for the optimization tasks, while $\Phi^{\mathrm{FTP}}(t)$
is constructed on top of $\Phi^{\mathrm{PBP}}(t)$ to further refine
the gate performance. The Pull-Back-Pulse $\Phi^{\mathrm{PBP}}(t)$
is expressed as

\begin{align}
 & \Phi^{\mathrm{PBP}}(t)=\nonumber \\
 & \Phi_{\mathrm{Idle}}+(\Phi_{i}-\Phi_{\mathrm{Idle}})\sin^{2}(\frac{\pi}{2}\cdot\dfrac{t}{t_{\mathrm{PB}}}),\;0\leq t\leq t_{\mathrm{PB}}.\label{eq:PBP_pulse_s1}\\
 & \Phi_{i},\;t_{\mathrm{PB}}\leq t\leq t_{\mathrm{PB}}+t_{\mathrm{FT}}.\label{eq:PBP_pulse_s2}\\
 & \Phi_{\mathrm{Idle}}+(\Phi_{i}-\Phi_{\mathrm{Idle}})\sin^{2}(\frac{\pi}{2}\cdot\dfrac{t-t_{\mathrm{PB}}-t_{\mathrm{FT}}}{t_{\mathrm{PB}}}+\frac{\pi}{2}),\label{eq:PBP_pulse_s3}\\
 & \;t_{\mathrm{PB}}+t_{\mathrm{FT}}\leq t\leq t_{f}.\nonumber 
\end{align}
It modulates the qubit frequency via Eq.\ (\ref{eq:freq_definition}),
pulling it from the idle frequency $\omega_{\mathrm{Idle}}$ (corresponding
to the idle flux $\Phi_{\mathrm{Idle}}$) to the interaction frequency
$\omega_{i}$ (corresponding to the interaction flux $\Phi_{i}$)
over a time interval $t_{\mathrm{PB}}$ {[}Eq.\ (\ref{eq:PBP_pulse_s1}){]},
maintaining the interaction frequency $\omega_{i}$ for a duration
$t_{\mathrm{FT}}$ {[}Eq.\ (\ref{eq:PBP_pulse_s2}){]}, and finally
pulling the frequency back to $\omega_{\mathrm{Idle}}$ over a second
interval $t_{\mathrm{PB}}$ {[}Eq.\ (\ref{eq:PBP_pulse_s3}){]}.
The total gate time is thus $t_{f}=t_{\mathrm{FT}}+2t_{\mathrm{PB}}$.
Here, we employ $\sin^{2}$ functions to connect the idle flux $\Phi_{\mathrm{Idle}}$
and the interaction flux $\Phi_{i}$ in Eqs.\ (\ref{eq:PBP_pulse_s1})
and (\ref{eq:PBP_pulse_s3}), ensuring that the first time derivative
$d\Phi^{\mathrm{PBP}}(t)/dt$ vanishes at the pulse-connection points
$t=$ $0$, $t_{\mathrm{PB}}$, $t_{\mathrm{PB}}+t_{\mathrm{FT}}$,
and $t_{f}$. This guarantees smooth transitions both within $\Phi^{\mathrm{PBP}}(t)$
and between $\Phi^{\mathrm{PBP}}(t)$ and the pulses of the preceding
and following gates. Additionally, we introduce the Pull-Back-Pulse
ratio, $\mathrm{PBPr}\equiv t_{f}/t_{\mathrm{PB}}$. A larger $\mathrm{PBPr}$
exacerbates filtering effects due to steeper rise and fall times associated
with a shorter $t_{\mathrm{PB}}$, whereas a smaller $\mathrm{PBPr}$
limits the Fine-Tune-Pulse tuning duration $t_{\mathrm{FT}}$, thereby
degrading optimization performance. In this work and its companion
paper \citep{Huang2026a}, we adopt $\mathrm{PBPr}=32$ for $\Phi^{\mathrm{PBP}}(t)$.

The Fine-Tune-Pulse $\Phi^{\mathrm{FTP}}(t)$ is activated at $t=t_{\mathrm{PB}}$
and deactivated at $t=t_{\mathrm{PB}}+t_{\mathrm{FT}}$, providing
pulse modulations around the interaction flux $\Phi_{i}$ in $\Phi^{\mathrm{PBP}}(t)$
{[}Eq.\ (\ref{eq:PBP_pulse_s2}){]}. It is constructed as a superposition
of $k_{\mathrm{max}}$ smooth flux pulses $\Phi_{k}^{\mathrm{FTP}}(t)$,
as expressed in Eq.\ (\ref{eq:FTP_pulse_sum}). Each $\Phi_{k}^{\mathrm{FTP}}(t)$
is initially defined by a sin function symmetric with respect to the
midpoint of the gate time, with frequency $(2k-1)\pi/t_{\mathrm{FT}}$
and an independent amplitude $a_{k}$ {[}Eqs.\ (\ref{eq:FTP_pulse_s1}-\ref{eq:FTP_pulse_s3}){]}.
To ensure smooth connections to $\Phi^{\mathrm{PBP}}(t)$ at $t=$$t_{\mathrm{PB}}$
and $t=t_{\mathrm{PB}}+t_{\mathrm{FT}}$, the sin functions in the
first and last $\pi/2$ segments of $\Phi_{k}^{\mathrm{FTP}}(t)$
are replaced by $\sin^{2}$ functions {[}Eqs.\ (\ref{eq:FTP_pulse_s1})
and (\ref{eq:FTP_pulse_s3}){]}, while the central segment retains
the original sin form {[}Eq.\ (\ref{eq:FTP_pulse_s2}){]}.

\begin{align}
 & \Phi^{\mathrm{FTP}}(t)=\Sigma_{k=1}^{k_{\mathrm{max}}}\Phi_{k}^{\mathrm{FTP}}(t)\label{eq:FTP_pulse_sum}\\
 & \Phi_{k}^{\mathrm{FTP}}(t)=\nonumber \\
 & a_{k}\cdot\sin^{2}\left((2k-1)\pi\frac{(t-t_{\mathrm{PB}})}{t_{\mathrm{FT}}}\right),\;t_{\mathrm{PB}}\leq t\leq t_{1,k}.\label{eq:FTP_pulse_s1}\\
 & a_{k}\cdot\sin\left((2k-1)\pi\frac{(t-t_{\mathrm{PB}})}{t_{\mathrm{FT}}}\right),\;t_{1,k}\leq t\leq t_{2,k}.\label{eq:FTP_pulse_s2}\\
 & a_{k}\cdot\sin^{2}\left((2k-1)\pi\frac{(t-t_{\mathrm{PB}})}{t_{\mathrm{FT}}}\right),\;t_{2,k}\leq t\leq t_{\mathrm{PB}}+t_{\mathrm{FT}}.\label{eq:FTP_pulse_s3}
\end{align}
Here $t_{1,k}=t_{\mathrm{PB}}+t_{\mathrm{\pi/2},k}$, $t_{2,k}=t_{\mathrm{PB}}+t_{\mathrm{FT}}-t_{\mathrm{\pi/2},k}$,
and $t_{\mathrm{\pi/2},k}=\frac{1}{2(2k-1)}t_{\mathrm{FT}}$. At $t=t_{1,k}$
and $t_{2,k}$, which are the connection points between the $\sin^{2}$
and $\sin^{1}$ segments of $\Phi_{k}^{\mathrm{FTP}}(t)$, the first
time derivative $d\Phi_{k}^{\mathrm{FTP}}(t)/dt$ vanishes, ensuring
that $\Phi_{k}^{\mathrm{FTP}}(t)$ is smooth within the pulse. Consequently,
the Fine-Tune-Pulse $\Phi^{\mathrm{FTP}}(t)$ in Eq.\ (\ref{eq:FTP_pulse_sum})
is smooth as well. Combined with the smooth Pull-Back-Pulse $\Phi^{\mathrm{PBP}}(t)$,
our designed flux pulse $\Phi(t)=\Phi^{\mathrm{FTP}}(t)+\Phi^{\mathrm{PBP}}(t)$
is smooth throughout the gate time $t_{f}$ and connects smoothly
to the flux pulses of the preceding and subsequent gates.

We treat the interaction fluxes $\Phi_{i,A/B}$ in the Pull-Back-Pulse
$\Phi_{A/B}^{\mathrm{PBP}}(t)$ and the amplitude sets $\left\{ a_{k,A/B}\right\} _{k=1}^{k_{\mathrm{max}}}$
in the Fine-Tune-Pulse $\Phi_{A/B}^{\mathrm{FTP}}(t)$ for qubits
$Q_{A/B}$ as the control parameters. These parameters are optimized
using the gradient-free Nelder-Mead (NM) algorithm \citep{NelderMead1965}
to minimize a cost function constructed for the target error contributions.
For more challenging optimization landscapes, corresponding to more
complex cost-function topographies, we employ the repeating-NM algorithm
\citep{HuangGoan2017} to escape local minima and achieve improved
performance.

To suppress intrinsic errors, we set the ideal gate infidelity $J_{1}$
in Eq.\ (\ref{eq:J1_definition}) as the cost function for optimization.
We refer to this procedure as the Qt-$n$ $J_{1}$ optimization, reflecting
that $J_{1}$ is evaluated by simulating the ideal propagator $U_{I}(t)$
of the ideal Hamiltonian $\mathcal{H}_{I}(t)$ {[}Eq.\ (\ref{eq:full_ideal_H}){]},
expressed in the subspace spanned by the Qt-0 to Qt-$n$ states. The
simulation uses the experimental system parameters $g_{AB}$, $\alpha_{A/B}$,
and $\omega_{0,A/B}$ as listed in Sec.\ \ref{subsec:Drift_H_intrinsic_error},
together with the idle qubit frequencies $\omega_{\mathrm{Idle},A/B}=2\pi\times5.5/4.7\,\mathrm{GHz}$,
all corresponding to the settings employed in the experimental CZ
gate implementation of Ref.\ \citep{BarendsKellyMegrantEtAl2014}.
We first apply the Qt-$n$ $J_{1}$ optimization to symmetric transmons
with $a_{E_{J}}=1$. Accordingly, we set $d=0$ in Eq.\ (\ref{eq:freq_definition})
for the conversion between the fluxes $\Phi_{A/B}$ and the qubit
frequencies $\Delta_{A/B}$. In Sec.\ \ref{subsec:Optimization_asymmetric_transmon},
we extend the optimization to asymmetric transmons with $a_{E_{J}}>1$.

For the Qt-2 $J_{1}$ optimization, aimed at suppressing the dominant
intrinsic errors discussed in Sec.\ \ref{subsec:Drift_H_intrinsic_error},
we set $\Phi_{A}(t)=\Phi_{A}^{\mathrm{PBP}}(t)+\Phi_{A}^{\mathrm{FTP}}(t)$
and $\Phi_{B}(t)=\Phi_{B}^{\mathrm{PBP}}(t)$. The Fine-Tune-Pulse
$\Phi_{A}^{\mathrm{FTP}}(t)$ contains $10$ control parameters, $\left\{ a_{k,A}\right\} _{k=1}^{k_{\mathrm{max}}=10}$,
while the Pull-Back-Pulse $\Phi_{A/B}^{\mathrm{PBP}}(t)$ are initialized
with the interaction fluxes $\Phi_{i,A/B}$, corresponding to the
interaction frequencies $\omega_{i,A/B}=2\pi\times4.7/4.7\,\mathrm{GHz}$.
Furthermore, following the experimental procedure, a calibrated single-qubit-Z
gate $U_{Z}$ {[}Eq.\ (\ref{eq:local_phase_compensated}){]} is inserted
after $U_{I,4\times4}(t_{f})$ to compensate for relative local phases.
In the simulation, we assume that $U_{Z}$ introduces no additional
error, since in experiment it can be implemented as virtual-Z gates
\citep{McKay2017}, which have zero duration and therefore contribute
no error when phase calibration and compensation are precise. For
the Qt-4 $J_{1}$ optimization, we enlarge the simulation subspace
and adopt the same optimization settings as in the Qt-2 case to further
reduce $J_{1}$. By contrast, for the auto-compensating (auto-c) Qt-4
$J_{1}$ optimization, which aims to compensate for relative local
phases via pulse modulation, the single-qubit-Z gate $U_{Z}$ is not
included in the simulation. To address the increased optimization
load associated with this additional task, the Fine-Tune-Pulse $\Phi_{B}^{\mathrm{FTP}}(t)$
for $Q_{B}$, with 10 control parameters $\left\{ a_{k,B}\right\} _{k=1}^{k_{\mathrm{max}}=10}$,
is also activated, so that $\Phi_{B}(t)=\Phi_{B}^{\mathrm{PBP}}(t)+\Phi_{B}^{\mathrm{FTP}}(t)$.
All other optimization settings remain identical to those of the Qt-4
$J_{1}$ optimization.

To explore lower $J_{1}$, we implement two additional optimization
schemes. First, to avoid local minima in the cost function and achieve
lower $J_{1}$, we implement the rp-Qt-4 $J_{1}$ optimization by
replacing the original NM algorithm in the Qt-4 $J_{1}$ optimization
with the repeating-NM algorithm \citep{HuangGoan2017}. Second, to
enhance the control resources for further lowering $J_{1}$, we increase
the number of control parameters in $\Phi_{A}^{\mathrm{FTP}}(t)$
from $10$ to $20$ when implementing the rp-Qt-4 $J_{1}$ optimization.
The resulting optimized ideal CZ gate infidelities $J_{1}$ for the
various schemes described above are shown in Fig.\ \ref{fig:J1_optimization}(a),
and the corresponding detailed intrinsic errors for selected schemes
are presented in Fig.\ \ref{fig:J1_op_detailed_analysis}. These
$J_{1}$ values and intrinsic errors are simulated using the ideal
Hamiltonian $\mathcal{H}_{I}(t)$ {[}Eq.\ (\ref{eq:full_ideal_H}){]}
with $7$ energy levels per qubit, encompassing all Qt-0 to Qt-6 states,
to ensure sufficient numerical convergence.

\subsection{Optimization results}

\label{subsec:Optimization_results}

\begin{figure}
\includegraphics[scale=0.52]{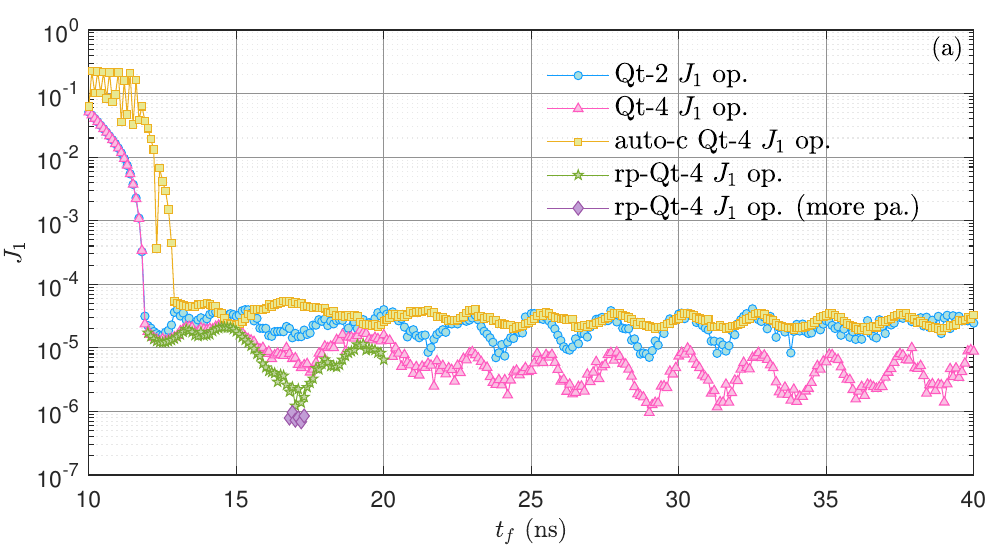}

\includegraphics[scale=0.52]{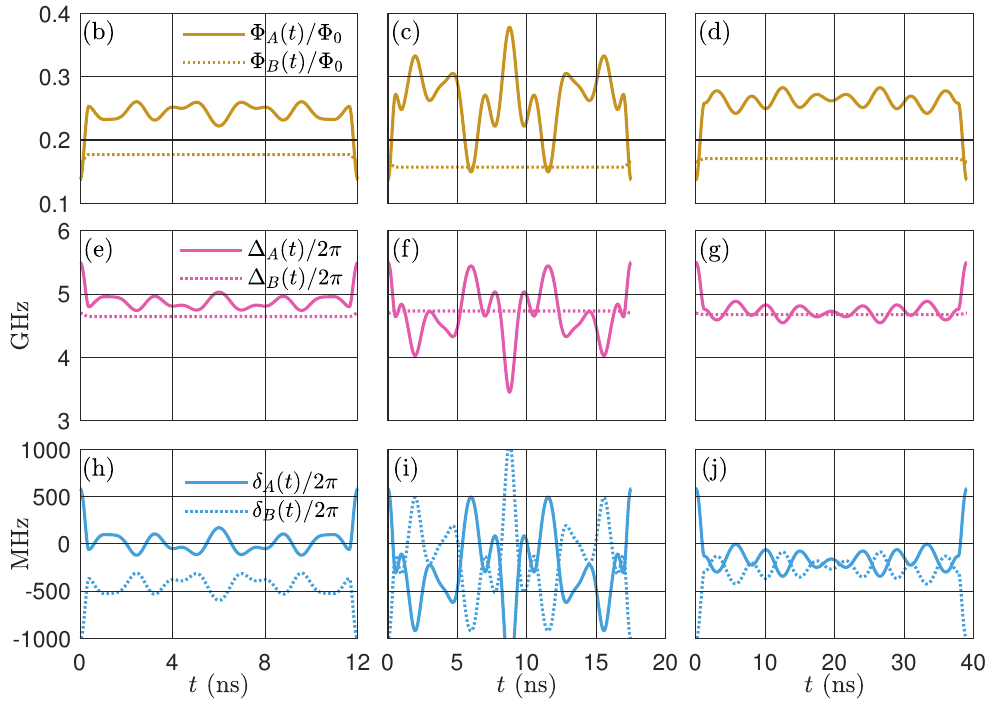}

\caption{Optimization results. (a) Optimized $J_{1}$ as a function of gate
time $t_{f}$ obtained from different optimization schemes: the Qt-2
$J_{1}$ optimization (blue-circle line), the Qt-4 $J_{1}$ optimization
(pink-triangle line), the auto-c Qt-4 $J_{1}$ optimization (yellow-square
line), the rp-Qt-4 $J_{1}$ optimization (green-pentagram line), and
the rp-Qt-4 $J_{1}$ optimization with an increased number of control
parameters in the Fine-Tune-Pulse $\Phi_{A}^{\mathrm{FTP}}(t)$ (purple-diamond
line). (b--d) Optimized flux pulses $\Phi_{A}(t)/\Phi_{0}$ (solid
yellow) and $\Phi_{B}(t)/\Phi_{0}$ (dashed yellow) as a function
of time $t$ for $t_{f}=12\,\mathrm{ns}$, $17.5\,\mathrm{ns}$, and
$39\,\mathrm{ns}$, respectively, obtained from the Qt-4 $J_{1}$
optimization. (e--g) Corresponding qubit frequencies $\Delta_{A}(t)$
(solid pink) and $\Delta_{B}(t)$ (dashed pink). (h--j) Corresponding
detunings $\delta_{A}(t)$ (solid blue) and $\delta_{B}(t)$ (dashed
blue).}

\label{fig:J1_optimization} 
\end{figure}

As shown in Fig.\ \ref{fig:J1_optimization}(a), for the Qt-2 and
Qt-4 $J_{1}$ optimizations, the minimum gate time of the optimized
CZ gate with $J_{1}<10^{-4}$ is $11.9\,\mathrm{ns}$, which is very
close to the theoretical speed limit $t_{f}=\pi/(\sqrt{2}g_{AB})\approx11.6\,\mathrm{ns}$.
By contrast, for the auto-c Qt-4 $J_{1}$ optimization, the minimum
gate time with $J_{1}<10^{-4}$ increases to $12.9\,\mathrm{ns}$,
reflecting an additional temporal overhead required to compensate
for the relative local phases. Through the Qt-4 $J_{1}$ optimization,
an optimized $J_{1}\approx1.80\times10^{-5}$ is achieved at $t_{f}=12\,\mathrm{ns}$,
using the optimized $\Phi_{A/B}(t)$ pulses shown in Fig.\ \ref{fig:J1_optimization}(b).
The corresponding qubit frequencies $\Delta_{A/B}(t)$ in Fig.\ \ref{fig:J1_optimization}(e)
are first pulled from the idle frequencies $\omega_{\mathrm{Idle},A/B}=2\pi\times5.5/4.7\,\mathrm{GHz}$
down to the optimized interaction frequencies $\omega_{i,A/B}\approx2\pi\times4.8/4.64\,\mathrm{GHz}$
by the Pull-Back-Pulse $\Phi_{A/B}^{\mathrm{PBP}}(t)$. Then, $\Delta_{A}(t)$
vibrates around $\omega_{i,A}$ by the Fine-Tune-Pulse $\Phi_{A}^{\mathrm{FTP}}(t)$
while $\Delta_{B}(t)$ remains fixed at $\omega_{i,B}$, before both
frequencies $\Delta_{A/B}(t)$ are pulled back to $\omega_{\mathrm{Idle},A/B}$
to complete the CZ gate. During the coupling period, the optimized
detunings $\delta_{A}(t)$ and $\delta_{B}(t)$ vibrate around $2\pi\times0\,\mathrm{MHz}$
and $-2\pi\times400\,\mathrm{MHz}$, respectively, to maximize the
$\left|1_{B}1_{A}\right\rangle \leftrightarrow\left|0_{B}2_{A}\right\rangle $
interaction, as shown in Fig.\ \ref{fig:J1_optimization}(h).

Throughout the $t_{f}$ region shown in Fig.\ \ref{fig:J1_optimization}(a),
values of $J_{1}\lesssim10^{-5}$ are attainable using the Qt-2 $J_{1}$
optimization, indicating that the dominant error contributions exceeding
$10^{-5}$---including the phase error, the Qt-2 leakage error, and
the $\left|0_{B}1_{A}\right\rangle \leftrightarrow\left|1_{B}0_{A}\right\rangle $
swap error identified in the drift-Hamiltonian case shown in Fig.\ \ref{fig:Drift_H_evolution}---can
all be effectively suppressed. For the Qt-4 $J_{1}$ optimization,
the Qt-3 and Qt-4 leakage pathways are also included to more fully
capture the leakage dynamics and more effectively synchronize their
induced errors toward their minima, allowing the minimum optimized
$J_{1}$ to reach $\sim10^{-6}$, representing roughly an order-of-magnitude
improvement compared with the Qt-2 $J_{1}$ optimization. For the
auto-c Qt-4 $J_{1}$ optimization, the minimum optimized $J_{1}$
increases to approximately $2\times10^{-5}$, reflecting the penalty
incurred by the additional task of auto-compensating for relative
local phases under limited control resources.

Zooming into Fig.\ \ref{fig:J1_optimization}(a), the optimized $J_{1}$
over the $t_{f}$ region can be divided into three distinct regimes,
each corresponding to a different error mechanism. In the first regime,
where the gate time $t_{f}$ is shorter than the minimum achievable
gate time with $J_{1}<10^{-4}$, the optimized $J_{1}$ increases
rapidly as $t_{f}$ is reduced. For both the Qt-2 and Qt-4 $J_{1}$
optimizations, the resulting ramping behavior of $J_{1}$ coincides
and is dominated by the Qt-2 leakage error, as shown in Fig.\ \ref{fig:J1_optimization}(a)
and Fig.\ \ref{fig:J1_op_detailed_analysis}. Under limited control
resources, shortening the gate time restricts the optimization's
ability to satisfy multiple synchronization tasks, forcing it to prioritize
the dominant ones. Consequently, the Qt-2 and Qt-4 $J_{1}$ optimizations
converge to the same optimal solution, independent of whether higher-level
states are included. In both cases, the Qt-2 leakage error is primarily
due to the $\left|1_{B}1_{A}\right\rangle $$\leftrightarrow$$\left|0_{B}2_{A}\right\rangle $
leakage pathway and exhibits behavior similar to the drift-Hamiltonian
case in Fig.\ \ref{fig:Drift_H_evolution}(d), where the Qt-2 leakage
error is also dominated by the same leakage channel and increases
toward its maximum as $t_{f}$ decreases from the first on-resonance
point at $t_{f}=1\pi/(\sqrt{2}g_{AB})\approx11.6\,\mathrm{ns}$. For
the auto-c Qt-4 $J_{1}$ optimization, the reduced gate time further
limits the additional synchronization task associated with phase compensation.
Consequently, the increasing optimized $J_{1}$ is driven jointly
by the Qt-2 leakage error and the phase error, as illustrated in Fig.\ \ref{fig:J1_op_detailed_analysis}.

In the second regime, where the gate time $t_{f}$ lies between the
minimum achievable gate time under the constraint $J_{1}<10^{-4}$
and $\sim20\,\mathrm{ns}$, we refer to this parameter range as the
hard region. Although the gate time in this regime exceeds that of
the first regime, multi-task synchronization remains challenging because
the available time is still insufficient. This increased optimization
difficulty manifests itself in more pronounced vibrations of the optimized
Fine-Tune-Pulse $\Phi_{A}^{\mathrm{FTP}}(t)$ required to satisfy
the synchronization conditions. These strong vibrations are evident
in Fig.\ \ref{fig:J1_optimization}(c) for the optimized $\Phi_{A}(t)$
at $t_{f}=17.5\,\mathrm{ns}$, corresponding to the lowest $J_{1}\approx4.14\times10^{-6}$
achieved by the Qt-4 $J_{1}$ optimization in the hard region. The
corresponding qubit frequencies $\Delta_{A/B}(t)$ and detunings $\delta_{A/B}(t)$
are illustrated in Figs.\ \ref{fig:J1_optimization}(f) and \ref{fig:J1_optimization}(i),
respectively. The minimum achievable $J_{1}$ in the hard region can
be further reduced to $1.22\times10^{-6}$ at $t_{f}=17\,\mathrm{ns}$
using the rp-Qt-4 $J_{1}$ optimization. Furthermore, increasing the
number of control parameters---specifically expanding $k_{\mathrm{max}}$
from $10$ to $20$ in $\left\{ a_{k,A}\right\} _{k=1}^{k_{\mathrm{max}}}$
for the Fine-Tune-Pulse $\Phi_{A}^{\mathrm{FTP}}(t)$ when implementing
the rp-Qt-4 $J_{1}$ optimization---allows $J_{1}$ to be suppressed
below $10^{-6}$, reaching $6.86\times10^{-7}$ at $t_{f}=17.2\,\mathrm{ns}$.

In the third regime, corresponding to gate times $t_{f}\gtrsim20\,\mathrm{ns}$,
a periodic-like structure in the optimized $J_{1}$ emerges. This
behavior is dominantly contributed by the jump-2-quanta leakage errors,
including the Qt-3 and Qt-4 leakage errors, and/or the $\left|0_{B}0_{A}\right\rangle \leftrightarrow\left|1_{B}1_{A}\right\rangle $
swap error, as illustrated in Fig.\ \ref{fig:J1_op_detailed_analysis}.
The observed periodicity originates from the presence of comparable
detuning $\delta_{A}$ and weak Fine-Tune-Pulse $\Phi_{A/B}^{\mathrm{FTP}}(t)$
vibration in this regime, resulting in system dynamics that closely
resemble evolution under the drift Hamiltonian. The emergence of comparable
$\delta_{A}$ and weak $\Phi_{A/B}^{\mathrm{FTP}}(t)$ vibrations
is associated with natural synchronization. Under specific combinations
of $\delta_{A}$ and $t_{f}$, the dominant error channels become
naturally synchronized, leading to local minima of the cost function
$J_{1}$ even in the absence of the Fine-Tune-Pulse $\Phi_{A/B}^{\mathrm{FTP}}(t)$.
Such natural synchronization points are identified at gate times $t_{f}\approx12.5$,
$18.5$, $23.0$, $27.1$, $32.7$, and $35.8\,\mathrm{ns}$, with
the corresponding detunings $\delta_{A}/2\pi\approx0$, $-61$, $-95$,
$-119$, $-197$, and $-4\,\mathrm{MHz}$, respectively, as obtained
by simulating $J_{1}$ using only the Pull-Back-Pulse $\Phi_{A/B}^{\mathrm{PBP}}(t)$.
The first and last synchronization points are close to the local minima
of $J_{1}$ at $t_{f}\approx12\,\mathrm{ns}$ and $36\,\mathrm{ns}$,
respectively, observed in the drift-Hamiltonian evolution with fixed
detuning $\delta_{A}=0$, as shown in Fig.\ \ref{fig:Drift_H_evolution}(a).
Among these natural synchronization conditions, the point at $t_{f}\approx32.7\,\mathrm{ns}$
with $\delta_{A}/2\pi\approx-197\,\mathrm{MHz}$ lies closest to the
initial optimization detuning $\delta_{A}/2\pi=-214\,\mathrm{MHz}$.
Consequently, the optimization procedure first steers the detuning
$\delta_{A}$ toward this natural synchronization point to reach a
local minimum of the cost function, and subsequently employs the Fine-Tune-Pulse
$\Phi_{A/B}^{\mathrm{FTP}}(t)$ to further suppress residual errors.
For gate time increasingly shifted from the natural synchronization
point at $t_{f}\approx32.7\,\mathrm{ns}$, stronger Fine-Tune-Pulse
$\Phi_{A/B}^{\mathrm{FTP}}(t)$ vibrations are required to compensate
for the mismatch in gate time. As a result, the weakest $\Phi_{A/B}^{\mathrm{FTP}}(t)$
vibration is observed near the natural synchronization point at $t_{f}\approx32.7\,\mathrm{ns}$.
Overall, the Fine-Tune-Pulse $\Phi_{A/B}^{\mathrm{FTP}}(t)$ vibration
remains substantially weaker in this regime compared to the hard region.
The reduced $\Phi_{A}^{\mathrm{FTP}}(t)$ vibrations for the optimized
CZ gate obtained by the Qt-4 $J_{1}$ optimization at $t_{f}=39\,\mathrm{ns}$
are shown in Fig.\ \ref{fig:J1_optimization}(d). The corresponding
optimized qubit frequencies $\Delta_{A/B}(t)$ and detunings $\delta_{A/B}(t)$
are illustrated in Figs.\ \ref{fig:J1_optimization}(g) and \ref{fig:J1_optimization}(j),
respectively.

\subsection{Detailed infidelity contribution analysis}

\label{subsec:Detailed_infidelity_contribution}

\begin{figure}
\includegraphics[scale=0.52]{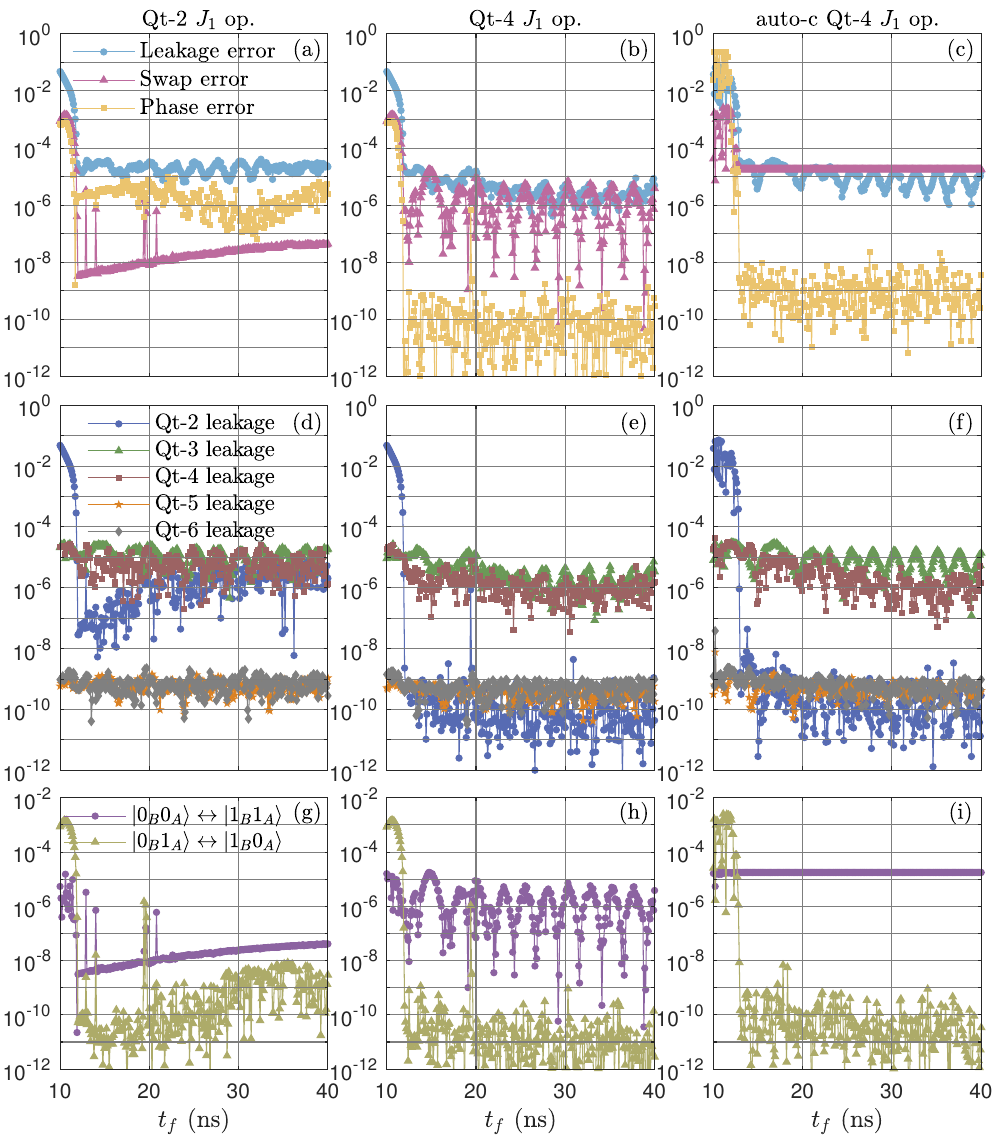}

\caption{Error contributions to the optimized $J_{1}$: The first column (panels
(a), (d), and (g)) for the Qt-2 $J_{1}$ optimization, the second
column (panels (b), (e), and (h)) for the Qt-4 $J_{1}$ optimization,
and the third column (panels (c), (f), and (i)) for the auto-c Qt-4
$J_{1}$ optimization. The first row (panels (a), (b), and (c)) shows
the individual contributions to $J_{1}$ from the leakage error (light-blue-circle
line), the swap error (light-purple-triangle line), and the phase
error (yellow-square line). The second row (panels (d), (e), and (f))
provides a breakdown of the leakage error into the contributions from
the Qt-2, Qt-3, Qt-4, Qt-5, and Qt-6 leakage errors, represented by
dark-blue-circle line, green-triangle line, brown-square line, orange-pentagram
line, and gray-diamond line, respectively. The third row (panels (g),
(h), and (i)) distinguishes the swap error into two components: the
$\left|0_{B}1_{A}\right\rangle \leftrightarrow\left|1_{B}0_{A}\right\rangle $
swap error (purple-circle line) and the $\left|0_{B}0_{A}\right\rangle \leftrightarrow\left|1_{B}1_{A}\right\rangle $
swap error (olive-triangle line).}

\label{fig:J1_op_detailed_analysis} 
\end{figure}

In this section, we examine in detail the error contributions associated
with the optimized $J_{1}$ obtained from the Qt-2 $J_{1}$ optimization,
the Qt-4 $J_{1}$ optimization, and the auto-c Qt-4 $J_{1}$ optimization,
which are discussed sequentially in the following paragraphs.

Following the Qt-2 $J_{1}$ optimization, we observe that for $t_{f}\geq12\,\mathrm{ns}$,
the dominant error contributions in the optimized $J_{1}$ output---including
the Qt-2 leakage error, swap error, and phase error---are suppressed
to the machine limit. The remaining residual error, on the order of
$10^{-10}$, is attributed to a jump-2-quanta leakage process mediated
by pathways from $\left|0_{B}0_{A}\right\rangle $ to Qt-2 states.
These results were obtained from simulations restricted to the subspace
spanned by the Qt-0 to Qt-2 states. When higher-level Qt-3 to Qt-6
states are included in the simulation, additional leakage pathways
become accessible. As shown in Fig.\ \ref{fig:J1_op_detailed_analysis}(a),
these pathways disrupt the synchronization established within the
truncated Qt-0 to Qt-2 subspace, resulting in a degradation of the
leakage error, swap error, and phase error reported for the truncated
subspace.

Specifically, the Qt-2 leakage error at $t_{f}=12\,\mathrm{ns}$ is
degraded to $3\times10^{-8}$ and subsequently increases with increasing
$t_{f}$ to the level of $10^{-6}$ to $10^{-5}$, where the Qt-3
and Qt-4 leakage errors exhibit oscillations, as shown in Fig.\ \ref{fig:J1_op_detailed_analysis}(d).
The rising Qt-2 leakage error interferes with the periodic Qt-3 and
Qt-4 leakage errors in the third regime, thereby causing the periodicity
of the leakage error to vanish for $t_{f}\gtrsim35\,\mathrm{ns}$,
as illustrated in Fig.\ \ref{fig:J1_op_detailed_analysis}(a). The
phase error for $t_{f}\geq12\,\mathrm{ns}$ degrades to $\sim10^{-6}$,
yet a local minimum of around $10^{-7}$ is observed near the natural
synchronization point at $t_{f}\approx32.7\,\mathrm{ns}$ {[}Fig.\ \ref{fig:J1_op_detailed_analysis}(a){]}.
Under this natural synchronization condition, the phase $\phi_{11}$
accumulates linearly toward the target, avoiding an abrupt increase
in the accumulation rate. Consequently, the phase error at this point
exhibits the lowest sensitivity to enhanced leakage pathways, thereby
yielding the minimum phase error.

The swap error, while degraded, stays mostly below $10^{-7}$ for
$t_{f}\geq12\,\mathrm{ns}$ {[}Fig.\ \ref{fig:J1_op_detailed_analysis}(a){]}.
Upon closer inspection of the detailed contributions to the swap error
in Fig.\ \ref{fig:J1_op_detailed_analysis}(g), the $\left|0_{B}0_{A}\right\rangle \leftrightarrow\left|1_{B}1_{A}\right\rangle $
swap error emerges as the dominant contribution, which increases with
$t_{f}$. A minor contribution from the $\left|0_{B}1_{A}\right\rangle \leftrightarrow\left|1_{B}0_{A}\right\rangle $
swap error exhibits a maximum at $t_{f}\approx35\,\mathrm{ns}$, which
is close to the natural synchronization point $t_{f}\approx32.7\,\mathrm{ns}$.
Consequently, when $t_{f}$ is in this vicinity, the optimized detuning
$\delta_{A}$ approaches the natural synchronization condition, $\delta_{A}/2\pi\approx-197\,\mathrm{MHz}$,
while the corresponding optimized $\Phi_{A/B}^{\mathrm{FTP}}(t)$
vibrations are substantially suppressed. These effects collectively
bring the system closer to the resonance condition for the $\left|0_{B}1_{A}\right\rangle \leftrightarrow\left|1_{B}0_{A}\right\rangle $
coupling, where $\delta_{01-10}/2\pi\approx0\,\mathrm{MHz}$ ($\delta_{A}/2\pi\approx-214\,\mathrm{MHz}$).
In this strong-coupling regime, the $\left|0_{B}1_{A}\right\rangle \leftrightarrow\left|1_{B}0_{A}\right\rangle $
swap error becomes highly sensitive to enhanced Qt-3 leakage pathways,
leading to the observed error peak in this region.

In summary, for the Qt-2 $J_{1}$ optimization, subspace truncation
in the simulation effectively reduces computational resources while
maintaining optimized $J_{1}$ below $10^{-4}$ for $t_{f}\geq11.9\,\mathrm{ns}$.
Additionally, the minimum $J_{1}$ can reach $7.01\times10^{-6}$
at $t_{f}=23.8\,\mathrm{ns}$, as shown in Fig.\ \ref{fig:J1_optimization}(a).
Although the Qt-3 to Qt-6 states are excluded from the simulation
for the Qt-2 $J_{1}$ optimization, modest improvements in the associated
Qt-3 to Qt-6 leakage errors are nonetheless observed {[}Fig.\ \ref{fig:J1_op_detailed_analysis}(d){]},
as compared to the drift Hamiltonian evolution results {[}Fig.\ \ref{fig:Drift_H_evolution}(d){]}.
This is attributed to the suppression of the $\left|0_{B}1_{A}\right\rangle \leftrightarrow\left|1_{B}0_{A}\right\rangle $
swap error and Qt-2 leakage error by the Qt-2 $J_{1}$ optimization.

The Qt-4 $J_{1}$ optimization, by incorporating additional Qt-3 to
Qt-4 states into the simulation, significantly improves the phase
error in Fig.\ \ref{fig:J1_op_detailed_analysis}(b), as well as
the same-quanta errors, which include the Qt-2 leakage error in Fig.\ \ref{fig:J1_op_detailed_analysis}(e)
and the $\left|0_{B}1_{A}\right\rangle \leftrightarrow\left|1_{B}0_{A}\right\rangle $
swap error in Fig.\ \ref{fig:J1_op_detailed_analysis}(h), when compared
to the Qt-2 $J_{1}$ optimization. These errors are reduced to below
$10^{-9}$ for nearly all gate times $t_{f}\geq12\,\mathrm{ns}$.
As a result, the optimized $J_{1}$ for $t_{f}\geq12\,\mathrm{ns}$
is primarily influenced by the jump-2-quanta errors, including the
Qt-3 and Qt-4 leakage errors depicted in Fig.\ \ref{fig:J1_op_detailed_analysis}(e),
as well as the $\left|0_{B}0_{A}\right\rangle \leftrightarrow\left|1_{B}1_{A}\right\rangle $
swap error shown in Fig.\ \ref{fig:J1_op_detailed_analysis}(h).
The periodic-like behavior of the optimized $J_{1}$ resulting from
the Qt-4 $J_{1}$ optimization, observed in the third regime for $t_{f}\apprge20\,\mathrm{ns}$
in Fig.\ \ref{fig:J1_optimization}(a), is solely attributable to
these jump-2-quanta errors.

The jump-2-quanta synchronization task is significantly more challenging
than the same-quanta synchronization task. In the case of the Qt-2
$J_{1}$ optimization, there is only one jump-2-quanta synchronization
task, namely the $\left|0_{B}0_{A}\right\rangle \leftrightarrow\left|1_{B}1_{A}\right\rangle $
swap error synchronization. However, for the Qt-4 $J_{1}$ optimization,
there are three such challenging tasks. Consequently, with the same
limited control resources, the $\left|0_{B}0_{A}\right\rangle \leftrightarrow\left|1_{B}1_{A}\right\rangle $
swap error in the Qt-2 $J_{1}$ optimization can be suppressed to
below $10^{-7}$ for almost all $t_{f}\geq12\,\mathrm{ns}$ as shown
in Fig.\ \ref{fig:J1_op_detailed_analysis}(g), whereas the three
jump-2-quanta errors in the Qt-4 $J_{1}$ optimization oscillate around
the order of $10^{-6}$. Nonetheless, the synchronization using additional
information from the Qt-3 to Qt-4 states in the Qt-4 $J_{1}$ optimization
still results in some local minima of the $\left|0_{B}0_{A}\right\rangle \leftrightarrow\left|1_{B}1_{A}\right\rangle $
swap error, shown in Fig.\ \ref{fig:J1_op_detailed_analysis}(h),
being lower than those observed at the corresponding $t_{f}$ by the
Qt-2 $J_{1}$ optimization. For the Qt-3 and Qt-4 leakage errors in
the third regime, on the order of $10^{-6}$, a half-order-of-magnitude
improvement is noticeable compared the orders of $10^{-6}$ to $10^{-5}$
range from the Qt-2 $J_{1}$ optimization in Fig.\ \ref{fig:J1_op_detailed_analysis}(d).
Furthermore, a reduction of approximately one order of magnitude is
seen relative to the order of $10^{-5}$ for the Qt-3 and Qt-4 leakage
errors resulting from the drift Hamiltonian evolution in Fig.\ \ref{fig:Drift_H_evolution}(d).
Additionally, a modest improvement is evident for the Qt-5 and Qt-6
leakage errors, which are suppressed to below $10^{-9}$ for nearly
all $t_{f}$ in Fig.\ \ref{fig:J1_op_detailed_analysis}(e), in comparison
to the Qt-2 $J_{1}$ optimization results in Fig.\ \ref{fig:J1_op_detailed_analysis}(d).

In summary, the Qt-4 $J_{1}$ optimization, by including additional
information from the Qt-3 to Qt-4 states, proves effective in reducing
gate infidelity. As shown in Fig.\ \ref{fig:J1_optimization}(a),
the optimized $J_{1}$ can be suppressed below $10^{-5}$ for $t_{f}\geq20.3\,\mathrm{ns}$,
oscillating between $10^{-5}$ and $10^{-6}$. Furthermore, the minimum
$J_{1}$ can even reach $9.60\times10^{-7}$ at $t_{f}=29\,\mathrm{ns}$.

For the auto-c Qt-4 $J_{1}$ optimization, an additional task is introduced
relative to the Qt-4 $J_{1}$ optimization, which involves minimizing
the phase error arising from the relative local phases. This task
eliminates the need to insert a calibrated single-qubit-Z gate $U_{Z}$
to compensate for relative local phases---a step required in both
Qt-2 and Qt-4 optimizations. However, the added synchronization task
in the auto-c Qt-4 $J_{1}$ optimization inevitably leads to an increased
load, which, in turn, results in a degradation of its performance
compared to the standard Qt-4 $J_{1}$ optimization. The modest degradation
of the same-quanta errors, including the Qt-2 leakage error in Fig.\ \ref{fig:J1_op_detailed_analysis}(f)
and the $\left|0_{B}1_{A}\right\rangle \leftrightarrow\left|1_{B}0_{A}\right\rangle $
swap error in Fig.\ \ref{fig:J1_op_detailed_analysis}(i), is noticeable
when compared to the corresponding errors in the Qt-4 $J_{1}$ optimization
shown in Figs.\ \ref{fig:J1_op_detailed_analysis}(e) and \ref{fig:J1_op_detailed_analysis}(h),
respectively. For the phase error in Fig.\ \ref{fig:J1_op_detailed_analysis}(c),
a degradation of approximately an order of magnitude is observed compared
to the phase error in Fig.\ \ref{fig:J1_op_detailed_analysis}(b)
resulting from the Qt-4 $J_{1}$ optimization. The phase error from
the auto-c Qt-4 $J_{1}$ optimization consists of both the relative
local phase error and the $\phi_{11}$ phase error, whereas the phase
errors from the Qt-2 and Qt-4 $J_{1}$ optimizations are solely due
to the $\phi_{11}$ phase error, as the relative local phase errors
are fully compensated by the calibrated single-qubit-Z gates. Despite
the observed degradation in these errors, they remain below $10^{-8}$
for nearly $t_{f}\gtrsim13\,\mathrm{ns}$, which does not significantly
affect the optimized $J_{1}$. In fact, the optimized $J_{1}$ is
predominantly influenced by the jump-2-quanta errors.

As illustrated in Fig.\ \ref{fig:J1_op_detailed_analysis}(f), the
Qt-4 leakage error for $t_{f}\gtrsim25\,\mathrm{ns}$ is on the order
of $10^{-6}$, exhibiting comparable performance to the jump-2-quanta
errors observed in the Qt-4 $J_{1}$ optimization, as shown in Figs.\ \ref{fig:J1_op_detailed_analysis}(e)
and \ref{fig:J1_op_detailed_analysis}(h). Additionally, the Qt-3
leakage error, also depicted in Fig.\ \ref{fig:J1_op_detailed_analysis}(f),
oscillates between the orders of $10^{-6}$ and $10^{-5}$ for $t_{f}\gtrsim25\,\mathrm{ns}$,
experiencing modest degradation. The $\left|0_{B}0_{A}\right\rangle \leftrightarrow\left|1_{B}1_{A}\right\rangle $
swap error in Fig.\ \ref{fig:J1_op_detailed_analysis}(i) further
degrades to $1.78\times10^{-5}$. The enhanced load in the auto-c
Qt-4 $J_{1}$ optimization causes the jump-2-quanta errors to degrade
in a manner that depends on the number of available jump-2-quanta
pathways. Specifically, the degradation decreases as the number of
available pathways increases. This can be attributed to the fact that
a greater number of pathways provides more degrees of freedom for
tuning the synchronization, thereby making the synchronization task
less challenging. For the $\left|0_{B}0_{A}\right\rangle \leftrightarrow\left|1_{B}1_{A}\right\rangle $
swap error, the degradation is most pronounced, as it is limited to
a single jump-2-quanta pathway, $\left|0_{B}0_{A}\right\rangle \leftrightarrow\left|1_{B}1_{A}\right\rangle $.
In contrast, the case of the Qt-3 leakage error benefits from two
dominant jump-2-quanta pathways, $\left|0_{B}1_{A}\right\rangle \leftrightarrow\left|1_{B}2_{A}\right\rangle $
and $\left|1_{B}0_{A}\right\rangle \leftrightarrow\left|2_{B}1_{A}\right\rangle $,
as illustrated in Fig.\ \ref{fig:Even_Odd_Quanta_Paths}(b), which
help mitigate the degradation. The Qt-4 leakage error for $t_{f}\gtrsim25\,\mathrm{ns}$
does not experience degradation, owing to the presence of the most
dominant jump-2-quanta pathways, $\left|1_{B}1_{A}\right\rangle \leftrightarrow\left|2_{B}2_{A}\right\rangle $,
$\left|1_{B}1_{A}\right\rangle \leftrightarrow\left|0_{B}2_{A}\right\rangle \leftrightarrow\left|1_{B}3_{A}\right\rangle $,
and $\left|1_{B}1_{A}\right\rangle \leftrightarrow\left|2_{B}0_{A}\right\rangle \leftrightarrow\left|3_{B}1_{A}\right\rangle $,
as shown in Fig.\ \ref{fig:Even_Odd_Quanta_Paths}(a).

In summary, for the auto-c Qt-4 $J_{1}$ optimization, although omitting
the additional phase compensation procedure introduces a performance
degradation penalty, the overall optimized $J_{1}$ for $t_{f}\geq12.9\,\mathrm{ns}$
remains below $6\times10^{-5}$, as shown in Fig.\ \ref{fig:J1_optimization}(a).
In the third regime, for $t_{f}\gtrsim20\,\mathrm{ns}$, the weak
periodic-like behavior of optimized $J_{1}$ is attributed to the
oscillating Qt-3 leakage error, with minima around $2\times10^{-5}$,
which are constrained by the constant $\left|0_{B}0_{A}\right\rangle \leftrightarrow\left|1_{B}1_{A}\right\rangle $
swap error.

\subsection{Filtering effects on the optimized CZ gates}

\label{subsec:Filtering_effects}

\begin{figure}[!h]
\includegraphics[scale=0.52]{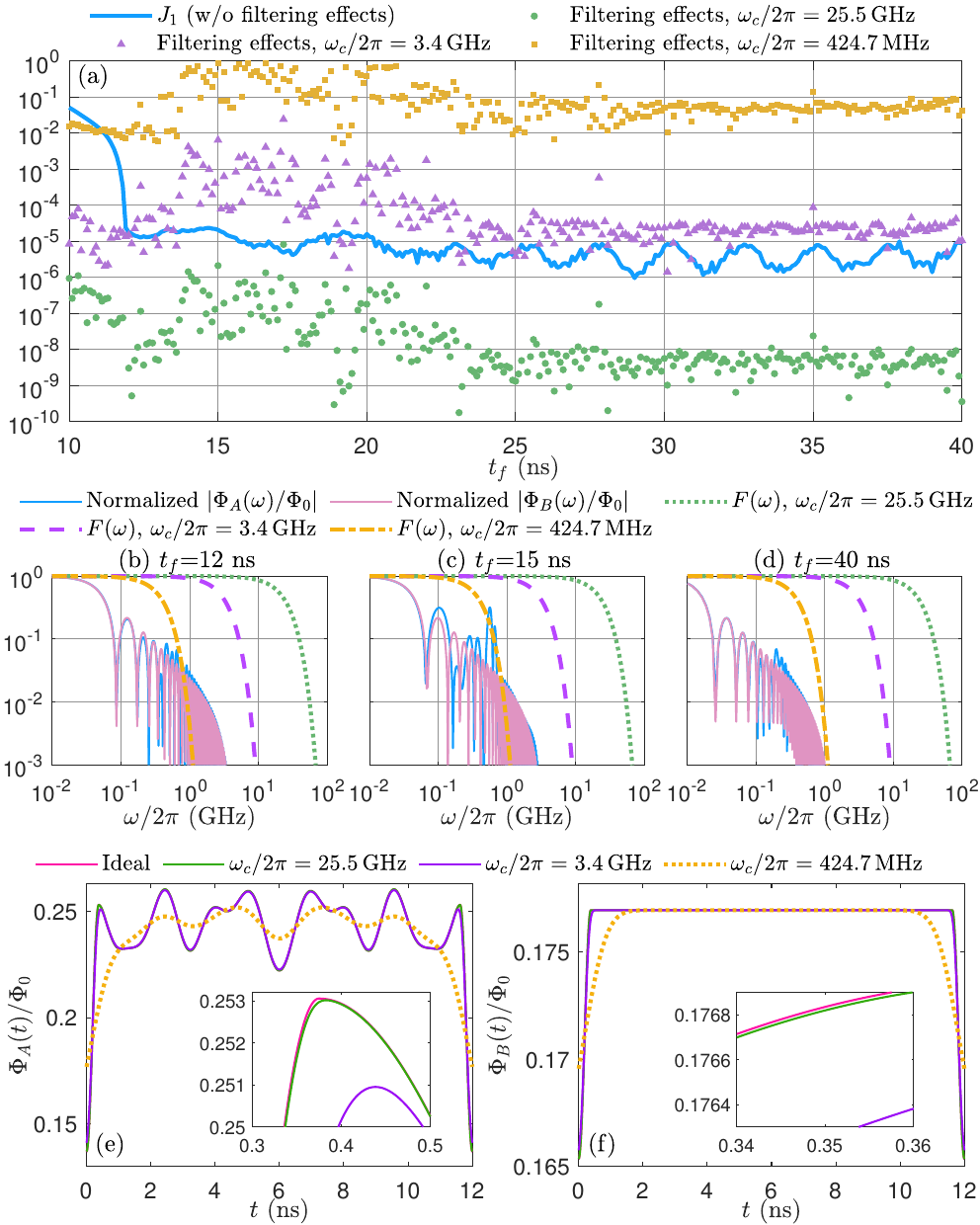}

\caption{Filtering effects on the optimized CZ gates via the Qt-4 $J_{1}$
optimization. Panel (a) shows the ideal $J_{1}$ (bold blue line)
obtained from the Qt-4 $J_{1}$ optimization without filtering effects.
The green circles, purple triangles, and yellow squares represent
the filtering effects on the optimized samples (bold blue line) from
Tektronix AWGs, modeled by a Gaussian filter $F(\omega)=\exp(-\omega^{2}/\omega_{c}^{2})$
with cutoff frequencies $\omega_{c}/2\pi=25.5\,\mathrm{GHz}$, $3.4\,\mathrm{GHz}$,
and $424.7\,\mathrm{MHz}$, respectively. Filtering effects are quantified
by the infidelity enhancement due to pulse distortion from the limited
bandwidth of the AWGs. Panels (b), (c), and (d) show the spectra of
the optimized samples from panel (a) for $t_{f}=12\,\mathrm{ns}$,
$15\,\mathrm{ns}$, and $40\,\mathrm{ns}$, respectively. The blue
and pink lines represent the normalized spectra $\left|\Phi_{A}(\omega)/\Phi_{0}\right|$
and $\left|\Phi_{B}(\omega)/\Phi_{0}\right|$, respectively. The green
dotted lines, purple dashed lines, and yellow dash-dotted lines represent
the Gaussian filters with cutoff frequencies $\omega_{c}/2\pi=25.5\,\mathrm{GHz}$,
$3.4\,\mathrm{GHz}$, and $424.7\,\mathrm{MHz}$, respectively. Panels
(e) and (f) depict the distorted pulses $\Phi_{A}(t)/\Phi_{0}$ and
$\Phi_{B}(t)/\Phi_{0}$, respectively, for the optimized sample with
$t_{f}=12\,\mathrm{ns}$. The magenta lines represent the ideal pulses
without filtering. The green lines, purple lines, and yellow dotted
lines represent the distorted pulses caused by AWGs with cutoff frequencies
$\omega_{c}/2\pi=25.5\,\mathrm{GHz}$, $3.4\,\mathrm{GHz}$, and $424.7\,\mathrm{MHz}$,
respectively. The inset in panel (e) shows an enlargement of the first
pulse peak at $t\approx0.4\,\mathrm{ns}$, and the inset in panel
(f) shows an enlargement of the pulse corner at $t\approx0.35\,\mathrm{ns}$. }

\label{fig:filtering_effects_and filtered_pulses} 
\end{figure}

In the preceding sections, we have demonstrated high-fidelity CZ gates
using optimized control flux pulses $\Phi_{A/B}(t)$. In experimental
practice, these ideal optimized pulses are programmed into arbitrary
waveform generators (AWGs), which synthesize the realistic pulses
used to control the qubits. However, due to the finite bandwidth of
AWGs, the high-frequency components of the ideal pulses are attenuated,
leading to distorted output waveforms. This distortion results in
a reduction in gate fidelity, diverging from the high-fidelity values
predicted by simulations based on the ideal pulses. Here, we refer
to this fidelity degradation as filtering effects. In this section,
we investigate the impact of these filtering effects on the high-fidelity
CZ gates developed in previous sections and recover the resulting
fidelity loss through additional fine-tuning optimization.

The filtered flux pulses $\Phi_{A/B}^{\mathrm{filt}}(t)$ are modeled
using the transfer function $\Phi_{A/B}^{\mathrm{filt}}(t)=\dfrac{1}{2\pi}\int_{-\infty}^{+\infty}d\omega e^{i\omega t}F(\omega)\Phi_{A/B}(\omega)$
\citep{TheisMotzoiWilhelm2016}, where $\Phi_{A/B}(\omega)$ represents
the ideal flux pulses $\Phi_{A/B}(t)$ in the frequency domain, obtained
through the Fourier transform $\Phi_{A/B}(\omega)=\int_{-\infty}^{+\infty}dte^{-i\omega t}\Phi_{A/B}(t)$.
The function $F(\omega)$, referred to as the filter, characterizes
the attenuation of high-frequency components, as determined by the
properties of the specific AWGs employed. In this study, we adopt
a Gaussian filter $F(\omega)=\exp(-\omega^{2}/\omega_{c}^{2})$ with
a cutoff frequency $\omega_{c}$, derived from the specifications
of commercial Tektronix AWGs \citep{MotzoiGambettaMerkelEtAl2011,Weaver1971}.
Using the optimized flux pulses $\Phi_{A/B}(t)$ from the Qt-4 $J_{1}$
optimization in Sec.\ \ref{subsec:Optimization_results}, we model
the corresponding filtered flux pulses $\Phi_{A/B}^{\mathrm{filt}}(t)$
by applying the transfer function with cutoff frequencies $\omega_{c}/2\pi=$
$424.7\,\mathrm{MHz}$, $3.4\,\mathrm{GHz}$, and $25.5\,\mathrm{GHz}$
for the Tektronix AWGs AWG5014, AWG5200, and AWG70000, respectively.
Finally, we use the modeled filtered flux pulses $\Phi_{A/B}^{\mathrm{filt}}(t)$
to calculate the degraded $J_{1}$ and demonstrate the resulting filtering
effects, specifically the increase in infidelity from the ideal $J_{1}$
to the degraded $J_{1}$, for the Tektronix AWGs with their associated
cutoff frequencies $\omega_{c}$, as shown in Fig.\ \ref{fig:filtering_effects_and filtered_pulses}(a).

To investigate the simulated filtering effects, we additionally present
the corresponding frequency spectra, $\Phi_{A/B}(\omega)$, derived
from the optimized flux pulses $\Phi_{A/B}(t)$ obtained through the
Qt-4 $J_{1}$ optimization, as shown in Figs.\ \ref{fig:filtering_effects_and filtered_pulses}(b),
\ref{fig:filtering_effects_and filtered_pulses}(c), and \ref{fig:filtering_effects_and filtered_pulses}(d)
for $t_{f}=12\,\mathrm{ns}$, $15\,\mathrm{ns}$, and $40\,\mathrm{ns}$,
respectively. As outlined in Sec.\ \ref{subsec:Control_scheme},
for the Qt-4 $J_{1}$ optimization scheme, the Fine-Tune-Pulse is
applied only to $Q_{A}$, such that $\Phi_{A}(t)=\Phi_{A}^{\mathrm{PBP}}(t)+\Phi_{A}^{\mathrm{FTP}}(t)$,
while $\Phi_{B}(t)=\Phi_{B}^{\mathrm{PBP}}(t)$. Consequently, the
frequency spectrum $\Phi_{B}(\omega)/\Phi_{0}$ corresponds solely
to the Pull-Back-Pulse $\Phi_{B}^{\mathrm{PBP}}(t)$, while the spectrum
$\Phi_{A}(\omega)/\Phi_{0}$ results from the combination of the Pull-Back-Pulse
and the Fine-Tune-Pulse, $\Phi_{A}^{\mathrm{PBP}}(t)+\Phi_{A}^{\mathrm{FTP}}(t)$.
Notably, for the normalized spectrum $\left|\Phi_{A}(\omega)/\Phi_{0}\right|$,
there are peaks surpassing those observed in the normalized $\left|\Phi_{B}(\omega)/\Phi_{0}\right|$
in the frequency range between approximately $0.1\,\mathrm{GHz}$
and $1\,\mathrm{GHz}$. These peaks correspond to the spectral components
of the sine function associated with specific Fine-Tune-Pulse elements,
$\Phi_{A,k}^{\mathrm{FTP}}(t)$, which exhibit stronger control parameters
$a_{k}$ as detailed in Eqs.\ (\ref{eq:FTP_pulse_sum})-(\ref{eq:FTP_pulse_s3}).

For $t_{f}=12\,\mathrm{ns}$, the dominant spectral region of $\Phi_{A/B}(\omega)$,
where the normalized $\left|\Phi_{A/B}(\omega)/\Phi_{0}\right|$ strength
exceeds $10^{-2}$, occurs at $\omega/2\pi\lesssim1.8\,\mathrm{GHz}$.
However, when the narrow-bandwidth filter $F(\omega)$ with $\omega_{c}/2\pi=424.7\,\mathrm{MHz}$
is applied to the transfer function, the dominant spectra of $\Phi_{A/B}(\omega)$
are attenuated to approximately $95\%$, $10\%$, $0.4\%$, and $10^{-8}$
of their original values at $\omega/2\pi\approx0.1\,\mathrm{GHz}$,
$0.6\,\mathrm{GHz}$, $1.0\,\mathrm{GHz}$, and $1.8\,\mathrm{GHz}$,
respectively, as shown in Fig.\ \ref{fig:filtering_effects_and filtered_pulses}(b).
Consequently, the spectra $\Phi_{A/B}(\omega)$ are nearly nullified
for $\omega/2\pi\gtrsim1\,\mathrm{GHz}$, significantly suppressed
for $0.1\lesssim\omega/2\pi\lesssim1\,\mathrm{GHz}$, and retained
only in the low-frequency region $\omega/2\pi\lesssim0.1\,\mathrm{GHz}$.
This attenuation leads to substantial distortion of the pulses $\Phi_{A/B}(t)$,
as illustrated in Figs.\ \ref{fig:filtering_effects_and filtered_pulses}(e)
and \ref{fig:filtering_effects_and filtered_pulses}(f). In contrast,
for wider-bandwidth filters $F(\omega)$ with $\omega_{c}/2\pi=3.4\,\mathrm{GHz}$
and $25.5\,\mathrm{GHz}$, the spectra $\Phi_{A/B}(\omega)$ at the
boundary of the dominant region $\omega/2\pi\approx1.8\,\mathrm{GHz}$
are attenuated to $75.6\%$ and $99.5\%$ of their original strengths,
respectively. As a result, only the higher-frequency components of
the dominant region are significantly affected. This leads to minor
shifts in the pulses, primarily at locations corresponding to the
higher-frequency components, such as the pulse beginning, end, peaks,
and corners, as shown in Figs.\ \ref{fig:filtering_effects_and filtered_pulses}(e)
and \ref{fig:filtering_effects_and filtered_pulses}(f) and their
insets. Therefore, as shown in Fig.\ \ref{fig:filtering_effects_and filtered_pulses}(a)
at $t_{f}=12\,\mathrm{ns}$, the significant pulse distortion caused
by the narrow-bandwidth filter $F(\omega)$ with $\omega_{c}/2\pi=424.7\,\mathrm{MHz}$
results in a filtering effect of $1.1\times10^{-2}$, while the minor
pulse distortion due to the wider-bandwidth filter $F(\omega)$ with
$\omega_{c}/2\pi=3.4\,\mathrm{GHz}$ yields a filtering effect of
$1.5\times10^{-5}$, which is comparable to the ideal $J_{1}\approx1.8\times10^{-5}$.
As the cutoff frequency is increased to $\omega_{c}/2\pi=25.5\,\mathrm{GHz}$,
the filtering effect is further reduced to $3.0\times10^{-9}$, more
than three orders of magnitude smaller than the ideal $J_{1}$, and
can thus be neglected.

The filtering effects manifest as a plateau-like bulge in the region
$14\lesssim t_{f}\lesssim21\,\mathrm{ns}$, as depicted in Fig.\ \ref{fig:filtering_effects_and filtered_pulses}(a).
This region corresponds to the hard region in the context of CZ gate
optimization, as discussed in Sec.\ \ref{subsec:Optimization_results}.
Within the hard region, effective synchronization requires strong
Fine-Tune-Pulse vibrations, exemplified by the optimized $\Phi_{A}(t)$
for $t_{f}=17.5\,\mathrm{ns}$ shown in Fig.\ \ref{fig:J1_optimization}(c).
Such strong Fine-Tune-Pulse vibrations in the time domain result in
pronounced peaks in the corresponding frequency spectrum. The normalized
spectrum $\left|\Phi_{A}(\omega)/\Phi_{0}\right|$ for $t_{f}=15\,\mathrm{ns}$,
shown in Fig.\ \ref{fig:filtering_effects_and filtered_pulses}(c),
exhibits significantly stronger peaks in the frequency range $0.1\lesssim\omega/2\pi\lesssim1\,\mathrm{GHz}$
compared to spectra at times outside the hard region, such as $t_{f}=12\,\mathrm{ns}$
and $t_{f}=40\,\mathrm{ns}$ in Figs.\ \ref{fig:filtering_effects_and filtered_pulses}(b)
and \ref{fig:filtering_effects_and filtered_pulses}(d), respectively.
The attenuation of these prominent spectral peaks due to filtering
leads to more substantial pulse distortion in the time domain, thereby
resulting in the characteristic plateau behavior of the filtering
effect observed in the hard region. On the plateau, a steep dip is
observed at $t_{f}\approx19.2\,\mathrm{ns}$. This point is close
to the natural synchronization position at $t_{f}\approx18.5\,\mathrm{ns}$,
where strong Fine-Tune-Pulse vibrations are not required for synchronization.
As a result, the filtering effect at $t_{f}\approx19.2\,\mathrm{ns}$
is significantly mitigated compared to other values of $t_{f}$ within
the hard region.

As $t_{f}$ increases, the frequency spectra $\Phi_{A/B}(\omega)$
are compressed towards the lower-frequency region. This occurs because,
in the Qt-4 $J_{1}$ optimization for all $t_{f}$, we set a fixed
Pull-Back-Pulse ratio ($\mathrm{PBPr=}t_{f}/t_{\mathrm{PB}}=32$)
in $\Phi_{A/B}(t)$. As $t_{f}$ increases, the rise and fall speeds
of the Pull-Back-Pulse slow down due to the longer $t_{\mathrm{PB}}$.
Additionally, the fixed number of control parameters---specifically
$k_{\mathrm{max}}=10$ for the Fine-Tune-Pulse in $\Phi_{A}(t)$---results
in a decrease in the oscillation frequencies of the sine functions
in the Fine-Tune-Pulse in Eqs.\ (\ref{eq:FTP_pulse_s1})-(\ref{eq:FTP_pulse_s3})
as $t_{f}$ increases. These decreasing speeds and frequencies in
the time domain shift the corresponding spectrum to the lower-frequency
region. For instance, as shown in Fig.\ \ref{fig:filtering_effects_and filtered_pulses}(d),
the dominant spectrum for $t_{f}=40\,\mathrm{ns}$, with normalized
strength greater than $10^{-2}$, is compressed to the lower-frequency
region $\omega/2\pi\lesssim0.5\,\mathrm{GHz}$, compared to the dominant
region $\omega/2\pi\lesssim1.8\,\mathrm{GHz}$ for $t_{f}=12\,\mathrm{ns}$
in Fig.\ \ref{fig:filtering_effects_and filtered_pulses}(b). As
$t_{f}$ increases, the compressed spectrum away from the cutoff frequency
of the filter $F(\omega)$ is expected to reduce pulse distortion,
thereby mitigating the filtering effects. However, the accumulation
of pulse distortion errors with increasing $t_{f}$ counteracts this
trend, enhancing filtering effects instead. Consequently, the comparable
filtering effects in the range $25\lesssim t_{f}\lesssim40\,\mathrm{ns}$
{[}Fig.\ \ref{fig:filtering_effects_and filtered_pulses}(a){]} can
be largely attributed to these two competing factors.

In summary, the influence of filtering effects on the optimized CZ
gates via the Qt-4 $J_{1}$ optimization is effectively negligible
when employing an AWG with a cutoff frequency exceeding $25.5\,\mathrm{GHz}$.
Specifically, for the Tektronix AWG70000 with $\omega_{c}/2\pi=$
$25.5\,\mathrm{GHz}$, as illustrated in Fig.\ \ref{fig:filtering_effects_and filtered_pulses}(a),
the filtering effects are more than two orders of magnitude smaller
than the corresponding ideal $J_{1}$ for nearly all $t_{f}$, except
within the hard region. Even in the hard region, the filtering effects
remain approximately one order of magnitude below the ideal $J_{1}$.
These results demonstrate that the predicted high-fidelity CZ gates
are within experimental reach using the optimized control pulses generated
by standard commercial AWGs with $\omega_{c}/2\pi\gtrsim$ $25.5\,\mathrm{GHz}$.

\begin{figure}
\includegraphics[scale=0.52]{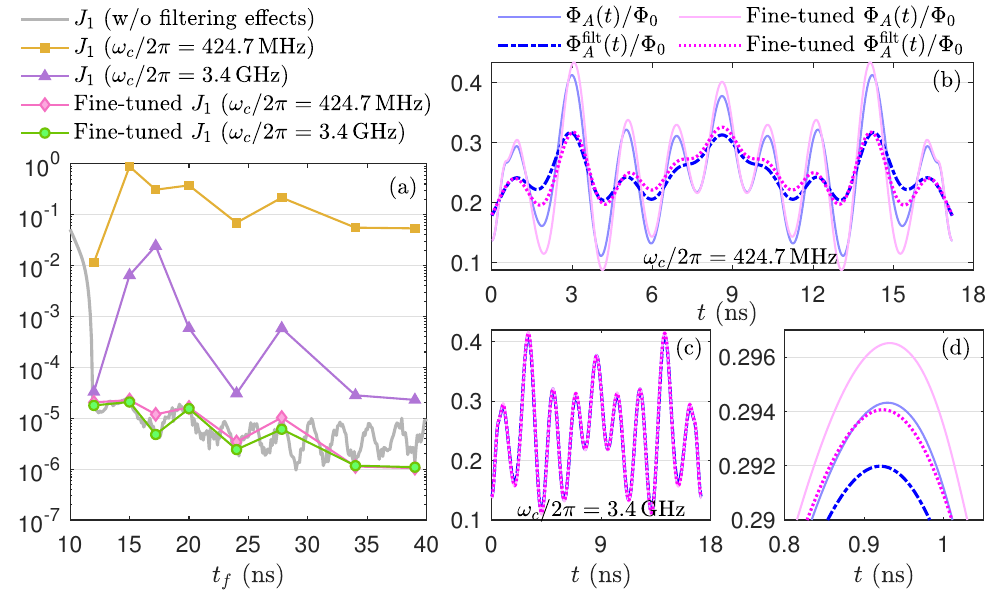}

\caption{Recovery of fidelity loss due to filtering effects through additional
fine-tuning optimization. Panel (a) displays the optimized $J_{1}$
as a function of $t_{f}$. The gray line represents the ideal $J_{1}$
obtained via the Qt-4 $J_{1}$ optimization without filtering effects,
while the yellow squares and purple triangles indicate the degraded
$J_{1}$ values, sampled from the conditions of the ideal $J_{1}$
(gray line) and affected by filtering effects from AWGs with cutoff
frequencies $\omega_{c}/2\pi=424.7\,\mathrm{MHz}$ and $3.4\,\mathrm{GHz}$,
respectively. The pink diamonds and green circles correspond to the
fine-tuned $J_{1}$ values for $\omega_{c}/2\pi=424.7\,\mathrm{MHz}$
and $3.4\,\mathrm{GHz}$, respectively, achieved by applying additional
fine-tuning optimization to the corresponding degraded samples (yellow
squares and purple triangles). Panels (b), (c), and (d) display the
control pulses for the sample with $t_{f}=17.2\,\mathrm{ns}$. In
these panels, the blue solid line and the blue dash-dotted line represent
the unfiltered optimized pulse $\Phi_{A}(t)/\Phi_{0}$ and the filtered
optimized pulse $\Phi_{A}^{\mathrm{filt}}(t)/\Phi_{0}$, respectively,
and the optimized pulse is obtained from the Qt-4 $J_{1}$ optimization.
The pink solid line and the pink dash-dotted line represent the unfiltered
fine-tuned pulse $\Phi_{A}(t)/\Phi_{0}$ and the filtered fine-tuned
pulse $\Phi_{A}^{\mathrm{filt}}(t)/\Phi_{0}$, respectively, and the
fine-tuned pulse results from the subsequent fine-tuning optimization
applied to the corresponding optimized sample from the Qt-4 $J_{1}$
optimization. Panel (b) corresponds to $\omega_{c}/2\pi=424.7\,\mathrm{MHz}$,
panel (c) corresponds to $\omega_{c}/2\pi=3.4\,\mathrm{GHz}$, and
panel (d) provides an enlarged view of the first peak in panel (c)
at $t\approx0.9\,\mathrm{ns}$.}

\label{fig:J1_finetune_optimzation} 
\end{figure}

For $\omega_{c}/2\pi\lesssim25.5\,\mathrm{GHz}$, filtering effects
are no longer negligible and must be explicitly taken into account.
To address this issue, we perform an additional fine-tuning optimization
following the Qt-4 $J_{1}$ optimization, with the aim of recovering
the fidelity loss due to the filtering effects. In this procedure,
the optimized control parameters obtained from the Qt-4 $J_{1}$ optimization
are used as the initial conditions for the refinement step. The cost
function $J_{1}$ is evaluated using the filtered control pulses $\Phi_{A/B}^{\mathrm{filt}}(t)$,
generated via the transfer function, rather than the original unfiltered
pulses $\Phi_{A/B}(t)$. In this way, the filtering is directly incorporated
into the optimization, allowing for effective mitigation of its impact.
All other optimization settings remain identical to those employed
in the Qt-4 $J_{1}$ optimization. With this additional fine-tuning
step, the filtering effects arising from AWGs with $\omega_{c}/2\pi=424.7\,\mathrm{MHz}$
and $3.4\,\mathrm{GHz}$ are effectively suppressed, such that the
corresponding fine-tuned $J_{1}$ values recover to the ideal unfiltered
level for most $t_{f}$ samples, as shown in Fig.\ \ref{fig:J1_finetune_optimzation}(a).

For $\omega_{c}/2\pi=3.4\,\mathrm{GHz}$, the pulse distortion is
not immediately apparent, as shown in Fig.\ \ref{fig:J1_finetune_optimzation}(c)
for $t_{f}=17.2\,\mathrm{ns}$. However, upon zooming in on the peak
near $t\approx0.9\,\mathrm{ns}$ in Fig.\ \ref{fig:J1_finetune_optimzation}(d),
it becomes evident that the unfiltered fine-tuned pulse $\Phi_{A}(t)$
is lifted to allow the filtered fine-tuned pulse $\Phi_{A}^{\mathrm{filt}}(t)$
to approach the original unfiltered pulse $\Phi_{A}(t)$, which was
obtained via the Qt-4 $J_{1}$ optimization, thereby recovering the
ideal unfiltered $J_{1}$. This explains why the fine-tuned $J_{1}$
values can recover to their respective ideal $J_{1}$ values, as shown
in Fig.\ \ref{fig:J1_finetune_optimzation}(a). In contrast, for
the lower bandwidth case with $\omega_{c}/2\pi=424.7\,\mathrm{MHz}$,
the filtered pulse $\Phi_{A}^{\mathrm{filt}}(t)$ exhibits significant
distortion compared to the unfiltered pulse $\Phi_{A}(t)$, as depicted
in Fig.\ \ref{fig:J1_finetune_optimzation}(b) for $t_{f}=17.2\,\mathrm{ns}$.
Consequently, the fine-tuning optimization does not restore the filtered
fine-tuned pulse $\Phi_{A}^{\mathrm{filt}}(t)$ to the original unfiltered
pulse $\Phi_{A}(t)$ to recover the corresponding ideal unfiltered
$J_{1}$. Instead, it adjusts the filtered fine-tuned pulse $\Phi_{A}^{\mathrm{filt}}(t)$
around the original filtered pulse $\Phi_{A}^{\mathrm{filt}}(t)$
to find another optimal $J_{1}$. This results in the optimally fine-tuned
$J_{1}$ values not necessarily restoring the corresponding ideal
unfiltered $J_{1}$. Specifically, at $t_{f}=17.2\,\mathrm{ns}$,
the fine-tuned $J_{1}\approx1.19\times10^{-5}$ is greater than the
ideal unfiltered $J_{1}\approx4.87\times10^{-6}$, as shown in Fig.\ \ref{fig:J1_finetune_optimzation}(a).
However, at $t_{f}=12\,\mathrm{ns}$, the fine-tuned $J_{1}\approx2.05\times10^{-5}$
nearly recovers to the ideal unfiltered $J_{1}\approx1.80\times10^{-5}$.
In conclusion, although the cutoff frequency is reduced to $\omega_{c}/2\pi=424.7\,\mathrm{MHz}$,
the fine-tuning optimization is still able to mitigate the fidelity
loss due to the filtering effects by more than three orders of magnitude
for most $t_{f}$, maintaining the fine-tuned $J_{1}\lesssim2\times10^{-5}$
for $t_{f}\geq12\,\mathrm{ns}$, as demonstrated in Fig.\ \ref{fig:J1_finetune_optimzation}(a).

\subsection{CZ gate optimization for asymmetric transmons}

\label{subsec:Optimization_asymmetric_transmon}

\begin{figure}
\includegraphics[scale=0.0624]{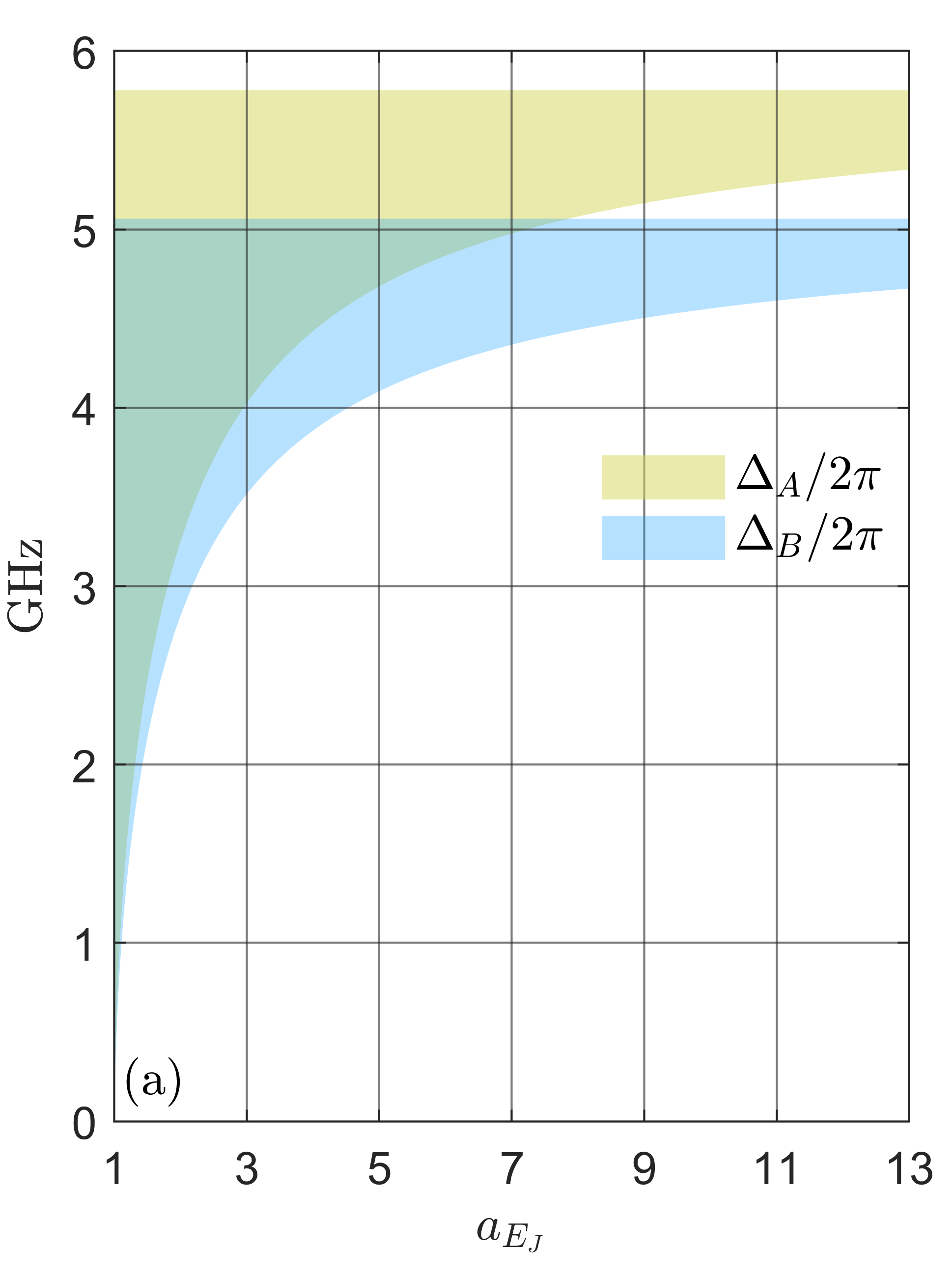}\includegraphics[scale=0.52]{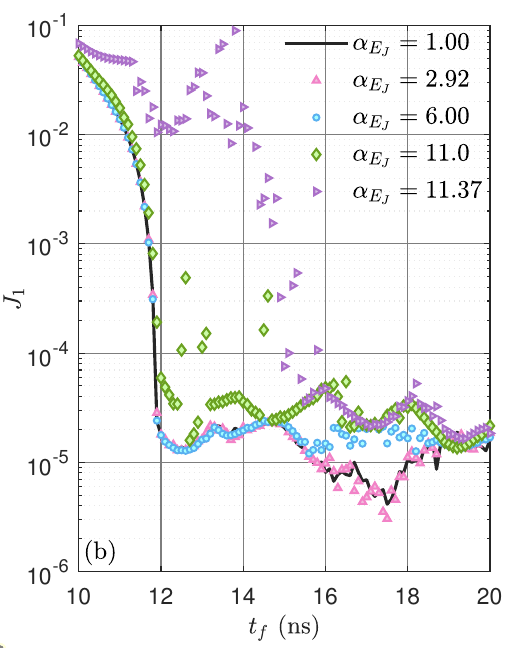}

\caption{Asymmetric transmons. (a) Tunable ranges of the qubit frequencies
$\Delta_{A}$ (olive shadow) and $\Delta_{B}$ (blue shadow) as a
function of $a_{E_{J}}$ from $1$ to $13$. (b) Optimized $J_{1}$
from the Qt-4 $J_{1}$ optimization as a function of $t_{f}$ ($10\,\mathrm{ns}$
to $20\,\mathrm{ns}$) for symmetric transmons with $a_{E_{J}}=1$
(black line) and asymmetric transmons with $a_{E_{J}}=2.92$ (pink
triangles), $a_{E_{J}}=6.00$ (blue circles), $a_{E_{J}}=11.0$ (green
diamonds), and $a_{E_{J}}=11.37$ (purple rightward triangles).}

\label{fig:J1op_aEJ_split} 
\end{figure}

In the preceding sections, we have concentrated on the control of
the CZ gate for symmetric transmons with $a_{E_{J}}=1$. In this section,
we extend the analysis to asymmetric transmons where $a_{E_{J}}>1$.
We begin by plotting the tunable ranges of $\Delta_{A/B}$ as a function
of $a_{E_{J}}$, as shown in Fig.\ \ref{fig:J1op_aEJ_split}(a).
To generate these plots, we apply Eq.\ (\ref{eq:freq_definition})
and vary the fluxes $\Phi_{A/B}/\Phi_{0}$ from 0 to $\pi/2$ for
each $a_{E_{J}}$, while maintaining the same system parameters $\omega_{0,A/B}$
and $\alpha_{A/B}$ as used for the symmetric transmons as discussed
in Sec.\ \ref{subsec:Error_analysis_ideal_CZ_gate}. It is evident
that, as $a_{E_{J}}$ increases, both the tunable ranges of $\Delta_{A}$
and $\Delta_{B}$ shrink, with the overlap region between them diminishing
accordingly. Notably, for $a_{E_{J}}\gtrsim7.86$, the overlap vanishes
entirely, and the separation between the two ranges continues to increase
as $a_{E_{J}}$ rises. As $a_{E_{J}}$ approaches $11.37$, the gap
between the minimum of $\Delta_{A}$ and the maximum of $\Delta_{B}$
begins to exceed $-\alpha_{A}=2\pi\times214\,\mathrm{MHz}$, resulting
in a minimum achievable detuning $\delta_{A}=(\Delta_{A}-\Delta_{B})-(-\alpha_{A})\gtrsim0$
for $a_{E_{J}}\gtrsim11.37$. In other words, for $a_{E_{J}}\gtrsim11.37$,
the condition for the strongest coupling, $\delta_{A}=0$, can no
longer be satisfied, thereby preventing the implementation of the
fastest CZ gate with gate time close to the speed limit $t_{f}=\pi/(\sqrt{2}g_{AB})\approx11.6\,\mathrm{ns}$.

We proceed to develop high-fidelity CZ gates for asymmetric transmons
using the Qt-4 $J_{1}$ optimization. The optimization settings are
identical to those employed for symmetric transmons in Sec.\ \ref{subsec:Control_scheme},
with the exception of the initial interaction frequencies $\omega_{i,A/B}$,
which are determined by $a_{E_{J}}$. For larger $a_{E_{J}}$, as
a result of the increased gap between the tunable ranges of $\Delta_{A/B}$
{[}see Fig.\ \ref{fig:J1op_aEJ_split}(a){]}, we set the initial
$\omega_{i,A}$ and $\omega_{i,B}$ to be near the minimum of $\Delta_{A}$
and the maximum of $\Delta_{B}$, respectively. This choice ensures
that $\delta_{A}$ approaches zero, thereby facilitating sufficient
coupling for the implementation of shorter gate-time CZ gates.

The optimization results for asymmetric transmons are shown in Fig.\ \ref{fig:J1op_aEJ_split}(b).
For $a_{E_{J}}\gtrsim11.37$, $\delta_{A}=0$ cannot be satisfied,
as discussed in the preceding paragraph. Consequently, the corresponding
optimized $\delta_{A}(t)$ cannot exhibit behavior similar to the
case of $a_{E_{J}}=1$ at $t_{f}=12\,\mathrm{ns}$, as shown in Fig.\ \ref{fig:J1_optimization}(h),
where it vibrates around zero detuning to maximize the coupling and
synchronize the leakage errors. As a result, for $a_{E_{J}}=11.37$,
the optimized $J_{1}$ at $t_{f}=12\,\mathrm{ns}$ degrades to $1.2\times10^{-2}$,
and the optimized $J_{1}\lesssim10^{-4}$ is achievable only for $t_{f}$
exceeding $\sim15\,\mathrm{ns}$. As $a_{E_{J}}$ decreases from $11.37$,
the gap between the tunable ranges of $\Delta_{A/B}$ narrows, and
eventually, the two ranges overlap, allowing the minimum achievable
$\delta_{A}$ to become more negative. This enables the optimized
$\delta_{A}(t)$ to vibrate around zero detuning with a larger amplitude,
leading to a greater reduction in $J_{1}$. For $a_{E_{J}}=11.0$,
the minimum achievable $\delta_{A}$ is $\sim-2\pi\times16\,\mathrm{MHz}$,
whereas for $a_{E_{J}}=6.0$, $\delta_{A}$ can reach a more negative
value, $\sim-2\pi\times0.4\,\mathrm{GHz}$. As a result, at $t_{f}=12\,\mathrm{ns}$,
the optimized $J_{1}$ for $a_{E_{J}}=11.0$ is limited to $\sim6\times10^{-5}$,
while the optimized $J_{1}$ for $a_{E_{J}}\lesssim6.0$ exhibits
performance comparable to that of symmetric transmons with $a_{E_{J}}=1$,
as shown in Fig.\ \ref{fig:J1op_aEJ_split}(b).

However, in the hard region $16\lesssim t_{f}\lesssim18\,\mathrm{ns}$,
the optimized $J_{1}$ for $a_{E_{J}}=6.0$ degrades compared to $a_{E_{J}}=1$,
as shown in Fig.\ \ref{fig:J1op_aEJ_split}(b). This degradation
occurs because, in the hard region, a stronger $\delta_{A}(t)$ vibration
amplitude is required to achieve the necessary synchronization. For
symmetric transmons with $a_{E_{J}}=1$, the optimized $J_{1}\approx4.1\times10^{-6}$
at $t_{f}=17.5\,\mathrm{ns}$ in the hard region is achieved with
a strong optimized $\delta_{A}(t)$ vibration, ranging approximately
from $-2\pi\times1.5\,\mathrm{GHz}$ to $2\pi\times0.5\,\mathrm{GHz}$,
as shown in Fig.\ \ref{fig:J1_optimization}(i). In contrast, for
$a_{E_{J}}=6.0$, the lower bound of $\delta_{A}$ is limited to $\sim-2\pi\times0.4\,\mathrm{GHz}$,
resulting in a degraded $J_{1}\approx1.49\times10^{-5}$. For $a_{E_{J}}=2.92$,
the lower bound of $\delta_{A}$ can be extended to $\sim-2\pi\times1.3\,\mathrm{GHz}$,
which is close to the minimum value of the optimized $\delta_{A}(t)$
at $t_{f}=17.5\,\mathrm{ns}$ for $a_{E_{J}}=1$. As a result, it
shows comparable $J_{1}$ performance to that of $a_{E_{J}}=1.0$,
not only in the hard region but also across all $t_{f}$, as shown
in Fig.\ \ref{fig:J1op_aEJ_split}(b). In summary, for asymmetric
transmons, comparable optimized $J_{1}$ performance to symmetric
transmons can be achieved once the required $\delta_{A}$ tunable
range is sufficient.

\section{Decoherence }

\label{sec:robust_against_noises}

In the Noisy Intermediate-Scale Quantum (NISQ) era, the depth of quantum
circuits is fundamentally constrained by decoherence, which limits
the reliable execution of quantum algorithms. Consequently, mitigating
the effects of decoherence on quantum gates has emerged as a central
challenge in this era. In this section, we propose high-fidelity and
robust control schemes for the CZ gate implementation, tailored to
mitigate the dominant decoherence mechanisms within the system. First,
in Sec.\ \ref{subsec:Noise_characterization_dephasing}, we characterize
the noise based on Ramsey dephasing experiments \citep{BarendsKellyMegrantEtAl2014}
to establish effective noise models. Subsequently, in Sec.\ \ref{subsec:Noise_characterization_from_CZ_infid_budget},
we incorporate the experimental error budget for the CZ gate reported
in the same reference \citep{BarendsKellyMegrantEtAl2014} to further
refine these models. Finally, in Sec.\ \ref{subsec:HF_robust_CZ_against_noise},
we develop high-fidelity and robust CZ gates optimized for the experimentally
characterized noise environment.

$ $

\subsection{Noise characterization from the Ramsey dephasing experiments}

\label{subsec:Noise_characterization_dephasing}

Decoherence in superconducting qubits is primarily governed by energy
relaxation and dephasing processes. In frequency-tunable transmons,
dephasing arises from two distinct noise mechanisms, which are characterized
by a fast dephasing time $T_{\phi,1}$ and a slow dephasing time $T_{\phi,2}$
\citep{BarendsKellyMegrantEtAl2014}. The fast dephasing time $T_{\phi,1}$
results from both energy relaxation, characterized by energy relaxation
time $T_{1}$, and white noise originating from the room-temperature
control electronics, denoted by a dephasing time $T_{\phi,1,\epsilon}$.
These contributions combine according to $1/T_{\phi,1}=1/(2T_{1})+1/T_{\phi,1,\epsilon}$
\citep{BarendsKellyMegrantEtAl2014}. The slow dephasing time $T_{\phi,2}$
is associated with flux noise exhibiting a $1/f$ spectral density.
Experimentally, $T_{\phi,1}$ and $T_{\phi,2}$ are extracted by fitting
the measured Ramsey decay envelope $V(t)$ to the form $\exp[-t/T_{\phi,1}-(t/T_{\phi,2})^{2}]$.

In Sec.\ \ref{subsec:Noise_models}, we construct noise models for
the room-temperature-control-electronics (RTCE) noise $\beta_{\epsilon}(t)$
and flux noise $\beta_{\phi}(t)$, which are associated with the dephasing
times $T_{\phi,1,\epsilon}$ and $T_{\phi,2}$, respectively. Using
these models, we derive analytical expressions for the estimated dephasing
times $T_{\phi,1}$ and $T_{\phi,2}$, and subsequently calibrate
the undetermined model parameters by fitting the estimated expressions
to the experimentally measured $T_{\phi,1}$ and $T_{\phi,2}$ over
a qubit frequency range from $4.15\,\mathrm{GHz}$ to $5.09\,\mathrm{GHz}$.
The experimental dephasing times, obtained from Ramsey measurements
on a qubit with $a_{E_{J}}=1.0$, $\omega_{0,A}/2\pi=5.238\,\mathrm{GHz}$,
and $\alpha_{A}/2\pi=-226\,\mathrm{MHz}$, are reported in FIG. S2
of Ref.\ \citep{BarendsKellyMegrantEtAl2014}. In the same reference,
the energy relaxation times $T_{1}$ measured over a frequency range
from $4.0\,\mathrm{GHz}$ to $5.8\,\mathrm{GHz}$ mostly lie between
$20\,\mu\mathrm{s}$ and $40\,\mu\mathrm{s}$; we adopt the median
value $T_{1}=30\,\mu\mathrm{s}$ for subsequent estimations and simulations.
Finally, in Sec\ \ref{subsec:Ramsey_dephasing_simulation}, the characterized
noise models are employed to simulate the Ramsey dephasing experiments
via the master equation, yielding simulated $T_{\phi,1}$ and $T_{\phi,2}$
to further validate the models. A comparison of the experimental dephasing
times, the estimated times with best-fit parameters, and the simulated
results is presented in Fig.\ \ref{fig:noise_characterication}(a).

\subsubsection{Noise models}

\label{subsec:Noise_models}

\begin{figure}
\includegraphics[scale=0.52]{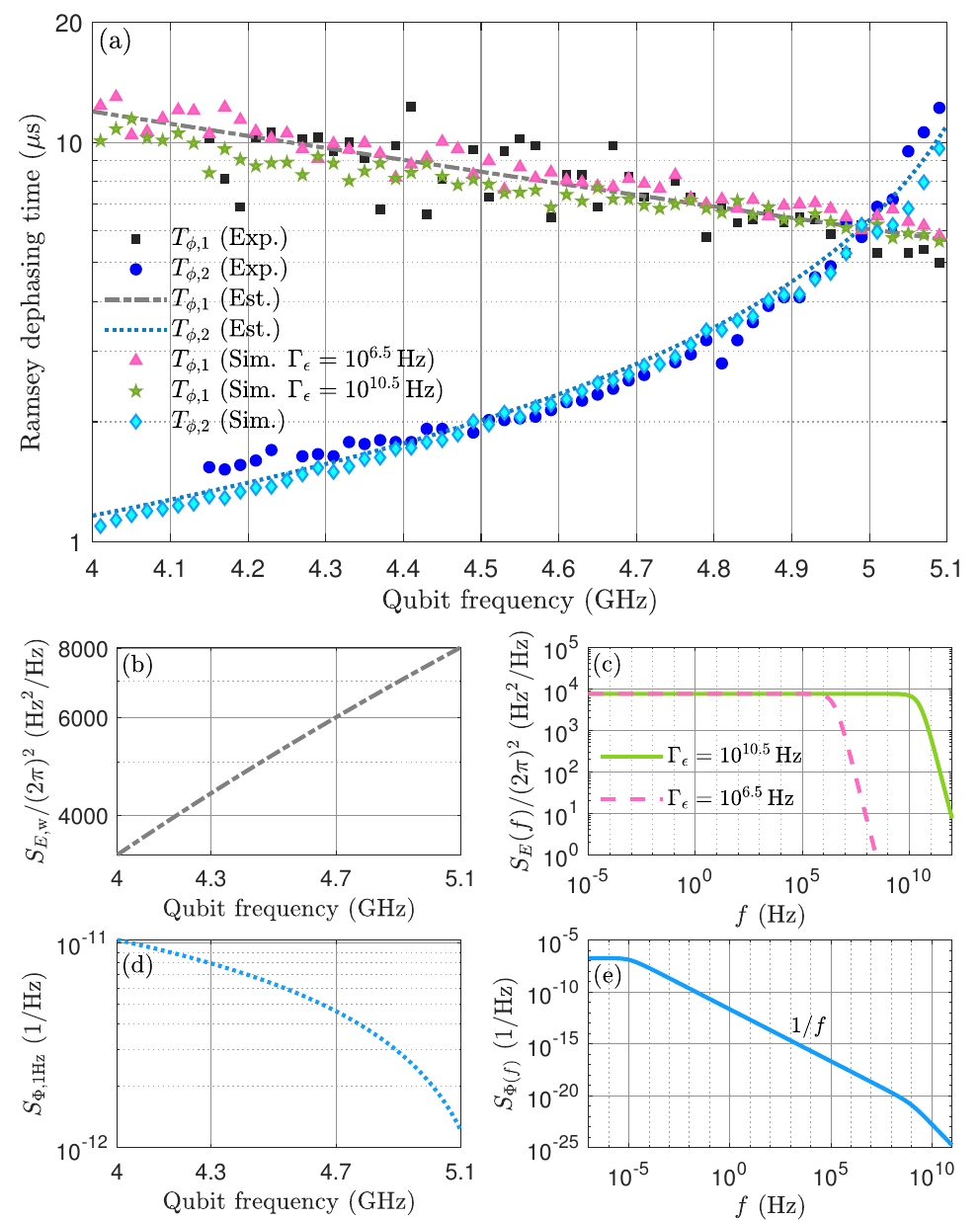}

\caption{Dephasing noise characterization. (a) Fast and slow dephasing times,
$T_{\phi,1}$ and $T_{\phi,2}$, as functions of the qubit frequency
in the range $4.0-5.1\,\mathrm{GHz}$. Black squares and blue circles
denote the experimental $T_{\phi,1}$ and $T_{\phi,2}$, respectively,
extracted from the Ramsey decay envelope $V(t)$ measured in the experiment
(see FIG.\ S2 of Ref.\ \citep{BarendsKellyMegrantEtAl2014}). The
gray dash-dot line and blue dotted line show the corresponding estimated
$T_{\phi,1}$ and $T_{\phi,2}$, obtained from the expressions $T_{\phi,1,\epsilon}$
in Eq.\ (\ref{eq:T_phi_1_e_est}) with the experimental $T_{1}=30\,\mu\mathrm{s}$,
and $T_{\phi,2}$ in Eq.\ (\ref{eq:T_phi_2_est}), derived from our
noise models with best-fit parameters $S_{\epsilon,\mathrm{w}}$,
$\eta_{\epsilon}$, and $S_{1/f}$. Pink triangles and green pentagrams
indicate simulated $T_{\phi,1}$, while cyan diamonds show simulated
$T_{\phi,2}$, all extracted from the simulated Ramsey envelope $V(t)$
generated using the same noise models and best-fit parameters. The
pink triangles and green pentagrams correspond to simulations with
$\Gamma_{\epsilon}=10^{6.5}\,\mathrm{Hz}$ and $\Gamma_{\epsilon}=10^{10.5}\,\mathrm{Hz}$,
respectively, which set the bandwidth over which the flat white-noise
spectrum extends in generating realizations of the RTCE noise $\beta_{\epsilon}(t)$.
In our noise models: (b) Constant spectrum $S_{E,\mathrm{w}}=\left[\Delta_{A}-\alpha_{A}\right]^{\eta_{\epsilon}}S_{\epsilon,\mathrm{w}}$
for the RTCE noise expressed in frequency units, $\beta_{\epsilon,\mathrm{freq}}(t)\equiv\beta_{\epsilon}(t)\left[\Delta_{A}-\alpha_{A}\right]^{\eta_{\epsilon}/2}$,
as a function of qubit frequency $\Delta_{A}$ from $4.0\,\mathrm{GHz}$
to $5.1\,\mathrm{GHz}$, evaluated with the best-fit parameters $S_{\epsilon,\mathrm{w}}$
and $\eta_{\epsilon}=3.73$. (d) $1/f$ spectrum at $1\,\mathrm{Hz}$,
$S_{\Phi,\mathrm{1Hz}}=(\Phi_{A}/\Phi_{0})^{2}S_{1/f}$ for the flux
noise $\beta_{\Phi}(t)\equiv(\Phi_{A}/\Phi_{0})\cdot\beta_{\phi}(t)$,
shown as a function of the qubit frequency $\Delta_{A}$ from $4.0\,\mathrm{GHz}$
to $5.1\,\mathrm{GHz}$ using the best-fit $S_{1/f}$. In our simulations:
(c) Corresponding spectra $S_{E}(f)=\left[\Delta_{A}-\alpha_{A}\right]^{\eta_{\epsilon}}S_{\epsilon}(f)$
at $\Delta_{A}/2\pi=5\,\mathrm{GHz}$ for the generated RTCE noise
$\beta_{\epsilon}(t)$ realizations. The green solid and pink dashed
curves correspond to $\Gamma_{\epsilon}=10^{10.5}\,\mathrm{Hz}$ and
$\Gamma_{\epsilon}=10^{6.5}\,\mathrm{Hz}$, respectively, illustrating
the extension of the flat white-noise spectrum to approximately $1\:\mathrm{MHz}$
and $10\:\mathrm{GHz}$. (e) Corresponding spectrum $S_{\Phi}(f)=(\Phi_{A}/\Phi_{0})^{2}S_{\phi}(f)$
at $\Delta_{A}/2\pi=5\:\mathrm{GHz}$ for the generated flux-noise
$\beta_{\phi}(t)$ realizations, exhibiting a $1/f$ behavior over
a frequency range from $\sim10^{-5}$ to $\sim10^{9}\,\mathrm{Hz}$. }

\label{fig:noise_characterication} 
\end{figure}

As illustrated in Fig.\ \ref{fig:noise_characterication}(a), the
fast dephasing time $T_{\phi,1}$ and the slow dephasing time $T_{\phi,2}$
exhibit a clear dependence on qubit frequency. Experimentally, $T_{\phi,1}$
decreases while $T_{\phi,2}$ increases with increasing qubit frequency.
Revisiting the qubit frequency in Eq.\ (\ref{eq:freq_definition}),
it can be expressed as $\Delta_{A}=\sqrt{8E_{C}E_{J\Sigma}}\cdot[(1-d^{2})\cdot\cos^{2}(\pi\Phi_{A}/\Phi_{0})+d^{2}]^{1/4}-E_{C}$,
using the relations $(\omega_{0,A}-\alpha_{A})=\sqrt{8E_{C,A}E_{J,A}}$
and $\alpha_{A}=-E_{C}$, where $\hbar E_{J\Sigma}=\hbar E_{J1}+\hbar E_{J2}$
denotes the total Josephson energy and $\hbar E_{C}$ is the charging
energy. We model RTCE noise $\beta_{\epsilon}(t)$ as entering the
system through the $E_{J\Sigma}$ channel in the form $E_{J\Sigma}\cdot[1+\beta_{\epsilon}(t)]$,
and flux noise $\beta_{\phi}(t)$ as accompanying the flux via $[\text{\ensuremath{\Phi_{A}}}(t)/\Phi_{0}]\cdot[1+\beta_{\phi}(t)]$.
Substituting $E_{J\Sigma}\cdot[1+\beta_{\epsilon}(t)]$ and $[\text{\ensuremath{\Phi_{A}}}(t)/\Phi_{0}]\cdot[1+\beta_{\phi}(t)]$
for $E_{J\Sigma}$ and $\text{\ensuremath{\Phi_{A}}}(t)/\Phi_{0}$,
respectively, in the ideal qubit frequency $\Delta_{A}$ of Eq.\ (\ref{eq:freq_definition}),
we extract the lowest-order noise contributions under the small-noise
approximation, which are then used to construct the corresponding
model noise Hamiltonians. For RTCE noise $\beta_{\epsilon}(t)$, the
lowest-order contribution is proportional to $(\Delta_{A}-\alpha_{A})$,
which enhances dephasing and thereby reduces $T_{\phi,1}$ as $\Delta_{A}$
increases, consistent with the experimental trend. To adjust the slope
of $T_{\phi,1}$ versus qubit frequency, we introduce an additional
exponential factor $\eta_{\epsilon}/2$ on the base $(\Delta_{A}-\alpha_{A})$,
yielding the model RTCE noise Hamiltonian $\mathcal{H}_{\epsilon,A}(t)$
as given in Eq.\ (\ref{eq:noise_H_RTCE_noise}). For flux noise $\beta_{\phi}(t)$,
the corresponding model Hamiltonian $\mathcal{H}_{\phi,A}(t)$ is
presented in Eq.\ (\ref{eq:noise_H_flux_noise}). Here, both $\Phi_{A}/\Phi_{0}$
and $d\Delta_{A}/d(\Phi_{A}/\Phi_{0})$ decrease with increasing $\Delta_{A}$,
mitigating dephasing and thereby extending $T_{\phi_{2}}$, in agreement
with the experimental observations. In addition, we model the RTCE
noise $\beta_{\epsilon}(t)$ with a white noise spectrum $S_{\epsilon}(\omega)=S_{\epsilon,\mathrm{w}}$
and the flux noise $\beta_{\phi}(t)$ with a $1/f$ noise spectrum
$S_{\phi}(\omega)=S_{1/f}/f$ , where the parameters $S_{\epsilon,\mathrm{w}}$
and $S_{1/f}$ are employed to calibrate the respective noise strengths.

\begin{align}
\mathcal{H}_{\epsilon,A}(t) & /\hbar=\beta_{\epsilon}(t)\left[\Delta_{A}(t)-\alpha_{A}\right]^{\eta_{\epsilon}/2}a_{A}^{\dagger}a_{A},\label{eq:noise_H_RTCE_noise}\\
\mathcal{H}_{\phi,A}(t) & /\hbar=\beta_{\phi}(t)\dfrac{d\Delta_{A}(t)}{d(\Phi_{A}(t)/\Phi_{0})}\frac{\Phi_{A}(t)}{\Phi_{0}}a_{A}^{\dagger}a_{A}.\label{eq:noise_H_flux_noise}
\end{align}

Next, we utilize the noise models developed above, parameterized by
$\eta_{\epsilon}$, $S_{\epsilon,\mathrm{w}}$, and $S_{1/f}$, to
derive analytical estimates for the dephasing times $T_{\phi,1}$
and $T_{\phi,2}$. We start from the Ramsey decay envelope, whose
definition and derivation are provided in Appendix\ \ref{appendix:estimated_dephasing_times}.
For $\left\langle \varphi^{2}(t)\right\rangle \ll1$, the Ramsey decay
envelope can be approximated as $V(t)\approx\exp\left[-\left\langle \varphi^{2}(t)\right\rangle /2\right]$,
where $\left\langle \varphi^{2}(t)\right\rangle $ denotes the ensemble
average of $\varphi^{2}(t)$. Here, $\varphi(t)=\int_{0}^{t}\left(\text{\ensuremath{\beta_{\epsilon,\mathrm{freq}}}}(t')+\beta_{\phi,\mathrm{freq}}(t')\right)dt'$
with $\beta_{\epsilon,\mathrm{freq}}(t)\equiv\beta_{\epsilon}(t)\left[\Delta_{A}-\alpha_{A}\right]^{\eta_{\epsilon}/2}$
and $\beta_{\phi,\mathrm{freq}}(t)\equiv\beta_{\phi}(t)\left[d\Delta_{A}/d(\Phi_{A}/\Phi_{0})\right]\left[\text{\ensuremath{\Phi_{A}}}/\Phi_{0}\right]$,
corresponding to the noise Hamiltonians $\mathcal{H}_{\epsilon,A}(t)$
and $\mathcal{H}_{\phi,A}(t)$ in Eqs.\ (\ref{eq:noise_H_RTCE_noise})
and (\ref{eq:noise_H_flux_noise}), respectively. Assuming $\beta_{\epsilon}(t)$
and $\text{\ensuremath{\beta_{\phi}}}(t)$ are statistically independent,
we have $\left\langle \varphi^{2}(t)\right\rangle =\left\langle \varphi_{\epsilon}^{2}(t)\right\rangle +\left\langle \varphi_{\phi}^{2}(t)\right\rangle $,
where $\varphi_{\epsilon}(t)\equiv\int_{0}^{t}\text{\ensuremath{\beta_{\epsilon,\mathrm{freq}}}}(t')dt'$
and $\varphi_{\phi}(t)\equiv\int_{0}^{t}\text{\ensuremath{\beta_{\phi,\mathrm{freq}}}}(t')dt'$.
Consequently, the Ramsey decay envelope can be expressed as

\begin{equation}
V(t)\approx\exp\left[-\left\langle \varphi_{\epsilon}^{2}(t)\right\rangle /2-\left\langle \varphi_{\phi}^{2}(t)\right\rangle /2\right].\label{eq:Ramsey_envelope}
\end{equation}
We then invoke the Wiener--Khinchin theorem, $C(t_{1},t_{2})=\left\langle \beta(t_{1})\beta(t_{2})\right\rangle =\frac{1}{2\pi}\int_{-\infty}^{\infty}S(\omega)e^{-i\omega(t_{1}-t_{2})}d\omega$,
where $C(t_{1},t_{2})$ and $S(\omega)$ denote the correlation function
and the power spectral density of $\beta(t)$, respectively, to derive

\begin{align}
 & \left\langle \varphi_{\epsilon}^{2}(t)\right\rangle =\nonumber \\
 & \dfrac{1}{2\pi}\left[\Delta_{A}-\alpha_{A}\right]^{\eta_{\epsilon}}\int_{-\infty}^{\infty}d\omega S_{\epsilon}(\omega)\dfrac{1}{(\dfrac{\omega}{2})^{2}}\sin^{2}(\dfrac{\omega}{2}t),\label{eq:phi_2_RTCE_noise}\\
 & \left\langle \varphi_{\phi}^{2}(t)\right\rangle =\nonumber \\
 & \dfrac{1}{2\pi}\left[\dfrac{d\Delta_{A}}{d(\Phi_{A}/\Phi_{0})}\dfrac{\Phi_{A}}{\Phi_{0}}\right]^{2}\int_{-\infty}^{\infty}d\omega S_{\phi}(\omega)\dfrac{1}{(\dfrac{\omega}{2})^{2}}\sin^{2}(\dfrac{\omega}{2}t).\label{eq:phi_2_flux_noise}
\end{align}
By substituting the noise spectra $S_{\epsilon}(\omega)=S_{\epsilon,\mathrm{w}}$
and $S_{\phi}(\omega)=S_{1/f}/f$, introduced above, into $\left\langle \varphi_{\epsilon}^{2}(t)\right\rangle $
and $\left\langle \varphi_{\phi}^{2}(t)\right\rangle $, respectively,
the Ramsey decay envelope in Eq.\ (\ref{eq:Ramsey_envelope}) can
be approximated as $V(t)\approx\exp[-t/T_{\phi,1,\epsilon}-(t/T_{\phi,2})^{2}]$,
yielding the estimated dephasing times

\begin{align}
T_{\phi,1,\epsilon} & =2/\left[(\Delta_{A}-\alpha_{A})^{\eta_{\epsilon}}\cdot S_{\epsilon,\mathrm{w}}\right],\label{eq:T_phi_1_e_est}\\
T_{\phi,2} & =1/\left\{ \left[\dfrac{d\Delta_{A}}{d(\Phi_{A}/\Phi_{0})}\dfrac{\Phi_{A}}{\Phi_{0}}\right]\sqrt{S_{1/f}\cdot C_{\mathrm{const}}}\right\} ,\label{eq:T_phi_2_est}
\end{align}
with the constant $C_{\mathrm{const}}\approx22.8039$. Upon further
including the fast dephasing contribution from energy relaxation with
characteristic time $T_{1}$, the Ramsey decay envelope becomes $V(t)\approx\exp[-t/T_{\phi,1}-(t/T_{\phi,2})^{2}]$,
where $T_{\phi,1}=2T_{1}\cdot T_{\phi,1,\epsilon}/(2T_{1}+T_{\phi,1,\epsilon})$.
Detailed derivations of the above expressions are also provided in
Appendix\ \ref{appendix:estimated_dephasing_times}.

We then adjust the parameters $\eta_{\epsilon}$ and $S_{\epsilon,\mathrm{w}}$
in the estimated $T_{\phi,1,\epsilon}$ {[}Eq.\ (\ref{eq:T_phi_1_e_est}){]}
with $T_{1}=30\,\mu\mathrm{s}$ to match the experimentally measured
$T_{\phi,1}$, and similarly tune the parameter $S_{1/f}$ in the
estimated $T_{\phi,2}$ {[}Eq.\ (\ref{eq:T_phi_2_est}){]} to reproduce
the experimental $T_{\phi,2}$. The resulting best-fit estimates for
$T_{\phi,1}$ and $T_{\phi,2}$ accurately capture the observed trends
of the experimental coherence times over the qubit frequency range
from $4.15\,\mathrm{GHz}$ to $5.09\,\mathrm{GHz}$, as illustrated
in Fig.\ \ref{fig:noise_characterication}(a). The corresponding
best-fit RTCE white noise spectrum, $S_{E,\mathrm{w}}=\left[\Delta_{A}-\alpha_{A}\right]^{\eta_{\epsilon}}S_{\epsilon,\mathrm{w}}$,
for the frequency-unit noise $\beta_{\epsilon,\mathrm{freq}}(t)\equiv\beta_{\epsilon}(t)\left[\Delta_{A}-\alpha_{A}\right]^{\eta_{\epsilon}/2}$
under the best-fit values of $S_{\epsilon,\mathrm{w}}$ and $\eta_{\epsilon}=3.73$,
is presented in Fig.\ \ref{fig:noise_characterication}(b). As $\Delta_{A}$
increases $4\,\mathrm{GHz}$ to $5.1\,\mathrm{GHz}$, the best-fit
$S_{E,\mathrm{w}}$ rises approximately from $(2\pi)^{2}\times3400$
to $(2\pi)^{2}\times8000\,\mathrm{Hz}^{2}/\mathrm{Hz}$, accounting
for the decreasing trend of $T_{\phi,1}$ through the relation $T_{\phi,1,\epsilon}=2/S_{E,\mathrm{w}}$
from Eq.\ (\ref{eq:T_phi_1_e_est}). Similarly, the best-fit flux-noise
spectrum at $1\,\mathrm{Hz}$, $S_{\Phi,\mathrm{1Hz}}=(\Phi_{A}/\Phi_{0})^{2}S_{1/f}$,
for the flux noise $\beta_{\Phi}(t)=(\Phi_{A}/\Phi_{0})\beta_{\phi}(t)$
under the best-fit $S_{1/f}$ is presented in Fig.\ \ref{fig:noise_characterication}(d).
With increasing $\Delta_{A}$ from $4\,\mathrm{GHz}$ to $5.1\,\mathrm{GHz}$,
the best-fit $S_{\Phi,\mathrm{1Hz}}$ decreases from $\sim10^{-11}$
to $\sim10^{-12}$, while the derivative $d\Delta_{A}/d(\Phi_{A}/\Phi_{0})$
also diminishes. Both effects contribute to the increasing trend of
$T_{\phi,2}$, consistent with the relation $T_{\phi,2}\propto1/\left\{ \left[d\Delta_{A}/d(\Phi_{A}/\Phi_{0})\right]\cdot\sqrt{S_{\Phi,\mathrm{1Hz}}}\right\} $
from Eq.\ (\ref{eq:T_phi_2_est}).

\subsubsection{Ramsey dephasing simulation}

\label{subsec:Ramsey_dephasing_simulation}

In this section, we employ the noise models characterized in Sec.\ \ref{subsec:Noise_models}
to generate realizations of the RTCE noise $\beta_{\epsilon}(t)$
and the flux noise $\beta_{\phi}(t)$ for simulating Ramsey dephasing
experiments under the experimental system parameters \citep{BarendsKellyMegrantEtAl2014}.
We then follow the experimental procedure described in Appendix\ \ref{appendix:estimated_dephasing_times}
to extract the simulated $T_{\phi,1}$ and $T_{\phi,2}$, providing
a further validation of the noise models.

To simulate the system dynamics, we employ the Lindblad-form master
equation

\begin{equation}
\dot{\rho}=-\dfrac{i}{\hbar}\left[\mathcal{H}_{A},\rho\right]+\dfrac{1}{T_{1}}\left\{ a_{A}\rho a_{A}^{\dagger}-\frac{1}{2}a_{A}^{\dagger}a_{A}\rho-\frac{1}{2}\rho a_{A}^{\dagger}a_{A}\right\} ,\label{eq:Lindblad_master_eq}
\end{equation}
incorporating an energy relaxation time $T_{1}=30\,\mu\mathrm{s}$
and the characterized RTCE noise $\beta_{\epsilon}(t)$ and flux noise
$\beta_{\phi}(t)$ within the realistic single-qubit Hamiltonian $\mathcal{H}_{A}(t)=\hbar\Delta_{A}^{\mathrm{\phi}}(t)a_{A}^{\dagger}a_{A}+\hbar(\alpha_{A}/2)\cdot a_{A}^{\dagger}a_{A}(a_{A}^{\dagger}a_{A}-1)+\mathcal{H}_{\epsilon,A}(t)$.
The RTCE noise $\beta_{\epsilon}(t)$ enters the system through the
model noise Hamiltonian $\mathcal{H}_{\epsilon,A}(t)$ in Eq.\ (\ref{eq:noise_H_RTCE_noise}).
The flux noise $\beta_{\phi}(t)$ modulates the flux $\text{\ensuremath{\Phi_{A}}}(t)/\Phi_{0}$
through $\Delta_{A}^{\mathrm{\phi}}(t)$, which is obtained by replacing
$\text{\ensuremath{\Phi_{A}}}(t)/\Phi_{0}$ with $[\text{\ensuremath{\Phi_{A}}}(t)/\Phi_{0}]\cdot[1+\beta_{\phi}(t)]$
in the ideal qubit frequency $\Delta_{A}(t)$ in Eq.\ (\ref{eq:freq_definition}).
Detailed discussions of the realizations of $\beta_{\epsilon}(t)$
and $\beta_{\phi}(t)$ are presented in the following paragraphs.
To obtain the simulated Ramsey decay envelope $V(t)$ from the ensemble-averaged
Ramsey fringes, we evolve the system dynamics according to Eq.\ (\ref{eq:Lindblad_master_eq})
for an ensemble of $5000$ realizations of $\beta_{\epsilon}(t)$
and $\beta_{\phi}(t)$. The resulting $V(t)$ is then fitted to the
form $\exp[-t/T_{\phi,1}-(t/T_{\phi,2})^{2}]$ to determine the simulated
$T_{\phi,1}$ and $T_{\phi,2}$ at a fixed qubit frequency $\Delta_{A}$.
This procedure is repeated for $\Delta_{A}/2\pi$ ranging from $4.01\,\mathrm{GHz}$
to $5.09\,\mathrm{GHz}$ in $0.02\,\mathrm{GHz}$ increments. The
corresponding simulated values of $T_{\phi,1}$ and $T_{\phi,2}$
are presented in Fig.\ \ref{fig:noise_characterication}(a).

For the RTCE noise $\beta_{\epsilon}(t)$, the white-noise behavior
with a flat spectrum $S_{\epsilon,\mathrm{w}}$ is characterized via
Ramsey experiments, as described in Sec.\ \ref{subsec:Noise_models}.
However, since these experiments are implemented on a timescale of
$\sim\mu\mathrm{s}$, they can only probe noise at frequencies below
$\sim\mathrm{MHz}$ ($\sim1/\mu\mathrm{s}$). That is, the Ramsey
experiments can only confirm a constant-spectrum behavior for $\beta_{\epsilon}(t)$
at frequencies $f\lesssim1\,\mathrm{MHz}$. To model $\beta_{\epsilon}(t)$
beyond this limit, we employ a single-$\Gamma_{OU}$ Ornstein--Uhlenbeck
(OU) process with spectrum $S_{\epsilon}(f)=(\sigma_{\epsilon}^{2}/\pi)\cdot\Gamma_{\epsilon}/(\Gamma_{\epsilon}^{2}+f^{2})$,
as introduced in Appendix\ \ref{appendix:noise_generation_OU}. The
parameter $\Gamma_{\epsilon}$ controls the frequency range over which
the spectrum approximates white noise. Specifically, for $f\gg\Gamma_{\epsilon}$,
$S_{\epsilon}(f)\approx(\sigma_{\epsilon}^{2}/\pi)\cdot\Gamma_{\epsilon}/f^{2}$,
exhibiting a $1/f^{2}$ decay, whereas for $f\ll\Gamma_{\epsilon}$,
$S_{\epsilon}(f)$ is approximately constant, $S_{\epsilon}(f)\approx\sigma_{\epsilon}^{2}/(\pi\Gamma_{\epsilon})=S_{\epsilon,\mathrm{w}}$.
The parameter $\sigma_{\epsilon}=\sqrt{\pi\Gamma_{\epsilon}S_{\epsilon,\mathrm{w}}}$
defines the standard deviation of the simulated $\beta_{\epsilon}(t)$.
For the dephasing simulations, we generate realizations of $\beta_{\epsilon}(t)$
with $\Gamma_{\epsilon}=10^{6.5}\,\mathrm{Hz}$ and $\Gamma_{\epsilon}=10^{10.5}\,\mathrm{Hz}$.
As shown in Fig.\ \ref{fig:noise_characterication}(c), for $\Gamma_{\epsilon}=10^{6.5}\,\mathrm{Hz}$,
the noise spectrum $S_{E}(f)=\left[\Delta_{A}-\alpha_{A}\right]^{\eta_{\epsilon}}S_{\epsilon}(f)$
at $\Delta_{A}/2\pi=5\,\mathrm{GHz}$ extends the flat spectral region
$S_{E,\mathrm{w}}\approx\left[\Delta_{A}-\alpha_{A}\right]^{\eta_{\epsilon}}S_{\epsilon,\mathrm{w}}$
up to $f\approx1\,\mathrm{MHz}$, consistent with the Ramsey detection
limit. In contrast, for $\Gamma_{\epsilon}=10^{10.5}\,\mathrm{Hz}$,
the constant spectrum extends well beyond this limit, up to $\sim10\,\mathrm{GHz}$.
The simulated $T_{\phi,1}$ for both $\Gamma_{\epsilon}=10^{6.5}\,\mathrm{Hz}$
and $\Gamma_{\epsilon}=10^{10.5}\,\mathrm{Hz}$ follows the trend
of the estimated $T_{\phi,1}$ across the qubit frequency range from
$\Delta_{A}/2\pi=4.01\,\mathrm{GHz}$ to $5.09\,\mathrm{GHz}$, with
distributions within the experimental variation, as shown in Fig.\ \ref{fig:noise_characterication}(a).
Decreasing $\Gamma_{\epsilon}$ to $10^{6}\,\mathrm{Hz}$ results
in the constant-spectrum region decaying below the Ramsey sensitivity
limit, producing simulated $T_{\phi,1}$ values that exceed the estimated
and experimental measurements. These results indicate that our noise
model with $\Gamma_{\epsilon}\gtrsim10^{6.5}\,\mathrm{Hz}$ reliably
captures the characteristics of $\beta_{\epsilon}(t)$. In Sec.\ \ref{subsec:Noise_characterization_from_CZ_infid_budget},
we further refine the characterization of $\beta_{\epsilon}(t)$ by
calibrating $\Gamma_{\epsilon}$ against the experimental CZ gate
error budget \citep{BarendsKellyMegrantEtAl2014}. Because the gate
operates on a timescale of $\sim40\,\mathrm{ns}$, the corresponding
spectral sensitivity extends to $\sim25\,\mathrm{MHz}$ ($\sim1/40\,\mathrm{ns}$),
significantly higher than that accessible via Ramsey measurements.

For superconducting qubits, the $1/f$ spectrum of flux noise has
been experimentally observed over a wide frequency range $f$, approximately
from $10^{-5}\,\mathrm{Hz}$ to $1\,\mathrm{GHz}$ \citep{Quintana2017}.
To generate flux noise $\beta_{\phi}(t)$ with a $1/f$ spectrum for
dephasing simulations, we employ a multi-$\Gamma_{OU}$ OU process
with a composite noise spectrum $S_{\phi}(f)=\Sigma_{j}(\sigma_{j}^{2}/\pi)\cdot\Gamma_{j}/(\Gamma_{j}^{2}+f^{2})$.
A detailed description of the multi-$\Gamma_{OU}$ OU process, including
the parameters $\sigma_{j}$ and $\Gamma_{j}$, is provided in Appendix\ \ref{appendix:noise_generation_OU}.
The resulting composite spectrum $S_{\Phi}(f)=(\Phi_{A}/\Phi_{0})^{2}S_{\phi}(f)$
at $\Delta_{A}/2\pi=5\,\mathrm{GHz}$ is shown in Fig.\ \ref{fig:noise_characterication}(e),
demonstrating that the $1/f$ behavior is reproduced over the frequency
range approximately from $10^{-5}\,\mathrm{Hz}$ to $1\,\mathrm{GHz}$,
in agreement with experimental observations. Outside this range, the
spectrum approaches a constant for $f\lesssim10^{-5}\,\mathrm{Hz}$
and transitions to a $1/f^{2}$ decay for $f\gtrsim1\,\mathrm{GHz}$.
In the simulation, we assume that the RTCE noise $\beta_{\epsilon}(t)$
and the flux noise $\beta_{\phi}(t)$ are independent, which is reflected
in the fact that the simulated $T_{\phi,2}$ is largely insensitive
to the choice of $\Gamma_{\epsilon}=10^{6.5}\,\mathrm{Hz}$ or $\Gamma_{\epsilon}=10^{10.5}\,\mathrm{Hz}$
used to generate $\beta_{\epsilon}(t)$. Figure\ \ref{fig:noise_characterication}(a)
displays the simulated $T_{\phi,2}$ for $\Gamma_{\epsilon}=10^{6.5}\,\mathrm{Hz}$,
which follows the trend of the estimated $T_{\phi,2}$ and agrees
well with the experimental $T_{\phi,2}$. These results further confirm
that our noise model accurately captures the characteristics of the
flux noise $\beta_{\phi}(t)$.

\subsection{Noise characterization from the CZ gate error budget}

\label{subsec:Noise_characterization_from_CZ_infid_budget}

In Sec.\ \ref{subsec:Noise_characterization_dephasing}, we characterize
the dephasing noises using Ramsey experiments within the framework
of our developed noise models. However, owing to the limited duration
of the Ramsey measurements, the parameter $\Gamma_{\epsilon}$, which
sets the bandwidth of the constant white-noise spectrum in the RTCE
noise $\beta_{\epsilon}(t)$, remains undetermined. To resolve this
issue, we first derive the infidelity contributions arising from the
characterized noises in Sec.\ \ref{subsec:J2_noises}. Subsequently,
in Sec.\ \ref{subsec:further_noise_characterization}, we employ
the simulated infidelity contributions to calibrate the parameter
$\Gamma_{\epsilon}$ by referring to the CZ gate error budget reported
in the same work \citep{BarendsKellyMegrantEtAl2014} that presented
the Ramsey experiments.

\subsubsection{Infidelity contributions from the characterized noises }

\label{subsec:J2_noises}

In Sec.\ \ref{subsec:Ramsey_dephasing_simulation}, we employ the
Lindblad master equation in Eq.\ (\ref{eq:Lindblad_master_eq}) to
simulate the dynamics of the single qubit $Q_{A}$ in the Ramsey experiment.
To evaluate the error contributions to the CZ gate infidelity, it
is necessary to simulate the two-qubit dynamics. Therefore, we extend
the right-hand side of the Lindblad master equation in Eq.\ (\ref{eq:Lindblad_master_eq})
by including the term $-(i/\hbar)[\mathcal{H}_{B},\rho]+(1/T_{1})\{a_{B}\rho a_{B}^{\dagger}-\frac{1}{2}a_{B}^{\dagger}a_{B}\rho-\frac{1}{2}\rho a_{B}^{\dagger}a_{B}\}$
associated with the second qubit $Q_{B}$ and the coupling term $-(i/\hbar)[\mathcal{H}_{I,AB},\rho]$
describing the $Q_{A}-Q_{B}$ interaction. In the simulations, we
assume identical energy relaxation times $T_{1}$ for both qubits.
To distinguish the noise channels acting on the two qubits, we introduce
subscripts $A$ and $B$ for the RTCE noise $\beta_{\epsilon}(t)$
and the flux noise $\beta_{\phi}(t)$, denoted as $\beta_{\epsilon,A/B}(t)$
and $\beta_{\phi,A/B}(t)$, representing noise entering the system
through $Q_{A}$ and $Q_{B}$, respectively. Analogous to the treatment
of $Q_{A}$ in Sec.\ \ref{subsec:Ramsey_dephasing_simulation}, the
RTCE noise $\beta_{\epsilon,B}(t)$ and the flux noise $\beta_{\phi,B}(t)$
enter the system through the realistic Hamiltonian of $Q_{B}$, $\mathcal{H}_{B}(t)=\hbar\Delta_{B}^{\mathrm{\phi}}(t)a_{B}^{\dagger}a_{B}+\hbar[\Delta_{B}(t)-\alpha_{B}]^{\eta_{\epsilon}/2}\beta_{\epsilon,B}(t)a_{B}^{\dagger}a_{B}+\hbar(\alpha_{B}/2)a_{B}^{\dagger}a_{B}(a_{B}^{\dagger}a_{B}-1)$.
The $\Delta_{B}^{\mathrm{\phi}}(t)$ is obtained by substituting $[\text{\ensuremath{\Phi_{B}}}(t)/\Phi_{0}]\cdot[1+\beta_{\phi,B}(t)]$
for $\text{\ensuremath{\Phi_{B}}}(t)/\Phi_{0}$ in the ideal qubit
frequency $\Delta_{B}(t)$ given in Eq.\ (\ref{eq:freq_definition}).
The $Q_{A}-Q_{B}$ coupling Hamiltonian is given by $\mathcal{H}_{I,AB}(t)=g_{AB}(-i)(a_{A}-a_{A}^{\dagger})(-i)(a_{B}-a_{B}^{\dagger})$.

To evaluate the gate infidelity, we first vectorize the two-qubit
Lindblad master equation to obtain the propagator $G(t)$ corresponding
to the vectorized density matrix $\mathrm{vec(\rho)}$; a detailed
derivation is provided in Appendix\ \ref{appendix:LindBlad_master_eq_infid}.
We then project $G(t)$ onto the subspace spanned by the two-qubit
computational basis states, yielding the projected propagator $G_{16\times16}(t)$.
The ensemble-averaged infidelity for the two-qubit gate is subsequently
defined in terms of $G_{16\times16}(t)$ as

\begin{equation}
\left\langle \mathcal{I}\right\rangle \equiv\left\langle 1-\dfrac{1}{16}\mathrm{Re}\left\{ \mathrm{Tr}\left[G_{T}^{\dagger}G_{16\times16}(t_{f})\right]\right\} \right\rangle ,\label{eq:ensemble_infid_master_eq}
\end{equation}
where $G_{T}=U{}_{T}^{\star}\otimes U_{T}$ denotes the target gate
for the propagator $G_{16\times16}(t_{f})$, and $\mathrm{Re}$ indicates
taking the real part of the quantity on which it acts. In our case,
$t_{f}\leq40\,\mathrm{ns}$, $T_{1}=30\,\mu\mathrm{s}$, and $t_{f}/T_{1}\lesssim10^{-3}$,
so the ensemble-averaged infidelity $\left\langle \mathcal{I}\right\rangle $
in Eq.\ (\ref{eq:ensemble_infid_master_eq}) can be approximated
as 
\begin{equation}
\left\langle \mathcal{I}\right\rangle \approx\left\langle 1-\dfrac{1}{16}\left|\mathrm{Tr}\left[U_{T}^{\dagger}U_{4\times4}(t_{f})\right]\right|^{2}\right\rangle +\dfrac{t_{f}}{T_{1}}.\label{eq:ensemble_infid_schrodinger_eq}
\end{equation}
Here, $U_{4\times4}(t_{f})$ denotes the realistic propagator in the
subspace spanned by the two-qubit computational basis states, obtained
by projecting the full propagator $U(t)$ simulated from the Schrödinger
equation $\dot{U}(t)=(-i/\hbar)\cdot\mathcal{H}(t)U(t)$ with the
Hamiltonian $\mathcal{H}(t)=\mathcal{H}_{A}(t)+\mathcal{H}_{B}(t)+\mathcal{H}_{I,AB}(t)$.
The detailed derivation of Eq.\ (\ref{eq:ensemble_infid_schrodinger_eq})
is also provided in Appendix\ \ref{appendix:LindBlad_master_eq_infid}.

Assuming that the RTCE noises $\beta_{\epsilon,A/B}(t)$ and the flux
noises $\beta_{\phi,A/B}(t)$ are small and statistically independent,
the ensemble-averaged infidelity $\left\langle \mathcal{I}\right\rangle $
in Eq.\ (\ref{eq:ensemble_infid_schrodinger_eq}) can be further
approximated and expanded as

\begin{equation}
\left\langle \mathcal{I}\right\rangle \approx J_{1}+\left\langle J_{2,T_{1}}\right\rangle +\left\langle J_{2,\epsilon}\right\rangle +\left\langle J_{2,\phi}\right\rangle ,\label{eq:ensemble_infid_J2}
\end{equation}
following the approach introduced in the robust control paper \citep{HuangGoan2017}.
Here, $J_{1}$ denotes the ideal gate infidelity defined in Eq.\ (\ref{eq:J1_definition})
and $\left\langle J_{2,T_{1}}\right\rangle \equiv t_{f}/T_{1}$ represents
the contribution from energy relaxation. The term $\left\langle J_{2,\epsilon}\right\rangle =\left\langle J_{2,\epsilon,A}\right\rangle +\left\langle J_{2,\epsilon,B}\right\rangle $
corresponds to the infidelity arising from the RTCE noises $\beta_{\epsilon,A}(t)$
and $\beta_{\epsilon,B}(t)$, associated with the individual contributions
$\left\langle J_{2,\epsilon,A}\right\rangle $ and $\left\langle J_{2,\epsilon,B}\right\rangle $,
respectively. Similarly, $\left\langle J_{2,\phi}\right\rangle =\left\langle J_{2,\phi,A}\right\rangle +\left\langle J_{2,\phi,B}\right\rangle $
accounts for the infidelity due to the flux noises $\beta_{\phi,A}(t)$
and $\beta_{\phi,B}(t)$, corresponding to $\left\langle J_{2,\phi,A}\right\rangle $
and $\left\langle J_{2,\phi,B}\right\rangle $, respectively. The
total infidelity can thus be expressed compactly as $\left\langle \mathcal{I}\right\rangle \approx J_{1}+\left\langle J_{2}\right\rangle $,
where $\left\langle J_{2}\right\rangle =\left\langle J_{2,T_{1}}\right\rangle +\left\langle J_{2,\epsilon}\right\rangle +\left\langle J_{2,\phi}\right\rangle $
represents the total contribution from decoherence in this work.

Next, we evaluate the gate infidelity contributions $\left\langle J_{2,\epsilon}\right\rangle $
and $\left\langle J_{2,\phi}\right\rangle $ arising from dephasing
noises using the formalism introduced in Ref.\ \citep{HuangGoan2017}.
For a two-qubit system, the infidelity contribution associated with
a noise source $\beta_{j}(t)$ can be expressed as 
\begin{align}
\left\langle J_{2,j}\right\rangle = & \int_{0}^{t_{f}}dt_{1}\int_{0}^{t_{1}}dt_{2}C_{j}(t_{1},t_{2})\nonumber \\
 & \cdot\left\{ \mathrm{Re}[\mathrm{Tr}\left(R_{j}(t_{1})R_{j}(t_{2})\right)_{4\times4}]/2-\right.\nonumber \\
 & \left.\mathrm{Tr}\left(R_{j,4\times4}(t_{1})\right)\cdot\mathrm{Tr}\left(R_{j,4\times4}(t_{2})\right)/8\right\} .\label{eq:J2_general_format}
\end{align}
The noise $\beta_{j}(t)$ enters the system through the noise Hamiltonian
$\mathcal{H}_{N,j}(t)=\hbar\beta_{j}(t)H_{N,j}(t)$, where $H_{N,j}(t)$
specifies the operator through which the noise $\beta_{j}(t)$ couples
to the system. In the expression for $\left\langle J_{2,j}\right\rangle $,
the statistical properties of the noise $\beta_{j}(t)$ are characterized
by the correlation function $C_{j}(t_{1},t_{2})=\left\langle \beta_{j}(t_{1})\beta_{j}(t_{2})\right\rangle $,
while the noise-system coupling form $H_{N,j}(t)$ is encoded in the
operator $R_{j}(t)=U_{I}^{\dagger}(t)H_{N,j}(t)U_{I}(t)$, where $U_{I}(t)$
denotes the ideal propagator. The matrices $R_{j.4\times4}(t)$ and
$\left(R_{j}(t_{1})R_{j}(t_{2})\right)_{4\times4}$ are obtained by
projecting $R_{j}(t)$ and $\left(R_{j}(t_{1})R_{j}(t_{2})\right)$,
respectively, onto the subspace spanned by the two-qubit computational
basis states. For the RTCE noises $\beta_{\epsilon,A/B}(t)$ and the
flux noises $\beta_{\phi,A/B}(t)$, the corresponding noise-system
coupling forms $H_{N,j}(t)$ can be expressed as

\begin{align}
H_{\epsilon,A/B}(t) & =\left[\Delta_{A/B}(t)-\alpha_{A/B}\right]^{\eta_{\epsilon}/2}a_{A/B}^{\dagger}a_{A/B},\label{eq:RTCE_noise_H}\\
H_{\phi,A/B}(t) & =\dfrac{d\Delta_{A/B}(t)}{d(\Phi_{A/B}(t)/\Phi_{0})}\dfrac{\Phi_{A/B}(t)}{\Phi_{0}}a_{A/B}^{\dagger}a_{A/B},\label{eq:Flux_noise_H}
\end{align}
which are obtained from the noise Hamiltonians $\mathcal{H}_{\epsilon,A}(t)=\hbar\beta_{\epsilon,A}(t)H_{\epsilon,A}(t)$
and $\mathcal{H}_{\phi,A}(t)=\hbar\beta_{\phi,A}(t)H_{\phi,A}(t)$
shown in Eqs.\ (\ref{eq:noise_H_RTCE_noise}) and (\ref{eq:noise_H_flux_noise}),
respectively. The noise Hamiltonians $\mathcal{H}_{\epsilon,B}(t)$
and $\mathcal{H}_{\phi,B}(t)$ are taken to have the same forms as
$\mathcal{H}_{\epsilon,A}(t)$ and $\mathcal{H}_{\phi,A}(t)$. The
matrices $R_{j}(t)$ for $\beta_{\epsilon,A/B}(t)$ and $\beta_{\phi,A/B}(t)$
are given by $R_{\epsilon,A/B}(t)=U_{I}^{\dagger}(t)H_{\epsilon,A/B}(t)U_{I}(t)$
and $R_{\phi,A/B}(t)=U_{I}^{\dagger}(t)H_{\phi,A/B}(t)U_{I}(t)$,
respectively. The corresponding noise correlation functions are $C_{\epsilon,A/B}(t_{1},t_{2})=\left\langle \beta_{\epsilon,A/B}(t_{1})\beta_{\epsilon,A/B}(t_{2})\right\rangle $
and $C_{\phi,A/B}(t_{1},t_{2})=\left\langle \beta_{\phi,A/B}(t_{1})\beta_{\phi,A/B}(t_{2})\right\rangle $.
These correlation functions can be derived from the characterized
noise spectra shown in Figs.\ \ref{fig:noise_characterication}(c)
and \ref{fig:noise_characterication}(e) via the Wiener--Khinchin
theorem. Finally, the gate infidelity contributions from fast and
slow dephasing can be evaluated by substituting the corresponding
matrices $R_{j}(t)$ and correlation functions $C_{j}(t_{1},t_{2})$
into Eq.\ (\ref{eq:J2_general_format}) for $\left\langle J_{2,j}\right\rangle $.
Specifically, the fast dephasing contributions $\left\langle J_{2,\epsilon,A/B}\right\rangle $
are obtained using $R_{\epsilon,A/B}(t)$ and $C_{\epsilon,A/B}(t_{1},t_{2})$,
while the slow dephasing contributions $\left\langle J_{2,\phi,A/B}\right\rangle $
are obtained using $R_{\phi,A/B}(t)$ and $C_{\phi,A/B}(t_{1},t_{2})$.

\subsubsection{CZ gate error budget }

\label{subsec:further_noise_characterization}

This section completes the characterization of dephasing in the system.
We use the experimental CZ gate error budget to calibrate the parameter
$\Gamma_{\epsilon}$ in our noise model. The experimental CZ gate
error budget \citep{BarendsKellyMegrantEtAl2014} is for a $40\,\mathrm{ns}$
gate with a measured overall infidelity of approximately $6\times10^{-3}$.
The gate is implemented by keeping $Q_{B}$ at its idle frequency,
$\Delta_{B}=\omega_{\mathrm{Idle},B}$, while tuning $\Delta_{A}(t)$
along a fast adiabatic trajectory using the control method of Ref.\ \citep{MartinisGeller2014}.
The realistic system parameters for this experimental CZ gate implementation
are specified in Sec.\ \ref{sec:High_fidelity_CZ_gate}, where they
are employed for both the simulation of the drift-Hamiltonian evolution
and the $J_{1}$ optimization. The error budget indicates that $55\%$
of the CZ gate infidelity is attributed to decoherence, i.e., $\left\langle J_{2}\right\rangle =\left\langle J_{2.T_{1}}\right\rangle +\left\langle J_{2,\epsilon}\right\rangle +\left\langle J_{2,\phi}\right\rangle \approx3.3\times10^{-3}$,
while the remaining $45\%$ arises from state leakage and control
error, i.e., $J_{1}\approx2.7\times10^{-3}$.

\begin{figure}
\includegraphics[scale=0.52]{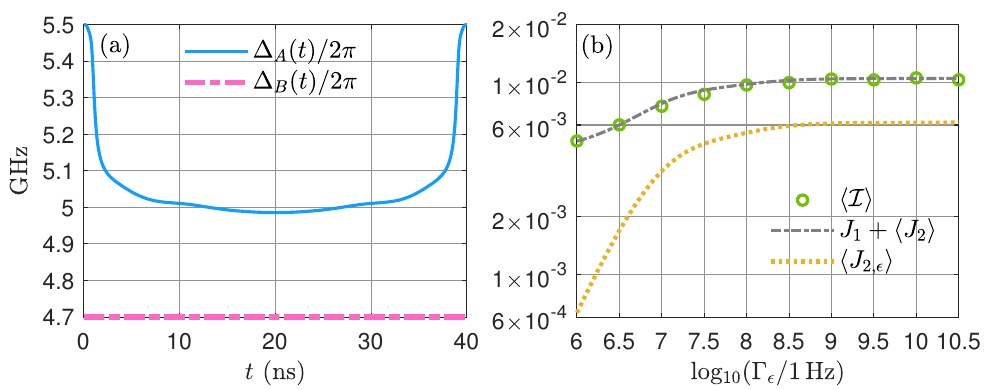}

\caption{Calibration of the parameter $\Gamma_{\epsilon}$. (a) Control pulses
$\Delta_{A}(t)$ (blue solid line) and $\Delta_{B}(t)$ (pink dash-dotted
line) constructed for calibrating $\Gamma_{\epsilon}$ based on the
experimental CZ gate error budget. (b) Corresponding ensemble-averaged
infidelity $\left\langle \mathcal{I}\right\rangle $ (green circles),
total infidelity $J_{1}+\left\langle J_{2}\right\rangle $ (gray dash-dotted
line), and RTCE noise contribution $\left\langle J_{2,\epsilon}\right\rangle $
(yellow dotted line) as a function of $\Gamma_{\epsilon}$ from $10^{6.0}\,\mathrm{Hz}$
to $10^{10.5}\,\mathrm{Hz}$ for the pulses shown in (a).}

\label{fig:noise_char_err_budget} 
\end{figure}

To calibrate the parameter $\Gamma_{\epsilon}$, we first reproduce
the experimental CZ gate with an intrinsic error $J_{1}$ close to
the experimental budget target. We then adjust $\Gamma_{\epsilon}$
in our model to modulate the RTCE noise contribution $\left\langle J_{2,\epsilon}\right\rangle $,
such that the total infidelity $J_{1}+\left\langle J_{2}\right\rangle $
matches the experimentally measured CZ gate infidelity. To reproduce
the experimental CZ gate, we use the system parameters \citep{BarendsKellyMegrantEtAl2014}
and control method \citep{MartinisGeller2014} from its experimental
implementation to construct the pulses $\Delta_{B}=\omega_{\mathrm{Idle},B}$
and $\Delta_{A}(t)=\Delta_{B}-\alpha_{A}+2\sqrt{2}g_{AB}\cdot\cot\theta(t)$,
with $\theta(t)=\theta_{i}+\sum_{n=1}^{n_{\mathrm{max}}}2\lambda_{n}\sin^{2}(\pi nt/t_{f})$,
where $n_{\mathrm{max}}=8$ and $t_{f}=40\,\mathrm{ns}$. By optimizing
the control parameters $\left\{ \lambda_{n}\right\} $ and adjusting
the pulse amplitude, the constructed pulses $\Delta_{A/B}(t)$ shown
in Fig.\ \ref{fig:noise_char_err_budget}(a) achieve an ideal CZ
gate infidelity $J_{1}\approx2.67\times10^{-3}$ with leakage error
$1.49\times10^{-3}$, in close agreement with the experimental budget
$J_{1}\approx2.7\times10^{-3}$ and leakage error $1.5\times10^{-3}$.
We then compute the decoherence contributions for our constructed
pulses using the method outlined in Sec.\ \ref{subsec:J2_noises}
and the characterized parameters of our developed noise models presented
in Sec.\ \ref{subsec:Noise_characterization_dephasing}. The energy
relaxation contribution is $\left\langle J_{2,T_{1}}\right\rangle \approx1.33\times10^{-3}$
for $t_{f}=40\,\mathrm{ns}$ and $T_{1}=30\,\mu\mathrm{s}$, while
the flux noise contribution is $\left\langle J_{2,\phi}\right\rangle \approx3.16\times10^{-4}$.
The RTCE noise contribution $\left\langle J_{2,\epsilon}\right\rangle $
is calculated by varying $\Gamma_{\epsilon}$ from $10^{6.0}\,\mathrm{Hz}$
to $10^{10.5}\,\mathrm{Hz}$, as shown in Fig.\ \ref{fig:noise_char_err_budget}(b).
$\left\langle J_{2,\epsilon}\right\rangle $ increases with $\Gamma_{\epsilon}$
and saturates at approximately $6\times10^{-3}$ for $\Gamma_{\epsilon}\gtrsim10^{8.5}\,\mathrm{Hz}$.
This saturation arises because the CZ gate, implemented with a duration
of $40\,\mathrm{ns}$, cannot probe noise at frequencies significantly
exceeding $\sim25\,\mathrm{MHz}$ ($\sim1/40\,\mathrm{ns}$). This
behavior can be further understood by transforming $\left\langle J_{2,\epsilon}\right\rangle $,
defined in the time domain in Eq.\ (\ref{eq:J2_general_format}),
to the frequency domain: $\left\langle J_{2,\epsilon}\right\rangle =(1/2\pi)\int_{-\infty}^{\infty}d\omega S_{\epsilon}(\omega)F_{\epsilon}(\omega)/\omega^{2}$
\citep{HuangGoan2017}. Here, we temporarily omit the subscripts $A/B$
for clarity. In this representation, $\left\langle J_{2,\epsilon}\right\rangle $
can be viewed as the integral of the effective overlap between the
RTCE noise spectrum $S_{\epsilon}(\omega)$ and the function $F_{\epsilon}(\omega)/\omega^{2}$.
The function $F_{\epsilon}(\omega)/\omega^{2}$ dominates over a frequency
range associated with the spectrum of the control pulses $\Delta_{A/B}(t)$.
When the cutoff frequency $\Gamma_{\epsilon}$ of the flat white-noise
spectrum in $S_{\epsilon}(\omega)$ is below this dominant region
of $F_{\epsilon}(\omega)/\omega^{2}$, increasing $\Gamma_{\epsilon}$
enlarges the effective overlap, leading to a rise in $\left\langle J_{2,\epsilon}\right\rangle $.
Conversely, once $\Gamma_{\epsilon}$ exceeds the dominant region
of $F_{\epsilon}(\omega)/\omega^{2}$, further increases in $\Gamma_{\epsilon}$
do not affect the overlap, and $\left\langle J_{2,\epsilon}\right\rangle $
saturates at a fixed value.

Next, we combine the ideal gate infidelity $J_{1}$ with the total
decoherence contribution $\left\langle J_{2}\right\rangle =\left\langle J_{2,T_{1}}\right\rangle +\left\langle J_{2,\epsilon}\right\rangle +\left\langle J_{2,\phi}\right\rangle $
to evaluate the total infidelity $J_{1}+\left\langle J_{2}\right\rangle $
as a function of $\Gamma_{\epsilon}$ in the range from $10^{6.0}\,\mathrm{Hz}$
to $10^{10.5}\,\mathrm{Hz}$, as shown in Fig.\ \ref{fig:noise_char_err_budget}(b).
In addition, we simulate the ensemble-averaged infidelity $\left\langle \mathcal{I}\right\rangle $
using Eq.\ (\ref{eq:ensemble_infid_schrodinger_eq}) over an ensemble
of $1000$ realizations of $\beta_{\epsilon,A/B}(t)$ and $\beta_{\phi,A/B}(t)$
to verify the calculation of $J_{1}+\left\langle J_{2}\right\rangle $.
The results exhibit excellent agreement, i.e., $\left\langle \mathcal{I}\right\rangle \approx J_{1}+\left\langle J_{2}\right\rangle $,
as illustrated in Fig.\ \ref{fig:noise_char_err_budget}(b), thereby
validating the method for evaluating the infidelity contributions
introduced in Sec.\ \ref{subsec:J2_noises}. The total infidelity
$J_{1}+\left\langle J_{2}\right\rangle $ follows the trend of $\left\langle J_{2,\epsilon}\right\rangle $
with respect to $\Gamma_{\epsilon}$, increasing initially with $\Gamma_{\epsilon}$
and subsequently saturating to a constant value of $\sim1.0\times10^{-2}$
for $\Gamma_{\epsilon}\gtrsim10^{8.5}\,\mathrm{Hz}$. Notably, at
$\Gamma_{\epsilon}=10^{6.5}\,\mathrm{Hz}$, the total infidelity is
consistent with the experimentally measured CZ gate infidelity of
approximately $6\times10^{-3}$. Furthermore, the model with $\Gamma_{\epsilon}=10^{6.5}\,\mathrm{Hz}$
also accurately reproduces the Ramsey dephasing experiments discussed
in Sec.\ \ref{subsec:Noise_characterization_dephasing}. Consequently,
we adopt $\Gamma_{\epsilon}=10^{6.5}\,\mathrm{Hz}$ in our noise model.
In Sec.\ \ref{subsec:HF_robust_CZ_against_noise}, these characterized
decoherence contributions, based on our validated noise models, will
be employed to construct high-fidelity and robust CZ gates against
the identified noise sources.

\subsection{High-fidelity and robust CZ gates against the noises}

\label{subsec:HF_robust_CZ_against_noise}

In Sec.\ \ref{sec:High_fidelity_CZ_gate}, we design the control
pulses $\Phi_{A/B}(t)=\Phi_{A/B}^{\mathrm{PBP}}(t)+\Phi_{A/B}^{\mathrm{FTP}}(t)$
to reduce the CZ gate time and suppress the intrinsic error $J_{1}$
through optimization. In Secs.\ \ref{subsec:Noise_characterization_dephasing}-\ref{subsec:Noise_characterization_from_CZ_infid_budget},
we develop validated noise models that accurately capture the dephasing
processes in the system. Based on these models, we quantify the decoherence-induced
contribution to the CZ gate infidelity, $\left\langle J_{2}\right\rangle =\left\langle J_{2,T_{1}}\right\rangle +\left\langle J_{2,\epsilon}\right\rangle +\left\langle J_{2,\phi}\right\rangle $,
which includes contributions from energy relaxation $\left\langle J_{2,T_{1}}\right\rangle $,
RTCE noise $\left\langle J_{2,\epsilon}\right\rangle $, and flux
noise $\left\langle J_{2,\phi}\right\rangle $. This framework enables
the formulation of optimal control strategies aimed at minimizing
the total infidelity, $J_{1}+\left\langle J_{2}\right\rangle $, thereby
realizing high-fidelity and robust CZ gates against the noises via
the optimized control pulses $\Phi_{A/B}(t)$.

The most effective strategy to mitigate the infidelity arising from
decoherence is to reduce the gate time $t_{f}$. However, a reduction
in $t_{f}$ amplifies intrinsic errors, such as leakage, leading to
an increase in $J_{1}$. For the Qt-4 $J_{1}$ optimization, the optimized
$J_{1}$ remains below $10^{-4}$ for $t_{f}\gtrsim11.9\,\mathrm{ns}$,
whereas for $t_{f}<11.9\,\mathrm{ns}$, $J_{1}$ increases sharply,
as shown in Fig.\ \ref{fig:J1_optimization}(a). In contrast, the
contribution from energy relaxation, $\left\langle J_{2,T_{1}}\right\rangle =t_{f}/T_{1}$,
decreases linearly with decreasing $t_{f}$. By considering both contributions
$J_{1}$ and $\left\langle J_{2,T_{1}}\right\rangle $ simultaneously,
we find a minimum of $J_{1}+\left\langle J_{2,T_{1}}\right\rangle \approx4.18\times10^{-4}$
at $t_{f}=12\,\mathrm{ns}$, close to the CZ gate speed limit $t_{f}=\pi/(\sqrt{2}g_{AB})\approx11.6\,\mathrm{ns}$,
with $J_{1}\approx1.8\times10^{-5}$ and $\left\langle J_{2,T_{1}}\right\rangle =4\times10^{-4}$.
That is, reducing the gate time below $12\,\mathrm{ns}$ incurs a
penalty in $J_{1}$ that outweighs the corresponding reduction in
$\left\langle J_{2,T_{1}}\right\rangle $. Consequently, we fix $t_{f}=12\,\mathrm{ns}$
to maintain the optimal $J_{1}+\left\langle J_{2,T_{1}}\right\rangle $
with $\left\langle J_{2,T_{1}}\right\rangle =4\times10^{-4}$, when
searching for optimal control strategies to suppress the RTCE noise
contribution $\left\langle J_{2,\epsilon}\right\rangle $ and flux
noise contribution $\left\langle J_{2,\phi}\right\rangle $.

Both the RTCE noise and the flux noise are sources of dephasing, meaning
that their corresponding noise Hamiltonians, $\mathcal{H}_{\epsilon,A/B}(t)$
and $\mathcal{H}_{\phi,A/B}(t)$ in Eqs.\ (\ref{eq:noise_H_RTCE_noise})
and (\ref{eq:noise_H_flux_noise}), are parallel to the qubit-frequency
terms $\Delta_{A/B}(t)\cdot a_{A/B}^{\dagger}a_{A/B}$ in the ideal
Hamiltonian $\mathcal{H}_{I}(t)$ in Eq.\ (\ref{eq:full_ideal_H}).
As a result, suppressing these dephasing noises by modulating the
control frequencies $\Delta_{A/B}(t)$ is relatively ineffective.
A more effective approach is to reduce the gate time $t_{f}$ and/or
operate the CZ gate at specific sweet spots, where the gate is less
sensitive to noise. Considering the optimal $J_{1}+\left\langle J_{2,T_{1}}\right\rangle $,
the gate time $t_{f}$ can only be reduced to $12\,\mathrm{ns}$.
For the RTCE noise, the noise-system coupling form $H_{\epsilon,A/B}(t)$
in Eq.\ (\ref{eq:RTCE_noise_H}) is proportional to $\left[\Delta_{A/B}-\alpha_{A/B}\right]^{\eta_{\epsilon}/2}$
where $\eta_{\epsilon}=3.73$; therefore, lowering $\Delta_{A/B}$
helps reduce the RTCE noise contribution $\left\langle J_{2,\epsilon}\right\rangle $.
In other words, the sweet spot for the RTCE noise corresponds to the
control flux region $\Phi_{A/B}/\Phi_{0}$ associated with smaller
$\Delta_{A/B}$. Similarly, for the flux noise, the sweet spot with
reduced flux noise contribution $\left\langle J_{2,\phi}\right\rangle $
occurs in the region of $\Phi_{A/B}/\Phi_{0}$ where $\Phi_{A/B}/\Phi_{0}\approx0$
and/or where the derivative $d\Delta_{A/B}/d(\Phi_{A/B}/\Phi_{0})\approx0$,
since the noise-system coupling form $H_{\phi,A/B}(t)$ in Eq.\ (\ref{eq:Flux_noise_H})
is proportional to the product of $\Phi_{A/B}/\Phi_{0}$ and $d\Delta_{A/B}/d(\Phi_{A/B}/\Phi_{0})$.
Accordingly, our general strategy is to set the Pull-Back-Pulse $\Phi_{A/B}^{\mathrm{PBP}}(t)$
close to the sweet spot and then modulate the Fine-Tune-Pulse $\Phi_{A/B}^{\mathrm{FTP}}(t)$
around the sweet spot to effectively suppress $\left\langle J_{2,\epsilon}\right\rangle $
and $\left\langle J_{2,\phi}\right\rangle $, while maintaining $J_{1}$
performance and keeping $\left\langle J_{2,T_{1}}\right\rangle =4\times10^{-4}$
through optimization under a gate time of $t_{f}=12\,\mathrm{ns}$.

Therefore, in Sec.\ \ref{subsec:CZ_against_noise_symmetric_transmons},
we first apply our strategy to construct high-fidelity and robust
CZ gates on symmetric transmons. In Sec.\ \ref{subsec:the_second_sweet_spot_analysis},
we investigate the sweet spots for asymmetric transmons, and in Sec.\ \ref{subsec:CZ_against_noise_asymmetric_transmons},
we construct the corresponding optimized CZ gates to further improve
performance. Finally, in Sec.\ \ref{subsec:robustness_against_noise},
we demonstrate the robust performance of our constructed CZ gates
against dephasing noises for both symmetric and asymmetric transmons,
and provide a concise summary of our results.

\subsubsection{CZ gate optimization for symmetric transmons}

\label{subsec:CZ_against_noise_symmetric_transmons}

\begin{figure}
\includegraphics[scale=0.52]{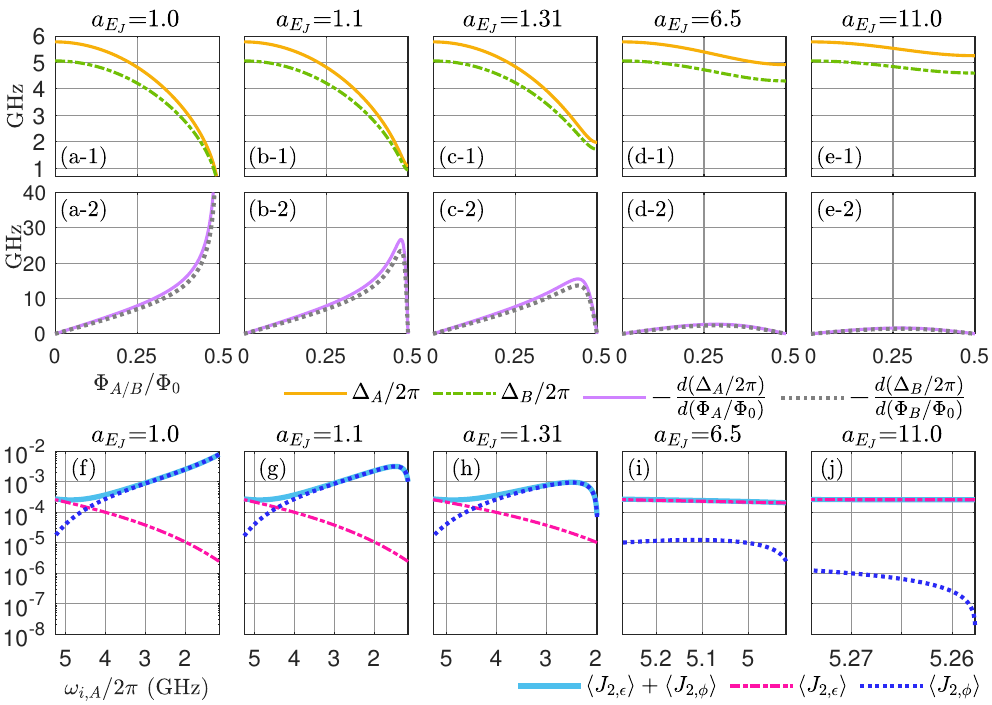}

\caption{Dephasing error contributions. (a-1)--(e-1) $Q_{A}$ frequency $\Delta_{A}/2\pi$
(yellow solid line) and $Q_{B}$ frequency $\Delta_{B}/2\pi$ (green
dash-dotted line) as functions of the corresponding fluxes $\Phi_{A}/\Phi_{0}$
and $\Phi_{B}/\Phi_{0}$ from $0$ to $0.5$ for symmetric transmons
($a_{E_{J}}=1.0$) and asymmetric transmons with $a_{E_{J}}=1.1$,
$1.31$, $6.5$, and $11.0$, respectively. (a-2)--(e-2) Derivatives
$-d(\Delta_{A}/2\pi)/d(\Phi_{A}/\Phi_{0})$ (purple solid line) and
$-d(\Delta_{B}/2\pi)/d(\Phi_{B}/\Phi_{0})$ (gray dotted line) as
functions of the corresponding fluxes from $0$ to $0.5$ for $a_{E_{J}}=1.0$,
$1.1$, $1.31$, $6.5$, and $11.0$, respectively. (f)--(j) Total
dephasing error contribution $\left\langle J_{2,\epsilon}\right\rangle +\left\langle J_{2,\phi}\right\rangle $
(cyan solid line), contribution from RTCE noise $\left\langle J_{2,\epsilon}\right\rangle $
(pink dash-dotted line), and contribution from flux noise $\left\langle J_{2,\phi}\right\rangle $
(blue dotted line) for $a_{E_{J}}=1.0$, $1.1$, $1.31$, $6.5$,
and $11.0$, simulated using the Pull-Back-Pulse $\Phi_{A/B}^{\mathrm{PBP}}(t)$
alone with $t_{f}=12\,\mathrm{ns}$ across the allowed $Q_{A}$ interaction
frequency $\omega_{i}$ region, satisfying the strongest coupling
condition $\delta_{A}=0$.}

\label{fig:J2_est_aEJ_split} 
\end{figure}

For symmetric transmons with $a_{E_{J}}=1.0$, as the flux $\Phi_{A/B}/\Phi_{0}$
is varied from $0$ to $0.5$, the qubit frequencies $\Delta_{A/B}$
decrease monotonically from their maximum values $\omega_{0,A/B}$,
as illustrated in Fig.\ \ref{fig:J2_est_aEJ_split}(a-1). Concurrently,
the corresponding derivatives $-d\Delta_{A/B}/d(\Phi_{A/B}/\Phi_{0})$
increase monotonically from zero, as shown in Fig.\ \ref{fig:J2_est_aEJ_split}(a-2).
Thus, for $a_{E_{J}}=1.0$, the sweet spot for the flux noise is located
at the point of maximal qubit frequencies, $\Delta_{A/B}=\omega_{0,A/B}$,
where $\Phi_{A/B}/\Phi_{0}=0$ and $d\Delta_{A/B}/d(\Phi_{A/B}/\Phi_{0})=0$.
However, this is in tension with the sweet spot for the RTCE noise,
which favors lower qubit frequencies $\Delta_{A/B}$. Therefore, it
is necessary to determine an optimal operating point that mitigates
both the RTCE noise and the flux noise. To this end, we simulate the
infidelity contributions from the RTCE noise, $\left\langle J_{2,\epsilon}\right\rangle $,
and from the flux noise, $\left\langle J_{2,\phi}\right\rangle $,
by varying the $Q_{A}$ interaction frequency $\omega_{i,A}$ under
the condition of maximal coupling $\delta_{A}=0$, using only the
Pull-Back-Pulse, $\Phi_{A/B}(t)=\Phi_{A/B}^{\mathrm{PBP}}(t)$ with
gate time $t_{f}=12\,\mathrm{ns}$, as shown in Fig.\ \ref{fig:J2_est_aEJ_split}(f).
The corresponding $Q_{B}$ interaction frequency is set as $\omega_{i,B}=\omega_{i,A}+\alpha_{A}$
to satisfy $\delta_{A}=0$. The total dephasing noise contribution,
$\left\langle J_{2,\epsilon}\right\rangle +\left\langle J_{2,\phi}\right\rangle $,
is primarily governed by $\left\langle J_{2,\epsilon}\right\rangle $
as $\Delta_{A/B}\rightarrow\omega_{0,A/B}$, whereas for smaller $\Delta_{A/B}$
it is dominated by $\left\langle J_{2,\phi}\right\rangle $. The combined
contribution $\left\langle J_{2,\epsilon}\right\rangle +\left\langle J_{2,\phi}\right\rangle $
reaches a minimum at $\omega_{i,A}/2\pi\approx4.9\,\mathrm{GHz}$,
which defines the corresponding operational sweet spot. Based on this,
we define the cost function $J_{1}+\left\langle J_{2,\epsilon}\right\rangle +\left\langle J_{2,\phi}\right\rangle $,
initialize the interaction frequencies at the sweet spot, $\omega_{i,A/B}/2\pi=4.9/4.686\,\mathrm{GHz}$,
in the Pull-Back-Pulse $\Phi_{A/B}^{\mathrm{PBP}}(t)$, and set the
number of control parameters $k_{\mathrm{max}}=10$ in the Fine-Tune-Pulse
$\Phi_{A}^{\mathrm{FTP}}(t)$ for subsequent optimization. The optimization
procedure follows the same protocol as the rp-Qt-4 $J_{1}$ optimization
described in Sec.\ \ref{sec:High_fidelity_CZ_gate}, thereby mitigating
the risk of convergence to local minima within the complex landscape
of the cost function $J_{1}+\left\langle J_{2,\epsilon}\right\rangle +\left\langle J_{2,\phi}\right\rangle $.

\begin{figure}
\includegraphics[scale=0.52]{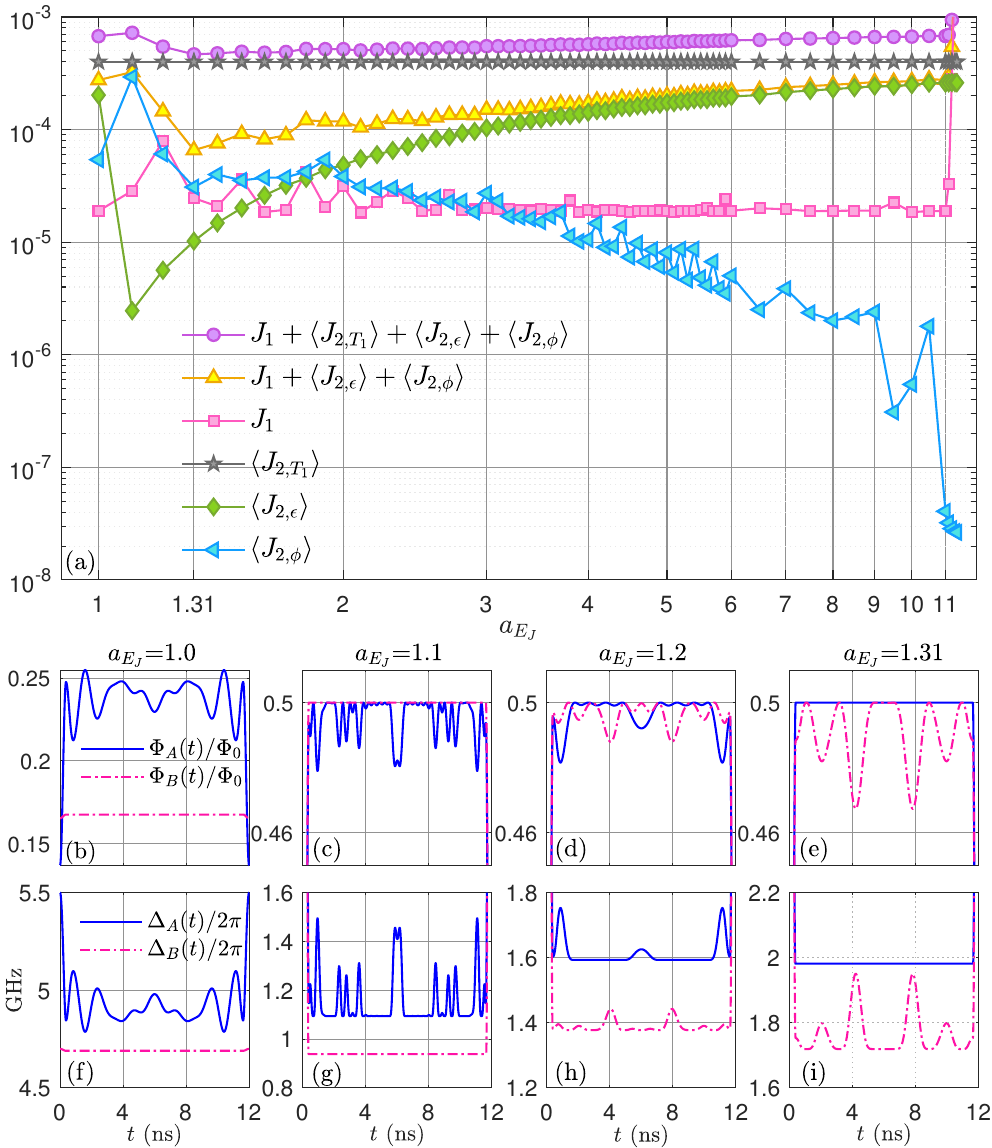}

\caption{Optimized performances for symmetric and asymmetric transmons. (a)
Infidelities and individual error contributions as functions of $a_{E_{J}}$
for the optimized CZ gates with $t_{f}=12\,\mathrm{ns}$, operating
at their respective sweet spots (the second sweet spots for weakly
asymmetric transmons). The total infidelity $J_{1}+\left\langle J_{2,T_{1}}\right\rangle +\left\langle J_{2,\epsilon}\right\rangle +\left\langle J_{2,\phi}\right\rangle $
is shown by the purple-circle line, the infidelity $J_{1}+\left\langle J_{2,\epsilon}\right\rangle +\left\langle J_{2,\phi}\right\rangle $
by the yellow-triangle line, the intrinsic error $J_{1}$ by the pink-square
line, the energy-relaxation contribution $\left\langle J_{2,T_{1}}\right\rangle $
by the gray-pentagram line, the RTCE-noise contribution $\left\langle J_{2,\epsilon}\right\rangle $
by the green-diamond line, and the flux-noise contribution $\left\langle J_{2,\phi}\right\rangle $
by the blue-leftward-triangle line. (b--e) Optimized flux pulses
$\Phi_{A}(t)/\Phi_{0}$ (solid blue line) and $\Phi_{B}(t)/\Phi_{0}$
(dash-dotted pink line) for $a_{E_{J}}=1.0$, $1.1$, $1.2$, and
$1.31$, respectively. (f--i) Corresponding optimized frequencies
$\Delta_{A}(t)$ (solid blue line) and $\Delta_{B}(t)$ (dash-dotted
pink line) for $a_{E_{J}}=1.0$, $1.1$, $1.2$, and $1.31$, respectively.}

\label{fig:noise_suppression_pulse} 
\end{figure}

The optimized total CZ gate infidelity, $J_{1}+\left\langle J_{2,T_{1}}\right\rangle +\left\langle J_{2,\epsilon}\right\rangle +\left\langle J_{2,\phi}\right\rangle $,
is $6.75\times10^{-4}$, as shown in Fig.\ \ref{fig:noise_suppression_pulse}(a)
at $a_{E_{J}}=1.0$. The corresponding optimized control flux pulses
$\Phi_{A/B}(t)/\Phi_{0}$ are presented in Fig.\ \ref{fig:noise_suppression_pulse}(b),
while the resulting optimized qubit frequencies $\Delta_{A/B}(t)$
are displayed in Fig.\ \ref{fig:noise_suppression_pulse}(f). The
optimized Pull-Back-Pulse $\Phi_{A/B}^{\mathrm{PBP}}(t)$ pull $\Delta_{A/B}(t)$
from the idle frequencies $\omega_{\mathrm{Idle},A/B}/2\pi=5.5/4.7\,\mathrm{GHz}$
to the sweet spot at $2\pi\times4.9/4.7\,\mathrm{GHz}$, thereby minimizing
$\left\langle J_{2,\epsilon}\right\rangle +\left\langle J_{2,\phi}\right\rangle $
as predicted in Fig.\ \ref{fig:J2_est_aEJ_split}(f), with optimized
values $\left\langle J_{2,\epsilon}\right\rangle \approx2.0\times10^{-4}$
and $\left\langle J_{2,\phi}\right\rangle \approx5.4\times10^{-5}$.
The optimized Fine-Tune-Pulse $\Phi_{A}^{\mathrm{FTP}}(t)$ modulates
$\Delta_{A}(t)$ around the sweet spot to synchronize intrinsic errors,
achieving an optimized $J_{1}\approx1.89\times10^{-5}$, comparable
to the $J_{1}\approx1.80\times10^{-5}$ obtained via the Qt-4 $J_{1}$
optimization described in Sec.\ \ref{subsec:Optimization_results}.
Therefore, for symmetric transmons with $a_{E_{J}}=1.0$, both the
ideal gate infidelity $J_{1}$ and the flux noise contribution $\left\langle J_{2,\phi}\right\rangle $
can be suppressed below $10^{-4}$. The CZ gate infidelity is primarily
limited by the energy relaxation contribution $\left\langle J_{2,T_{1}}\right\rangle =4\times10^{-4}$
and the RTCE noise contribution $\left\langle J_{2,\epsilon}\right\rangle \approx2.0\times10^{-4}$,
as shown in Fig.\ \ref{fig:noise_suppression_pulse}(a) at $a_{E_{J}}=1.0$.
Since $\left\langle J_{2,T_{1}}\right\rangle =t_{f}/T_{1}$, further
reduction requires a longer $T_{1}$, because the gate time $t_{f}=12\,\mathrm{ns}$
is already close to the CZ gate speed limit, $\pi/(\sqrt{2}g_{AB})\approx11.6\,\mathrm{ns}$.
Any attempt to reduce $t_{f}$ further leads to increased leakage
errors that outweigh the reduction in $\left\langle J_{2,T_{1}}\right\rangle $.
To further suppress $\left\langle J_{2,\epsilon}\right\rangle $ and
$\left\langle J_{2,\phi}\right\rangle $, the sweet spot for asymmetric
transmons will be investigated in the following section to enhance
overall gate performance.

We note that, for the optimized infidelities presented here in Sec\ \ref{subsec:HF_robust_CZ_against_noise}
and in the companion paper \citep{Huang2026a}, the filtering effects---simulated
using the Tektronix AWG70000, as described in Sec.\ \ref{subsec:Filtering_effects}---have
been fully incorporated.

\subsubsection{Sweet spots for asymmetric transmons}

\label{subsec:the_second_sweet_spot_analysis}

For asymmetric transmons with $a_{E_{J}}>1$, as the flux $\Phi_{A/B}/\Phi_{0}$
increases from $0$ to $0.5$, the derivative $-d\Delta_{A/B}/d(\Phi_{A/B}/\Phi_{0})$
initially rises from zero at $\Phi_{A/B}/\Phi_{0}=0$, reaches the
maximum, and subsequently decreases back to zero at $\Phi_{A/B}/\Phi_{0}=0.5$,
as illustrated in Figs.\ \ref{fig:J2_est_aEJ_split}(b-2) to \ref{fig:J2_est_aEJ_split}(e-2)
for $a_{E_{J}}=1.1$, $1.31$, $6.5$ and $11.0$, respectively. This
behavior indicates the existence of a second flux-noise sweet spot
for asymmetric transmons at $\Phi_{A/B}/\Phi_{0}=0.5$, where the
derivative $d\Delta_{A/B}/d(\Phi_{A/B}/\Phi_{0})=0$. At this second
sweet spot, the qubit frequency reaches the minimum, corresponding
to the minimum RTCE noise contribution $\left\langle J_{2,\epsilon}\right\rangle $,
thereby simultaneously favoring mitigation of RTCE noise. Consequently,
for asymmetric transmons, a common sweet spot for both the flux and
RTCE noises could occur near $\Phi_{A/B}/\Phi_{0}=0.5$.

In addition, the tunable range of $\Delta_{A/B}$, given by $(\omega_{0,A/B}-\alpha_{A/B})\cdot(1-\sqrt{d})$,
decreases as $a_{E_{J}}$ increases, i.e., as $d=(a_{E_{J}}-1)/(a_{E_{J}}+1)\rightarrow1$,
as shown in Figs.\ \ref{fig:J2_est_aEJ_split}(a-1) to \ref{fig:J2_est_aEJ_split}(e-1)
for $a_{E_{J}}=1.0$, $1.1$, $1.31$, $6.5$ and $11.0$, respectively.
This reduction in the $\Delta_{A/B}$ tunable range raises the minimum
$\Delta_{A/B}$, thereby increasing the minimum RTCE noise contribution
$\left\langle J_{2,\epsilon}\right\rangle $. By contrast, the reduced
tunable range lowers both the overall derivative $d\Delta_{A/B}/d(\Phi_{A/B}/\Phi_{0})$
and its peak value, as illustrated in Figs.\ \ref{fig:J2_est_aEJ_split}(b-2)
to \ref{fig:J2_est_aEJ_split}(e-2) for $a_{E_{J}}=1.1$, $1.31$,
$6.5$ and $11.0$, favoring the mitigation of the flux noise instead.
To further determine an optimal control strategy, we simulate $\left\langle J_{2,\epsilon}\right\rangle $
and $\left\langle J_{2,\phi}\right\rangle $ under the strongest coupling
condition, $\delta_{A}=0$, for asymmetric transmons with $a_{E_{J}}=1.1$,
$1.31$, $6.5$ and $11.0$, as shown in Figs.\ \ref{fig:J2_est_aEJ_split}(g)
to \ref{fig:J2_est_aEJ_split}(j). For these simulations, we use $\Phi_{A/B}(t)=\Phi_{A/B}^{\mathrm{PBP}}(t)$
with $t_{f}=12\,\mathrm{ns}$ and $\omega_{i,B}=\omega_{i,A}+\alpha_{A}$,
consistent with the settings for $a_{E_{J}}=1.0$ in Fig.\ \ref{fig:J2_est_aEJ_split}(f).

The overall $\left\langle J_{2,\phi}\right\rangle $ decreases with
increasing $a_{E_{J}}$, as illustrated in Figs.\ \ref{fig:J2_est_aEJ_split}(f)
to \ref{fig:J2_est_aEJ_split}(j), which correlates with the reduction
in the overall derivative $-d\Delta_{A/B}/d(\Phi_{A/B}/\Phi_{0})$
for $a_{E_{J}}=1.0$, $1.1$, $1.31$, $6.5$ and $11.0$, as shown
in Figs.\ \ref{fig:J2_est_aEJ_split}(a-2) to \ref{fig:J2_est_aEJ_split}(e-2),
respectively. For $a_{E_{J}}=1.0$, $\left\langle J_{2,\phi}\right\rangle $
increases monotonically as $\omega_{i,A}$ decreases {[}Fig.\ \ref{fig:J2_est_aEJ_split}(f){]},
consistent with the monotonically increasing derivative $-d\Delta_{A/B}/d(\Phi_{A/B}/\Phi_{0})$
depicted in Fig.\ \ref{fig:J2_est_aEJ_split}(a-2). This behavior
indicates that, for symmetric transmons, only one flux-noise sweet
spot exists at $\Phi_{A/B}/\Phi_{0}\approx0$. In contrast, for weakly
asymmetric transmons with smaller $a_{E_{J}}$, such as $a_{E_{J}}=1.1$
and $a_{E_{J}}=1.31$, $\left\langle J_{2,\phi}\right\rangle $ initially
increases with decreasing $\Delta_{A}$, reaching a maximum corresponding
to the peak in $-d\Delta_{A/B}/d(\Phi_{A/B}/\Phi_{0})$, and subsequently
decreases as $\omega_{i,A}$ approaches the minimum of $\Delta_{A}$,
where $-d\Delta_{A/B}/d(\Phi_{A/B}/\Phi_{0})\rightarrow0$. Consequently,
weakly asymmetric transmons exhibit two flux-noise sweet spots, located
at $\Phi_{A/B}/\Phi_{0}\approx0$ and $\Phi_{A/B}/\Phi_{0}\approx0.5$.
Notably, $\left\langle J_{2,\phi}\right\rangle $ at the second sweet
spot $\Phi_{A/B}/\Phi_{0}\approx0.5$ exceeds that at the first sweet
spot $\Phi_{A/B}/\Phi_{0}\approx0$, as shown in Figs.\ \ref{fig:J2_est_aEJ_split}(g)
and \ref{fig:J2_est_aEJ_split}(h) for $a_{E_{J}}=1.1$ and $a_{E_{J}}=1.31$,
respectively. This asymmetry arises because, under the strongest coupling
condition $\delta_{A}=0$, it is generally impossible for both $Q_{A}$
and $Q_{B}$ to reside simultaneously at their respective sweet spots.
Furthermore, the peak of $-d\Delta_{A/B}/d(\Phi_{A/B}/\Phi_{0})$
is located closer to the second sweet spot at $\Phi_{A/B}/\Phi_{0}\approx0.5$,
as seen in Figs.\ \ref{fig:J2_est_aEJ_split}(b-2) and \ref{fig:J2_est_aEJ_split}(c-2)
for $a_{E_{J}}=1.1$ and $a_{E_{J}}=1.31$. Consequently, a small
shift in $\Phi_{A/B}/\Phi_{0}$ from the second sweet spot results
in a more pronounced enhancement of $\left\langle J_{2,\phi}\right\rangle $,
owing to the steep increase of $-d\Delta_{A/B}/d(\Phi_{A/B}/\Phi_{0})$,
in contrast to the situation at the first sweet spot.

For weakly asymmetric transmons with $a_{E_{J}}\gtrsim1.2$, under
the condition $\delta_{A}=0$ and with decreasing $\omega_{i,A}$,
when $Q_{A}$ reaches the minimum of $\Delta_{A}$ (i.e., the second
sweet spot at $\Phi_{A}/\Phi_{0}=0.5$), $Q_{B}$ has not yet reached
the minimum of $\Delta_{B}$, remaining at $\Phi_{B}/\Phi_{0}<0.5$
and thus shifted from its second sweet spot. Conversely, for $a_{E_{J}}\lesssim1.2$,
when $Q_{B}$ arrives at its second sweet spot $\Phi_{B}/\Phi_{0}=0.5$,
$Q_{A}$ still resides at $\Phi_{A}/\Phi_{0}<0.5$. It is only when
$a_{E_{J}}\approx1.2$ that both $Q_{A}$ and $Q_{B}$ can simultaneously
reach their respective second sweet spots and satisfy $\delta_{A}=0$.
Consequently, for $a_{E_{J}}=1.31$ (above $1.2$), $Q_{A}$ can be
pulled to its second sweet spot $\Phi_{A}/\Phi_{0}=0.5$ using the
Pull-Back-Pulse $\Phi_{A}^{\mathrm{PBP}}(t)$, yielding a low $\left\langle J_{2,\phi,A}\right\rangle \approx3.2\times10^{-7},$
whereas $Q_{B}$ can only be brought close to $\Phi_{B}/\Phi_{0}=0.5$
via the Pull-Back-Pulse $\Phi_{B}^{\mathrm{PBP}}(t)$, resulting in
a higher $\left\langle J_{2,\phi,B}\right\rangle \approx7.0\times10^{-5}$,
which dominates the overall $\left\langle J_{2,\phi}\right\rangle $
{[}Fig.\ \ref{fig:J2_est_aEJ_split}(h){]}. By contrast, for $a_{E_{J}}=1.1$
(below $1.2$), $Q_{B}$ can be pulled to $\Phi_{B}/\Phi_{0}=0.5$
to obtain a low $\left\langle J_{2,\phi,B}\right\rangle \approx6.2\times10^{-7}$,
while $Q_{A}$ can only be brought close to $\Phi_{A}/\Phi_{0}=0.5$,
giving a higher $\left\langle J_{2,\phi,A}\right\rangle \approx1.0\times10^{-3}$,
which dominates $\left\langle J_{2,\phi}\right\rangle $ {[}Fig.\ \ref{fig:J2_est_aEJ_split}(g){]}.
Notably, $\left\langle J_{2,\phi,B}\right\rangle $ for $a_{E_{J}}=1.1$
and $\left\langle J_{2,\phi,A}\right\rangle $ for $a_{E_{J}}=1.31$
are comparable, lying within the same order of magnitude, whereas
$\left\langle J_{2,\phi,A}\right\rangle $ for $a_{E_{J}}=1.1$ exceeds
$\left\langle J_{2,\phi,B}\right\rangle $ for $a_{E_{J}}=1.31$ by
more than an order of magnitude. This behavior arises because, as
$a_{E_{J}}$ approaches 1.0, the peak magnitude of $-d\Delta_{A/B}/d(\Phi_{A/B}/\Phi_{0})$
increases and its position shifts closer to $\Phi_{A/B}/\Phi_{0}=0.5$
{[}Figs.\ \ref{fig:J2_est_aEJ_split}(a-2) to \ref{fig:J2_est_aEJ_split}(e-2){]}.
Consequently, the sensitivity of $\left\langle J_{2,\phi}\right\rangle $
to small variations in $\Phi_{A/B}/\Phi_{0}$ at the second sweet
spot increases, leading to higher $\left\langle J_{2,\phi,A/B}\right\rangle $.

For strongly asymmetric transmons with larger $a_{E_{J}}$, the separation
between the tunable ranges of $\Delta_{A}$ and $\Delta_{B}$ {[}see
Fig.\ \ref{fig:J1op_aEJ_split}(a){]} requires, in order to satisfy
the strongest coupling condition $\delta_{A}=0$, that $Q_{A}$ operates
near the minimum of $\Delta_{A}$, i.e., the second sweet spot at
$\Phi_{A}/\Phi_{0}=0.5$, while $Q_{B}$ operates near the maximum
of $\Delta_{B}$, i.e., the first sweet spot at $\Phi_{B}/\Phi_{0}=0$.
In this configuration, $\left\langle J_{2,\phi,B}\right\rangle \propto[\Phi_{B}/\Phi_{0}]^{2}[d\Delta_{B}/d(\Phi_{B}/\Phi_{0})]^{2}$
with $\Phi_{B}/\Phi_{0}\approx0$ and $-d\Delta_{B}/d(\Phi_{B}/\Phi_{0})\approx0$
is generally much smaller than $\left\langle J_{2,\phi,A}\right\rangle \propto[\Phi_{A}/\Phi_{0}]^{2}[d\Delta_{A}/d(\Phi_{A}/\Phi_{0})]^{2}$,
for which only $-d\Delta_{A}/d(\Phi_{A}/\Phi_{0})\approx0$. Consequently,
as shown in Fig.\ \ref{fig:J2_est_aEJ_split}(j) for strongly asymmetric
transmons with $a_{E_{J}}=11.0$, the total $\left\langle J_{2,\phi}\right\rangle $,
dominated by $\left\langle J_{2,\phi,A}\right\rangle $, decreases
monotonically with decreasing $\omega_{i,A}$, i.e., as $\Phi_{A}/\Phi_{0}\rightarrow0.5$.
When $Q_{A}$ reaches $\Phi_{A}/\Phi_{0}=0.5$ while $Q_{B}$ moves
away from $\Phi_{B}/\Phi_{0}=0$, the overall $\left\langle J_{2,\phi}\right\rangle $
attains its minimum value, $\left\langle J_{2,\phi}\right\rangle \approx1.98\times10^{-8}$,
which is ultimately limited by $\left\langle J_{2,\phi,B}\right\rangle \approx1.54\times10^{-8}$.

For both symmetric and asymmetric transmons, $\left\langle J_{2,\epsilon}\right\rangle $,
with $\left\langle J_{2,\epsilon,A/B}\right\rangle \propto\left[\Delta_{A/B}-\alpha_{A/B}\right]^{\eta_{\epsilon}}$,
decreases monotonically as $\omega_{i,A}$ decreases, i.e., as $\Phi_{A/B}/\Phi_{0}$
approaches its sweet spot at $\Phi_{A/B}/\Phi_{0}=0.5$, as shown
in Figs.\ \ref{fig:J2_est_aEJ_split}(f) to \ref{fig:J2_est_aEJ_split}(j).
However, the reduction in $\left\langle J_{2,\epsilon}\right\rangle $
becomes smaller with increasing $a_{E_{J}}$ due to the shrinking
tunable ranges of $\Delta_{A/B}$, as illustrated in Figs.\ \ref{fig:J2_est_aEJ_split}(a-1)
to \ref{fig:J2_est_aEJ_split}(e-1). Within the tunable ranges of
$\Delta_{A/B}$ under the condition $\delta_{A}=0$, for $a_{E_{J}}=1.1$,
$\left\langle J_{2,\epsilon}\right\rangle $ decreases from $2.61\times10^{-4}$
to $2.43\times10^{-6}$, achieving an improvement of two orders of
magnitude. By contrast, for $a_{E_{J}}=11.0$, $\left\langle J_{2,\epsilon}\right\rangle $
decreases only from $2.61\times10^{-4}$ to $2.58\times10^{-4}$,
yielding almost no improvement.

Finally, we analyze the total dephasing contribution, $\left\langle J_{2,\epsilon}\right\rangle +\left\langle J_{2,\phi}\right\rangle $,
for asymmetric transmons and compare it with symmetric transmons.
For symmetric transmons, there exists only one sweet spot at $\omega_{i,A}/2\pi\approx4.9\,\mathrm{GHz}$,
where $\left\langle J_{2,\epsilon}\right\rangle +\left\langle J_{2,\phi}\right\rangle \approx2.6\times10^{-4}$
with $\left\langle J_{2,\epsilon}\right\rangle \approx2.0\times10^{-4}$
and $\left\langle J_{2,\phi}\right\rangle \approx5.7\times10^{-5}$,
as illustrated in Fig.\ \ref{fig:J2_est_aEJ_split}(f). For strongly
asymmetric transmons, both $\left\langle J_{2,\epsilon}\right\rangle $
and $\left\langle J_{2,\phi}\right\rangle $ decrease as $\omega_{i,A}$
decreases, resulting in the only sweet spot at $\Phi_{A}/\Phi_{0}=0.5$,
where $\left\langle J_{2,\epsilon}\right\rangle +\left\langle J_{2,\phi}\right\rangle $
is minimized and predominantly determined by $\left\langle J_{2,\epsilon}\right\rangle $.
For instance, as shown in Fig.\ \ref{fig:J2_est_aEJ_split}(j) for
$a_{E_{J}}=11.0$, the minimum occurs at $\Phi_{A}/\Phi_{0}=0.5$
with $\left\langle J_{2,\epsilon}\right\rangle +\left\langle J_{2,\phi}\right\rangle \approx2.58\times10^{-4}$.
Although this minimum is comparable to that of symmetric transmons,
the corresponding $\left\langle J_{2,\phi}\right\rangle \approx1.98\times10^{-8}$
is more than three orders of magnitude smaller than for $a_{E_{J}}=1.0$,
reflecting potentially enhanced robustness against the flux noise.
For weakly asymmetric transmons, two sweet spots for total dephasing
emerge. The first sweet spot is near $\Phi_{A/B}/\Phi_{0}\approx0$,
coinciding with the sweet spot for symmetric transmons at $\omega_{i,A}/2\pi\approx4.9\,\mathrm{GHz}$.
This agreement arises because the behaviors of $\Delta_{A/B}$ and
$-d\Delta_{A/B}/d(\Phi_{A/B}/\Phi_{0})$ are comparable in this region
for both symmetric and weakly asymmetric transmons, as illustrated
in Figs.\ \ref{fig:J2_est_aEJ_split}(a-1) to \ref{fig:J2_est_aEJ_split}(c-1)
and \ref{fig:J2_est_aEJ_split}(a-2) to \ref{fig:J2_est_aEJ_split}(c-2).
Consequently, the $\left\langle J_{2,\epsilon}\right\rangle +\left\langle J_{2,\phi}\right\rangle $,
$\left\langle J_{2,\epsilon}\right\rangle $, $\left\langle J_{2,\phi}\right\rangle $
for asymmetric transmons at this first sweet spot are comparable to
those for symmetric transmons, as shown in Figs.\ \ref{fig:J2_est_aEJ_split}(f)
to \ref{fig:J2_est_aEJ_split}(h). The second sweet spot for weakly
asymmetric transmons occurs at $\Phi_{A/B}/\Phi_{0}\approx0.5$, where
the local minimum of $\left\langle J_{2,\epsilon}\right\rangle +\left\langle J_{2,\phi}\right\rangle $
is dominated by $\left\langle J_{2,\phi}\right\rangle $. As $a_{E_{J}}$
approaches $1.0$, the increasing $\left\langle J_{2,\phi}\right\rangle $
causes $\left\langle J_{2,\epsilon}\right\rangle +\left\langle J_{2,\phi}\right\rangle $
at this second sweet spot to exceed that of the first sweet spot.
For example, as shown in Fig.\ \ref{fig:J2_est_aEJ_split}(g) for
$a_{E_{J}}=1.1$, $\left\langle J_{2,\epsilon}\right\rangle +\left\langle J_{2,\phi}\right\rangle \approx1.02\times10^{-3}$
at the second sweet spot, nearly an order of magnitude higher than
at the first. However, the corresponding $\left\langle J_{2,\epsilon}\right\rangle \approx2.43\times10^{-6}$,
approximately two orders of magnitude lower than at the first sweet
spot, indicating potentially enhanced robustness against the RTCE
noise. As $a_{E_{J}}$ increases away from $1.0$, the decreasing
$\left\langle J_{2,\phi}\right\rangle $ reduces $\left\langle J_{2,\epsilon}\right\rangle +\left\langle J_{2,\phi}\right\rangle $
at the second sweet spot, although the increase in $\left\langle J_{2,\epsilon}\right\rangle $
is unavoidable. For $a_{E_{J}}=1.31$, the total dephasing at the
second sweet spot, occurring at $\omega_{i,A}/2\pi\approx1.98\,\mathrm{GHz}$,
is reduced to $\left\langle J_{2,\epsilon}\right\rangle +\left\langle J_{2,\phi}\right\rangle \approx8.04\times10^{-5}$,
approximately half an order of magnitude lower than at the first sweet
spot, as illustrated in Fig.\ \ref{fig:J2_est_aEJ_split}(h). The
corresponding $\left\langle J_{2,\phi}\right\rangle \approx7.03\times10^{-5}$
is slightly degraded relative to the first sweet spot, while $\left\langle J_{2,\epsilon}\right\rangle \approx1.02\times10^{-5}$
improves by more than an order of magnitude compared with the first
sweet spot, but increases by nearly an order of magnitude relative
to the second sweet spot for $a_{E_{J}}=1.1$. In the next section,
we systematically vary $a_{E_{J}}$ to identify the optimal total
dephasing contribution, $\left\langle J_{2,\epsilon}\right\rangle +\left\langle J_{2,\phi}\right\rangle $,
by optimizing CZ gate performance on asymmetric transmons at their
respective sweet spots.

\subsubsection{CZ gate optimization for asymmetric transmons}

\label{subsec:CZ_against_noise_asymmetric_transmons}

We optimize CZ gates with $t_{f}=12\,\mathrm{ns}$ for asymmetric
transmons with $a_{E_{J}}=1.1$ to $11.36$, defining the cost function
as $J_{1}+\left\langle J_{2,\epsilon}\right\rangle +\left\langle J_{2,\phi}\right\rangle $.
The optimization is carried out using the rp-Qt-4 optimization protocol
introduced in Sec.\ \ref{sec:High_fidelity_CZ_gate}, with initial
interaction frequencies $\omega_{i,A/B}$ set near their respective
sweet spots, as discussed in Sec.\ \ref{subsec:the_second_sweet_spot_analysis}.
Notably, weakly asymmetric transmons are optimized at their second
sweet spots, since their performance at the first sweet spots is comparable
to that of symmetric transmons. The resulting optimized infidelities
are presented in Fig.\ \ref{fig:noise_suppression_pulse}(a). The
optimized total infidelities, defined as $J_{1}+\left\langle J_{2,T_{1}}\right\rangle +\left\langle J_{2,\epsilon}\right\rangle +\left\langle J_{2,\phi}\right\rangle $
and dominated by the energy relaxation contribution $\left\langle J_{2,T_{1}}\right\rangle =4\times10^{-4}$,
are suppressed below $10^{-3}$ for all but the strongly asymmetric
transmons with $a_{E_{J}}\gtrsim11.2$. In this large-$a_{E_{J}}$
regime, the strongest coupling condition $\delta_{A}=0$ cannot be
achieved for the implementation of a $12\,\mathrm{ns}$ CZ gate, as
discussed in Sec.\ \ref{subsec:Optimization_asymmetric_transmon},
leading to a steep increase in $J_{1}$ for $a_{E_{J}}\gtrsim11.1$
and, consequently, a deterioration of the overall infidelity. For
$a_{E_{J}}\lesssim11.1$, the optimized $J_{1}$ remains $\sim2\times10^{-5}$,
except for certain weakly asymmetric transmons, where optimization
is more challenging due to the increased sensitivity of $\left\langle J_{2,\phi}\right\rangle $
to flux variations at the second sweet spot as $a_{E_{J}}\rightarrow1$,
as discussed in Sec.\ \ref{subsec:the_second_sweet_spot_analysis}.
In particular, for $a_{E_{J}}=1.2$, the optimized $J_{1}$ rises
to $7.90\times10^{-5}$. For asymmetric transmons, the optimized $\left\langle J_{2,\phi}\right\rangle $
decreases with increasing $a_{E_{J}}$, reflecting the reduction in
the overall derivative $-d\Delta_{A/B}/d(\Phi_{A/B}/\Phi_{0})$. The
sharp rise of $\left\langle J_{2,\phi}\right\rangle $ for $a_{E_{J}}\lesssim1.31$
is attributed to the enhanced $\left\langle J_{2,\phi}\right\rangle $
sensitivity at the second sweet spot. Conversely, the optimized $\left\langle J_{2,\epsilon}\right\rangle $
increases with increasing $a_{E_{J}}$ due to the narrowing of the
tunable ranges of $\Delta_{A/B}$.

Excluding the dominant constant contribution, $\left\langle J_{2,T_{1}}\right\rangle =4\times10^{-4}$,
the minimum of the optimized infidelity $J_{1}+\left\langle J_{2,\epsilon}\right\rangle +\left\langle J_{2,\phi}\right\rangle \approx6.57\times10^{-5}$
at $a_{E_{J}}=1.31$ is clearly identified in Fig.\ \ref{fig:noise_suppression_pulse}(a).
This value represents an improvement of more than half an order of
magnitude compared with the optimized $J_{1}+\left\langle J_{2,\epsilon}\right\rangle +\left\langle J_{2,\phi}\right\rangle \approx2.75\times10^{-4}$
for symmetric transmons with $a_{E_{J}}=1.0$. The infidelity reduction
is primarily attributable to the optimized $\left\langle J_{2,\epsilon}\right\rangle \approx1.02\times10^{-5}$
for $a_{E_{J}}=1.31$, which is more than an order of magnitude smaller
than the optimized $\left\langle J_{2,\epsilon}\right\rangle \approx2.02\times10^{-4}$,
the dominant term in $J_{1}+\left\langle J_{2,\epsilon}\right\rangle +\left\langle J_{2,\phi}\right\rangle $,
for $a_{E_{J}}=1.0$. In addition, the optimized $\left\langle J_{2,\phi}\right\rangle \approx3.09\times10^{-5}$
for $a_{E_{J}}=1.31$ is also reduced relative to $\left\langle J_{2,\phi}\right\rangle \approx5.41\times10^{-5}$
for $a_{E_{J}}=1.0$, while the optimized $J_{1}\approx2.46\times10^{-5}$
for $a_{E_{J}}=1.31$ is slightly degraded compared to $J_{1}\approx1.89\times10^{-5}$
for $a_{E_{J}}=1.0$.

Under the condition of comparable optimized infidelity $J_{1}+\left\langle J_{2,\epsilon}\right\rangle +\left\langle J_{2,\phi}\right\rangle $
relative to that for $a_{E_{J}}=1.0$, the most robust CZ gates against
the flux noise and the RTCE noise are identified in strongly asymmetric
transmons with $a_{E_{J}}=11.0$ and weakly asymmetric transmons with
$a_{E_{J}}=1.1$, respectively, as shown in Fig.\ \ref{fig:noise_suppression_pulse}(a).
For $a_{E_{J}}=11.0$, the optimized $\left\langle J_{2,\phi}\right\rangle \approx4.07\times10^{-8}$,
which is over three orders of magnitude lower than the corresponding
value for $a_{E_{J}}=1.0$. Meanwhile, the optimized $J_{1}\approx1.88\times10^{-5}$,
$\left\langle J_{2,\epsilon}\right\rangle \approx2.59\times10^{-4}$,
and $J_{1}+\left\langle J_{2,\epsilon}\right\rangle +\left\langle J_{2,\phi}\right\rangle \approx2.77\times10^{-4}$
remain comparable to those for $a_{E_{J}}=1.0$. In contrast, for
$a_{E_{J}}=1.1$, the optimized $\left\langle J_{2,\epsilon}\right\rangle \approx2.46\times10^{-6}$,
approximately two orders of magnitude lower than the value for $a_{E_{J}}=1.0$.
However, the optimization is hindered by the strong peaks in $-d\Delta_{A/B}/d(\Phi_{A/B}/\Phi_{0})$
approaching the second sweet spots as $a_{E_{J}}\rightarrow1$, resulting
in a degradation of the optimized $\left\langle J_{2,\phi}\right\rangle \approx2.93\times10^{-4}$
and $J_{1}\approx2.86\times10^{-5}$ relative to $a_{E_{J}}=1.0$.
Nevertheless, the total infidelity $J_{1}+\left\langle J_{2,\epsilon}\right\rangle +\left\langle J_{2,\phi}\right\rangle \approx3.24\times10^{-4}$
remains comparable to that for $a_{E_{J}}=1.0$.

We next specify the pulse designs for asymmetric transmons with varying
$a_{E_{J}}$, which achieve the optimized infidelities shown in Fig.\ \ref{fig:noise_suppression_pulse}(a).
As discussed in Sec.\ \ref{subsec:the_second_sweet_spot_analysis},
for $a_{E_{J}}\approx1.2$, both $Q_{A}$ and $Q_{B}$ can simultaneously
reside at their respective second sweet spots, $\Phi_{A}/\Phi_{0}=0.5$
and $\Phi_{B}/\Phi_{0}=0.5$, while also satisfying the strongest
coupling condition $\delta_{A}=0$. In contrast, as $a_{E_{J}}$ deviates
from $1.2$, these conditions cannot be satisfied simultaneously.
For $a_{E_{J}}\gtrsim1.2$, when $Q_{A}$ reaches its sweet spot at
$\Phi_{A}/\Phi_{0}=0.5$, $Q_{B}$ remains detuned at $\Phi_{B}/\Phi_{0}<0.5$
under the constraint $\delta_{A}=0$. Conversely, for $a_{E_{J}}\lesssim1.2$,
when $Q_{B}$ attains its sweet spot at $\Phi_{B}/\Phi_{0}=0.5$,
$Q_{A}$ is still located at $\Phi_{A}/\Phi_{0}<0.5$. Accordingly,
the pulse design strategies for asymmetric transmons are classified
into three distinct regimes. For $a_{E_{J}}\gtrsim1.2$, we set the
initial $Q_{A}$ interaction frequency $\omega_{i,A}$ at the minimum
of $\Delta_{A}$, i.e., by tuning the initial flux to $\Phi_{A}/\Phi_{0}=0.5$
in the Pull-Back-Pulse $\Phi_{A}^{\mathrm{PBP}}(t)$. The initial
$Q_{B}$ interaction frequency is then set to $\omega_{i,B}=\omega_{i,A}+\alpha_{A}$
by tuning the corresponding initial flux to $\Phi_{B}/\Phi_{0}<0.5$
in the Pull-Back-Pulse $\Phi_{B}^{\mathrm{PBP}}(t)$, ensuring the
condition $\delta_{A}=0$ is satisfied. To bring the overall flux
pulses closer to the sweet spot $\Phi_{A/B}/\Phi_{0}=0.5$, only the
$Q_{B}$ Fine-Tune-Pulse $\Phi_{B}^{\mathrm{FTP}}(t)$ is applied.
We then optimize the ten control parameters in $\Phi_{B}^{\mathrm{FTP}}(t)$,
together with the floating frequencies $\omega_{i,A}$ and $\omega_{i,B}$,
using the rp-Qt-4 optimization protocol with the cost function $J_{1}+\left\langle J_{2,\epsilon}\right\rangle +\left\langle J_{2,\phi}\right\rangle $.
For $a_{E_{J}}=1.2$, the initial interaction frequencies $\omega_{i,A}$
and $\omega_{i,B}$ are set at the minima of $\Delta_{A}$ and $\Delta_{B}$,
corresponding to $\Phi_{A/B}/\Phi_{0}=0.5$. In this case, both Fine-Tune-Pulse
$\Phi_{A}^{\mathrm{FTP}}(t)$ and $\Phi_{B}^{\mathrm{FTP}}(t)$ are
applied, each with $10$ control parameters, to address the challenging
optimization as $a_{E_{J}}\rightarrow1$. For $a_{E_{J}}\lesssim1.2$
(and conversely for $a_{E_{J}}\gtrsim1.2$), the initial $\omega_{i,B}$
is set at the minimum of $\Delta_{B}$, while the initial $\omega_{i,A}=\omega_{i,B}-\alpha_{A}$
under $\delta_{A}=0$, and only the $Q_{A}$ Fine-Tune-Pulse $\Phi_{A}^{\mathrm{FTP}}(t)$
is applied with additional control parameters to accommodate the increased
optimization difficulty. We show the optimized flux pulses $\Phi_{A/B}(t)/\Phi_{0}$
for $a_{E_{J}}=1.1$, $1.2$, and $1.31$, corresponding to the three
regimes, in Figs.\ \ref{fig:noise_suppression_pulse}(c), \ref{fig:noise_suppression_pulse}(d),
and \ref{fig:noise_suppression_pulse}(e), respectively. The corresponding
optimized qubit frequencies $\Delta_{A/B}(t)$ are shown in Figs.\ \ref{fig:noise_suppression_pulse}(g),
\ref{fig:noise_suppression_pulse}(h), and \ref{fig:noise_suppression_pulse}(i),
respectively.

For $a_{E_{J}}=1.31$, as shown in Fig.\ \ref{fig:noise_suppression_pulse}(e),
the optimized $\Phi_{A}(t)/\Phi_{0}=\Phi_{A}^{\mathrm{PBP}}(t)/\Phi_{0}$
is pulled to the sweet spot $\Phi_{A}/\Phi_{0}=0.5$, thereby minimizing
$\left\langle J_{2,\epsilon,A}\right\rangle $ and $\left\langle J_{2,\phi,A}\right\rangle $.
In contrast, the optimized $\Phi_{B}(t)/\Phi_{0}=\Phi_{B}^{\mathrm{PBP}}(t)/\Phi_{0}+\Phi_{B}^{\mathrm{FTP}}(t)/\Phi_{0}$
oscillates approximately between its sweet spot $\Phi_{B}/\Phi_{0}=0.5$
and $0.47$, which suppresses $\left\langle J_{2,\epsilon,B}\right\rangle $
and $\left\langle J_{2,\phi,B}\right\rangle $ while simultaneously
synchronizing intrinsic error contributions to their minima, yielding
the optimized $J_{1}\approx2.46\times10^{-5}$. Compared with the
calculated $\left\langle J_{2,\epsilon}\right\rangle $ and $\left\langle J_{2,\phi}\right\rangle $
at the second sweet spot---obtained using only the Pull-Back-Pulse
$\Phi_{A/B}(t)=\Phi_{A/B}^{\mathrm{PBP}}(t)$ at $\Phi_{A}/\Phi_{0}=0.5$
under $\delta_{A}=0$ as shown in Fig.\ \ref{fig:J2_est_aEJ_split}(h)---the
optimized $\left\langle J_{2,\epsilon}\right\rangle \approx1.02\times10^{-5}$
closely matches the calculated $\left\langle J_{2,\epsilon}\right\rangle $,
whereas the optimized $\left\langle J_{2,\phi}\right\rangle \approx3.09\times10^{-5}$
is reduced by more than half from the calculated $\left\langle J_{2,\phi}\right\rangle \approx7.03\times10^{-5}$.
The application of the same $\Phi_{A}(t)/\Phi_{0}=\Phi_{A}^{\mathrm{PBP}}(t)/\Phi_{0}$
control at $\Phi_{A}/\Phi_{0}=0.5$ renders the optimized $\left\langle J_{2,\epsilon,A}\right\rangle $
and $\left\langle J_{2,\phi,A}\right\rangle $ identical to the calculated
values. Consequently, the near invariance of $\left\langle J_{2,\epsilon}\right\rangle $
can be attributed to the identical $\left\langle J_{2,\epsilon,A}\right\rangle $,
which dominates $\left\langle J_{2,\epsilon}\right\rangle $ owing
to $\Delta_{A}>\Delta_{B}$, as illustrated in Fig.\ \ref{fig:noise_suppression_pulse}(i).
By contrast, $\left\langle J_{2,\phi}\right\rangle $ is governed
by $\left\langle J_{2,\phi,B}\right\rangle $, as $Q_{A}$ remains
at the second sweet spot $\Phi_{A}/\Phi_{0}=0.5$, minimizing $\left\langle J_{2,\phi,A}\right\rangle $
to a level below $\left\langle J_{2,\phi,B}\right\rangle $. Therefore,
the reduction in the optimized $\left\langle J_{2,\phi}\right\rangle $
primarily originates from the further suppression of $\left\langle J_{2,\phi,B}\right\rangle $,
demonstrating the efficacy of the Fine-Tune-Pulse $\Phi_{B}^{\mathrm{FTP}}(t)$.
Overall, our pulse design---employing the Pull-Back-Pulse $\Phi_{A/B}^{\mathrm{PBP}}(t)$
to bring $Q_{A}$ and $Q_{B}$ to the second sweet spots against dephasing
noises, combined with the Fine-Tune-Pulse $\Phi_{B}^{\mathrm{FTP}}(t)$
to synchronize intrinsic errors and further mitigate dephasing---effectively
improves the infidelity $J_{1}+\left\langle J_{2,\epsilon}\right\rangle +\left\langle J_{2,\phi}\right\rangle $.

For $a_{E_{j}}=1.2$, both qubits $Q_{A}$ and $Q_{B}$ can nearly
reside at their respective second flux-noise sweet spots, $\Phi_{A}/\Phi_{0}=0.5$
and $\Phi_{B}/\Phi_{0}=0.5$, simultaneously under $\delta_{A}=0$,
thereby minimizing their error contributions with approximately equal
amplitudes, i.e., $\left\langle J_{2,\phi,A}\right\rangle \approx\left\langle J_{2,\phi,B}\right\rangle $.
However, achieving synchronization of the intrinsic errors requires
inducing flux oscillations, which displace $Q_{A}$ and/or $Q_{B}$
from their sweet spots. We identify the optimal pulse design that
keeps $\left\langle J_{2,\phi}\right\rangle $ as close as possible
to its minimum while suppressing $J_{1}$, thereby minimizing the
total infidelity $J_{1}+\left\langle J_{2,\epsilon}\right\rangle +\left\langle J_{2,\phi}\right\rangle $,
by simultaneously applying the Pull-Back-Pulse $\Phi_{A/B}^{\mathrm{PBP}}(t)$
to both qubits. As shown in Fig.\ \ref{fig:noise_suppression_pulse}(d),
the optimized flux pulses $\Phi_{A/B}(t)/\Phi_{0}$ oscillate approximately
between the sweet spot $0.5$ and $0.48$, achieving a minimized infidelity
$J_{1}+\left\langle J_{2,\epsilon}\right\rangle +\left\langle J_{2,\phi}\right\rangle \approx1.45\times10^{-4}$,
with contributions $J_{1}\approx7.90\times10^{-5}$, $\left\langle J_{2,\phi}\right\rangle \approx6.07\times10^{-5}$,
and $\left\langle J_{2,\epsilon}\right\rangle \approx5.64\times10^{-6}$,
as shown in Fig.\ \ref{fig:noise_suppression_pulse}(a).

For $a_{E_{J}}=1.1$, the strong peak in $-d\Delta_{A/B}/d(\Phi_{A/B}/\Phi_{0})$
is very close to the second flux-noise sweet spot, $\Phi_{A/B}/\Phi_{0}=0.5$,
as shown in Fig.\ \ref{fig:J2_est_aEJ_split}(b-2). Consequently,
even minor deviations from the sweet spot result in substantial increases
in $\left\langle J_{2,\phi}\right\rangle $. To minimize the time
spent away from the sweet spots while allowing synchronization of
intrinsic errors, we fix $\Phi_{B}/\Phi_{0}=0.5$ in the Pull-Back-Pulse
$\Phi_{B}^{\mathrm{PBP}}(t)$, set the initial $\Phi_{A}/\Phi_{0}=0.4934$
in the Pull-Back-Pulse $\Phi_{A}^{\mathrm{PBP}}(t)$ under $\delta_{A}=0$,
and apply the Fine-Tune-Pulse $\Phi_{A}^{\mathrm{FTP}}(t)$ alone
with enhanced control parameters up to $30$. As illustrated in Fig.\ \ref{fig:noise_suppression_pulse}(c),
$\Phi_{B}(t)/\Phi_{0}$ remains pinned at $0.5$, while the optimized
$\Phi_{A}(t)/\Phi_{0}$, despite some high-frequency oscillations,
stays close to $0.5$ for most of the duration. This pulse design
yields a minimized total infidelity $J_{1}+\left\langle J_{2,\epsilon}\right\rangle +\left\langle J_{2,\phi}\right\rangle \approx3.24\times10^{-4}$,
with contributions $J_{1}\approx2.86\times10^{-5}$, $\left\langle J_{2,\phi}\right\rangle \approx2.93\times10^{-4}$,
and $\left\langle J_{2,\epsilon}\right\rangle \approx2.46\times10^{-6}$
, as shown in Fig.\ \ref{fig:noise_suppression_pulse}(a).

\subsubsection{Robust performance against noises}

\label{subsec:robustness_against_noise}

\begin{figure}
\includegraphics[scale=0.52]{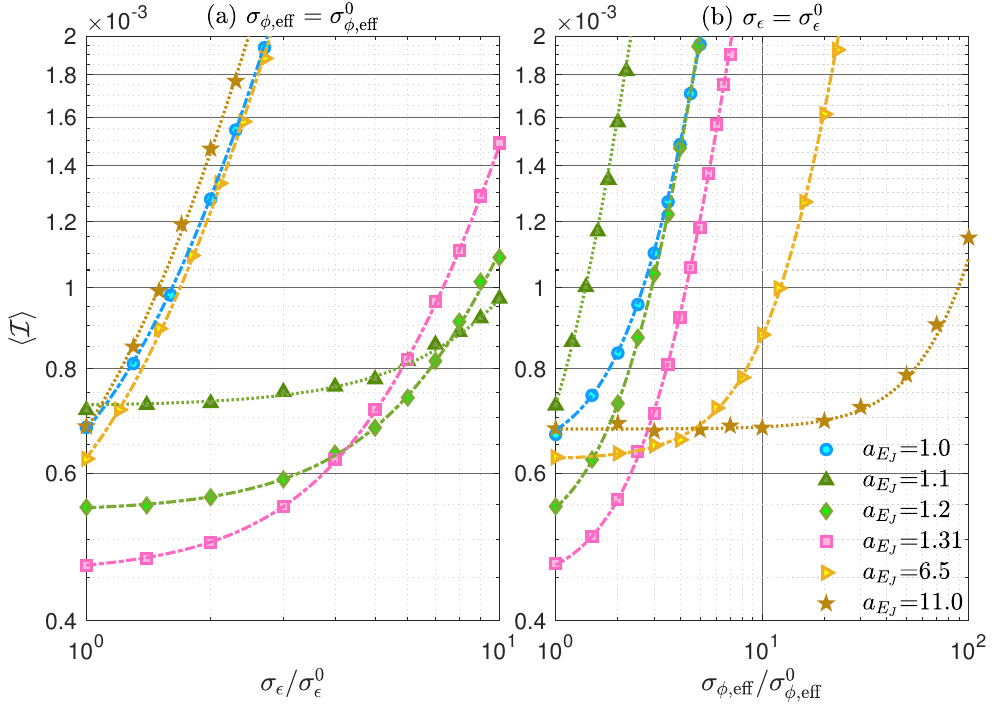}

\caption{Robustness of the optimized CZ gates against (a) RTCE noise $\beta_{\epsilon}(t)$
and (b) flux noise $\beta_{\phi}(t)$. In panel (a), the ensemble-averaged
infidelity $\left\langle \mathcal{I}\right\rangle $ is shown as a
function of the standard deviation $\sigma_{\epsilon}$ of the RTCE
noise, with the flux noise standard deviation fixed at its characterized
value, $\sigma_{\phi,\mathrm{eff}}=\sigma_{\phi,\mathrm{eff}}^{0}$.
In panel (b), $\left\langle \mathcal{I}\right\rangle $ is plotted
as a function of $\sigma_{\phi,\mathrm{eff}}$ at the characterized
$\sigma_{\epsilon}=\sigma_{\epsilon}^{0}$. The ensemble-averaged
infidelities $\left\langle \mathcal{I}\right\rangle $ are represented
by blue circles, dark-green triangles, bright-green diamonds, pink
squares, bright-yellow rightward triangles, and dark-yellow pentagrams
for the optimized CZ gates with $a_{E_{J}}=1.0$, $1.1$, $1.2$,
$1.31$, $6.5$, and $11.0$, respectively. The corresponding lines
are calculated using the total infidelity $J_{1}+\left\langle J_{2,T_{1}}\right\rangle +\left\langle J_{2,\epsilon}\right\rangle +\left\langle J_{2,\phi}\right\rangle $.}

\label{fig:robustness_against_noise} 
\end{figure}

To assess the robustness of the optimized CZ gates against the dephasing
noises, we compute the ensemble-averaged infidelity $\left\langle \mathcal{I}\right\rangle $
defined in Eq.\ (\ref{eq:ensemble_infid_schrodinger_eq}) using the
realistic Hamiltonian with the optimized control pulses $\Phi_{A/B}(t)/\Phi_{0}$,
constructed in Secs.\ \ref{subsec:CZ_against_noise_symmetric_transmons}
and \ref{subsec:CZ_against_noise_asymmetric_transmons}. The averaging
is performed over an ensemble of $5000$ realizations of the RTCE
noise $\beta_{\epsilon}(t)$ and the flux noise $\beta_{\phi}(t)$,
characterized in Secs.\ \ref{subsec:Noise_characterization_dephasing}
and \ref{subsec:Noise_characterization_from_CZ_infid_budget}. The
resulting values of $\left\langle \mathcal{I}\right\rangle $ are
shown as markers in Fig.\ \ref{fig:robustness_against_noise}, while
the total infidelity $J_{1}+\left\langle J_{2.T_{1}}\right\rangle +\left\langle J_{2,\epsilon}\right\rangle +\left\langle J_{2,\phi}\right\rangle $,
introduced in Sec.\ \ref{subsec:J2_noises}, is plotted as lines
for comparison. The total infidelity $J_{1}+\left\langle J_{2.T_{1}}\right\rangle +\left\langle J_{2,\epsilon}\right\rangle +\left\langle J_{2,\phi}\right\rangle $
provides an accurate prediction of the behavior of the ensemble-averaged
infidelity $\left\langle \mathcal{I}\right\rangle $. We define $\sigma_{\epsilon}^{0}$
as the characterized value of the standard deviation $\sigma_{\epsilon}$
associated with the RTCE noise $\beta_{\epsilon}(t)$, where $\sigma_{\epsilon}=\sqrt{\pi\Gamma_{\epsilon}S_{\epsilon,\mathrm{w}}}$.
The quantity $\sigma_{\epsilon}^{0}$ is determined by the characterized
white noise spectrum $S_{\epsilon,\mathrm{w}}$ and the corresponding
cutoff frequency $\Gamma_{\epsilon}$, as introduced in Secs.\ \ref{subsec:Noise_characterization_dephasing}
and \ref{subsec:Noise_characterization_from_CZ_infid_budget}. In
contrast, $\sigma_{\phi,\mathrm{eff}}^{0}$ denotes the characterized
value of the standard deviation $\sigma_{\phi,\mathrm{eff}}$ for
the flux noise $\beta_{\phi}(t)$. The quantity $\sigma_{\phi,\mathrm{eff}}$
is defined in Appendix\ \ref{appendix:noise_generation_OU}, and
$\sigma_{\phi,\mathrm{eff}}^{0}$ is determined by the characterized
$S_{1/f}$ spectrum and its distribution, as described in Sec.\ \ref{subsec:Noise_characterization_dephasing}.
To assess robustness against RTCE noise, we vary $\sigma_{\epsilon}$
from its characterized value $\sigma_{\epsilon}^{0}$ up to $10\cdot\sigma_{\epsilon}^{0}$,
while fixing $\sigma_{\phi,\mathrm{eff}}=\sigma_{\phi,\mathrm{eff}}^{0}$.
The resulting performance of the optimized CZ gates for $a_{E_{J}}=1.0$,
$1.1$, $1.2$. $1.31$, $6.5$, and $11.0$ is shown in Fig.\ \ref{fig:robustness_against_noise}(a).
Conversely, to evaluate robustness against flux noise, we vary $\sigma_{\phi,\mathrm{eff}}$
from $\sigma_{\phi,\mathrm{eff}}^{0}$ to $100\cdot\sigma_{\phi,\mathrm{eff}}^{0}$
while keeping $\sigma_{\epsilon}=\sigma_{\epsilon}^{0}$ , as presented
in Fig.\ \ref{fig:robustness_against_noise}(b).

For symmetric transmons with $a_{E_{J}}=1.0$, the optimized CZ gate
constructed in Sec.\ \ref{subsec:CZ_against_noise_symmetric_transmons}
exhibits an ensemble-averaged infidelity $\left\langle \mathcal{I}\right\rangle \approx6.80\times10^{-4}$
at the characterized noise strengths $\sigma_{\epsilon}=\sigma_{\epsilon}^{0}$
and $\sigma_{\phi,\mathrm{eff}}=\sigma_{\phi,\mathrm{eff}}^{0}$.
Imposing an infidelity threshold of $10^{-3}$, the gate tolerates
RTCE noise up to $\sigma_{\epsilon}\approx1.6\cdot\sigma_{\epsilon}^{0}$
with $\sigma_{\phi,\mathrm{eff}}=\sigma_{\phi,\mathrm{eff}}^{0}$
, as shown in Fig.\ \ref{fig:robustness_against_noise}(a), and flux
noise up to $\sigma_{\phi,\mathrm{eff}}\approx2.7\cdot\sigma_{\phi,\mathrm{eff}}^{0}$
with $\sigma_{\epsilon}=\sigma_{\epsilon}^{0}$, as shown in Fig.\ \ref{fig:robustness_against_noise}(b).
This threshold of $10^{-3}$ is adopted throughout the paper as the
criterion for quantifying robustness. For asymmetric transmons with
$a_{E_{J}}=1.31$, corresponding to the highest-fidelity CZ gate discussed
in Sec.\ \ref{subsec:CZ_against_noise_asymmetric_transmons}, the
ensemble-averaged infidelity at $\sigma_{\epsilon}=\sigma_{\epsilon}^{0}$
and $\sigma_{\phi,\mathrm{eff}}=\sigma_{\phi,\mathrm{eff}}^{0}$ is
reduced to $4.66\times10^{-4}$. Concurrently, the tolerances to RTCE
and flux noise are enhanced to $\sigma_{\epsilon}\approx7.3\cdot\sigma_{\epsilon}^{0}$
and $\sigma_{\phi,\mathrm{eff}}\approx4.3\cdot\sigma_{\phi,\mathrm{eff}}^{0}$,
respectively, representing a marked improvement over the symmetric
case with $a_{E_{J}}=1.0$. To further enhance robustness against
the RTCE noise, CZ gates should be operated on more weakly asymmetric
transmons with $a_{E_{J}}\rightarrow1$, where the RTCE noise contribution
$\left\langle J_{2,\epsilon}\right\rangle $ is minimized. In contrast,
further enhancing robustness against the flux noise requires operating
on more strongly asymmetric transmons with $a_{E_{J}}\gg1.31$, where
the flux noise contribution $\left\langle J_{2,\phi}\right\rangle $
is minimized, as detailed in Sec.\ \ref{subsec:the_second_sweet_spot_analysis}.
However, the optimized $\left\langle J_{2,\epsilon}\right\rangle $
and $\left\langle J_{2,\phi}\right\rangle $ exhibit competing trends
as functions of $a_{E_{J}}$, as shown in Fig.\ \ref{fig:noise_suppression_pulse}(a).
This trade-off indicates that enhancing robustness against the RTCE
noise inevitably degrades the robustness against the flux noise, and
vice versa.

For weakly asymmetric transmons with $a_{E_{J}}=1.1$, the optimized
CZ gate exhibits the lowest $\left\langle J_{2,\epsilon}\right\rangle $
as presented in Sec.\ \ref{subsec:CZ_against_noise_asymmetric_transmons},
corresponding to the highest tolerance against the RTCE noise, $\sigma_{\epsilon}>10\cdot\sigma_{\epsilon}^{0}$,
as shown in Fig.\ \ref{fig:robustness_against_noise}(a). This represents
an approximate sixfold improvement over the tolerance $\sigma_{\epsilon}\approx1.6\cdot\sigma_{\epsilon}^{0}$
for $a_{E_{J}}=1.0$. However, the ensemble-averaged infidelity, $\left\langle \mathcal{I}\right\rangle \approx7.14\times10^{-4}$
at $\sigma_{\epsilon}=\sigma_{\epsilon}^{0}$ and $\sigma_{\phi,\mathrm{eff}}=\sigma_{\phi,\mathrm{eff}}^{0}$,
and the flux-noise tolerance, $\sigma_{\phi,\mathrm{eff}}\approx1.4\cdot\sigma_{\phi,\mathrm{eff}}^{0}$
{[}Fig.\ \ref{fig:robustness_against_noise}(b){]}, are both degraded
compared to $\left\langle \mathcal{I}\right\rangle \approx6.80\times10^{-4}$
and $\sigma_{\phi,\mathrm{eff}}\approx2.7\cdot\sigma_{\phi,\mathrm{eff}}^{0}$
for $a_{E_{J}}=1.0$. By slightly reducing robustness against the
RTCE noise through increasing $a_{E_{J}}$ to $a_{E_{J}}=1.2$, the
ensemble-averaged infidelity can be lowered to $\left\langle \mathcal{I}\right\rangle \approx5.46\times10^{-4}$,
while the flux-noise tolerance increases to $\sigma_{\phi,\mathrm{eff}}\approx2.9\cdot\sigma_{\phi,\mathrm{eff}}^{0}$
relative to $a_{E_{J}}=1.0$. The corresponding reduction in the RTCE
noise tolerance to $\sigma_{\epsilon}\approx9\cdot\sigma_{\epsilon}^{0}$
still represents an over fivefold improvement in robustness compared
to the $a_{E_{J}}=1.0$ case.

For strongly asymmetric transmons with $a_{E_{J}}=11.0$, the optimized
$\left\langle J_{2,\phi}\right\rangle $ is suppressed by over three
orders of magnitude relative to $a_{E_{J}}=1.0$ as presented in Sec.\ \ref{subsec:CZ_against_noise_asymmetric_transmons}.
This substantial suppression yields a significantly enhanced flux
noise tolerance of $\sigma_{\phi\cdot\mathrm{eff}}\approx90\cdot\sigma_{\phi\cdot\mathrm{eff}}^{0}$,
representing a thirtyfold increase over $\sigma_{\phi\cdot\mathrm{eff}}\approx2.7\cdot\sigma_{\phi,\mathrm{eff}}^{0}$
for $a_{E_{J}}=1.0$, as shown in Fig.\ \ref{fig:robustness_against_noise}(b).
Furthermore, due to the mitigated infidelity $J_{1}+\left\langle J_{2,\epsilon}\right\rangle +\left\langle J_{2,\phi}\right\rangle $,
the ensemble-averaged infidelity $\left\langle \mathcal{I}\right\rangle \approx6.82\times10^{-4}$
at $\sigma_{\epsilon}=\sigma_{\epsilon}^{0}$ and $\sigma_{\phi,\mathrm{eff}}=\sigma_{\phi,\mathrm{eff}}^{0}$
and the RTCE noise tolerance $\sigma_{\epsilon}\approx1.5\cdot\sigma_{\epsilon}^{0}$
remain comparable to $\left\langle \mathcal{I}\right\rangle \approx6.80\times10^{-4}$
and $\sigma_{\epsilon}\approx1.6\cdot\sigma_{\epsilon}^{0}$ for $a_{E_{J}}=1.0$.
However, as $a_{E_{J}}$ is reduced from $11.0$ to $6.5$, the flux
noise tolerance is significantly degraded to $\sigma_{\phi\cdot\mathrm{eff}}\approx12\cdot\sigma_{\phi\cdot\mathrm{eff}}^{0}$,
while the improvements in both $\left\langle \mathcal{I}\right\rangle \approx6.24\times10^{-4}$
and $\sigma_{\epsilon}\approx1.7\cdot\sigma_{\epsilon}^{0}$ are marginal.
This behavior stems from the rapid increase in optimized $\left\langle J_{2,\phi}\right\rangle $
compared to the more gradual decrease in optimized $\left\langle J_{2,\epsilon}\right\rangle $
as $a_{E_{J}}$ decreases within this regime, as illustrated in Fig.\ \ref{fig:noise_suppression_pulse}(a).

In summary, we reduce the gate duration to $t_{f}=12\,\mathrm{ns}$,
approaching the CZ speed limit $t_{f}=\pi/(\sqrt{2}g_{AB})\approx11.6\,\mathrm{ns}$,
thereby suppressing decoherence-induced errors $\left\langle J_{2,T_{1}}\right\rangle +\left\langle J_{2,\epsilon}\right\rangle +\left\langle J_{2,\phi}\right\rangle $
and nearly saturating the improvement achievable solely through gate
time reduction. Simultaneously, the increase in intrinsic error $J_{1}$
associated with shorter gate times is mitigated via synchronization
enabled by the engineered control pulses $\Phi_{A/B}(t)=\Phi_{A/B}^{\mathrm{PBP}}(t)+\Phi_{A/B}^{\mathrm{FTP}}(t)$.
Furthermore, we find that the total dephasing contribution $\left\langle J_{2,\epsilon}\right\rangle +\left\langle J_{2,\phi}\right\rangle $
can be further suppressed by operating CZ gates on either symmetric
or asymmetric transmons at their respective sweet spots, where sensitivity
to dephasing noise is minimized. Finally, the minimum total infidelity
$J_{1}+\left\langle J_{2,T_{1}}\right\rangle +\left\langle J_{2,\epsilon}\right\rangle +\left\langle J_{2,\phi}\right\rangle \approx4.66\times10^{-4}$
is identified at $a_{E_{J}}=1.31$, as shown in Fig.\ \ref{fig:noise_suppression_pulse}(a).
This lowest total infidelity is primarily limited by the energy relaxation
contribution, $\left\langle J_{2,T_{1}}\right\rangle =4\times10^{-4}$
for $T_{1}=30\,\mu\mathrm{s}$. By incorporating tantalum for fabrication
as suggested in \citep{Place2021}, $T_{1}$ could potentially reach
$360\,\mu\mathrm{s}$, thereby reducing $\left\langle J_{2,T_{1}}\right\rangle $
to $3.33\times10^{-5}$ and resulting in a total infidelity $J_{1}+\left\langle J_{2,T_{1}}\right\rangle +\left\langle J_{2,\epsilon}\right\rangle +\left\langle J_{2,\phi}\right\rangle \approx9.9\times10^{-5}$
for $a_{E_{J}}=1.31$. Furthermore, if the RTCE noise from room-temperature
control electronics is eliminated via cryogenic control \citep{Bardin2019,Pauka2021,Xue2021}
or other methods, $\left\langle J_{2,\epsilon}\right\rangle $ can
be neglected. In this scenario, we can employ the most robust condition
against the flux noise at $a_{E_{J}}=11.0$, which yields the lowest
$\left\langle J_{2,\phi}\right\rangle \approx4.07\times10^{-8}$ and
the highest flux-noise tolerance of $\sigma_{\phi\cdot\mathrm{eff}}\approx90\cdot\sigma_{\phi\cdot\mathrm{eff}}^{0}$.
This leads to a further reduced total infidelity $J_{1}+\left\langle J_{2,T_{1}}\right\rangle +\left\langle J_{2,\phi}\right\rangle \approx5.22\times10^{-5}$,
limited by $\left\langle J_{2,T_{1}}\right\rangle \approx3.33\times10^{-5}$
for $T_{1}=360\,\mu\mathrm{s}$ and $J_{1}\approx1.88\times10^{-5}$.

\section{Conclusion}

\label{sec:Conclusion}

We have developed a comprehensive optimal-control framework for high-fidelity
and robust CZ gates in directly coupled, frequency-tunable transmon
qubits that systematically addresses both intrinsic Hamiltonian errors
and realistic decoherence mechanisms. By engineering smooth, single-shot
flux pulses composed of Pull-Back-Pulse and Fine-Tune-Pulse components,
we achieve $12\text{-ns}$ CZ gates---near the theoretical speed
limit of $11.6\,\mathrm{ns}$---with intrinsic error suppressed to
the $2\times10^{-5}$ level. Our framework incorporates experimentally
calibrated noise models based on Ramsey coherence measurements and
CZ-gate error budgets, capturing both white noise from room-temperature
control electronics and $1/f$ flux noise. These models enable identification
of optimal operating regimes---dephasing sweet spots---where sensitivity
to dephasing noise is substantially suppressed, providing a principled
approach to gate optimization. The optimized $12\text{-ns}$ CZ gate
using transmons with an optimal asymmetry of $a_{E_{J}}=1.31$, identified
under the experimentally characterized noise conditions reported in
Ref.\ \citep{BarendsKellyMegrantEtAl2014}, achieves a total infidelity
as low as $4.66\times10^{-4}$, limited primarily by the energy-relaxation
contribution of $4\times10^{-4}$ for $T_{1}=30\,\mu\mathrm{s}$.
With state-of-the-art coherence times ($T_{1}=360\,\mu\mathrm{s}$),
the projected total infidelity decreases to $9.9\times10^{-5}$.

A key finding is that transmon asymmetry provides tunable access to
distinct noise-resilient operating regimes. Transmons with the optimal
asymmetry ($a_{E_{J}}=1.31$), operating near the joint sweet spot
($\Phi/\Phi_{0}=0.5$), achieve a balance between fast- and slow-dephasing
errors. The corresponding optimized gate maintains a total infidelity
below $10^{-3}$ under either a sevenfold increase in fast-dephasing
noise strength or a fourfold increase in slow-dephasing noise strength
relative to the experimentally characterized levels. Strongly asymmetric
transmons ($a_{E_{J}}=11.0$) provide substantial flux-noise tolerance---up
to $90\times$ the experimentally characterized level---whereas weakly
asymmetric transmons ($a_{E_{J}}=1.1$) tolerate fast-dephasing noise
up to $10\times$ the experimentally characterized level. More broadly,
these results demonstrate that transmon asymmetry can serve as a tunable
design parameter for tailoring gate robustness to different noise
conditions.

The unified treatment of intrinsic and decoherence errors established
here provides a broadly applicable methodology for superconducting
quantum processors. The framework also accommodates realistic AWG
bandwidth limitations through filtering-aware post-optimization fine-tuning,
recovering high-fidelity gate performance at a representative cutoff
frequency of $424.7\,\mathrm{MHz}$. Beyond direct-coupling architectures,
the principles of sweet-spot identification, smooth physics-informed
pulse engineering, hierarchical intrinsic error decomposition, asymmetry-assisted
gate design, and modular error decomposition for optimization can
be extended to other coupling schemes, including tunable couplers
and resonator-mediated designs. The combination of near-speed-limit
operation, sub-$10^{-3}$ gate infidelity, and multi-fold noise tolerance
demonstrates the potential of the proposed framework for practical,
robust gate implementation in current and near-term superconducting
hardware.

However, achieving fault-tolerant quantum computing at scale requires
addressing an additional critical challenge: robustness against systematic
parameter uncertainties. Imperfect knowledge and device-to-device
variations in coupling strength ($g_{AB}$), anharmonicities ($\alpha_{A/B}$),
and zero-flux qubit frequencies ($\omega_{0,A/B}$), as well as temporal
drift in these parameters due to environmental fluctuations and aging
effects, can significantly degrade gate fidelity even when decoherence
is well controlled. In large-scale processors with thousands to millions
of qubits, continuous recalibration becomes impractical, necessitating
control strategies that maintain high fidelity across realistic parameter
distributions. A companion paper (Part $\mathrm{II}$) \citep{Huang2026a}
extends this framework to incorporate robust optimization against
such uncertainties, demonstrating that dynamical decoupling through
strong detuning modulation and operation at parameter-insensitive
sweet spots can provide resilience to parameter uncertainties comparable
to the noise robustness demonstrated here. Together, these complementary
approaches---decoherence-suppressing control (Part $\mathrm{I}$)
and parameter-robust control (Part $\mathrm{II}$)---provide a framework
for understanding and mitigating both decoherence and systematic parameter
uncertainties in superconducting quantum processors, thereby establishing
a pathway toward fast, high-fidelity, and robust two-qubit gates suitable
for large-scale fault-tolerant quantum computation.
\begin{acknowledgments}
H.-S. Goan acknowledges support from the NSTC, Taiwan, under Grants
No. NSTC 113-2112-M-002-022-MY3, No. NSTC 113-2119M-002-021, No. NSTC
114-2119-M-002-018, No. NSTC 114-2119-M-002-017-MY3, and from the
National Taiwan University under Grants No. NTU-CC-115L8937, No. NTUCC115L893704
and No. NTU-CC-115L8512. H.-S. Goan is also grateful for the support
of the ``Center for Advanced Computing and Imaging in Biomedicine
(NTU-115L900702)'' through the Featured Areas Research Center Program
within the framework of the Higher Education Sprout Project by the
Ministry of Education (MOE), Taiwan, the support of Taiwan Semiconductor
Research Institute (TSRI) through the Joint Developed Project (JDP)
and the support from the Physics Division, National Center for Theoretical
Sciences, Taiwan. 
\end{acknowledgments}

\appendix

\section{Error contributions to ideal CZ gate infidelity}

\label{appendix:error_contributions}

In this Appendix, we present a derivation of the intrinsic error contributions
to the ideal CZ gate infidelity $J_{1}$, encompassing the leakage
error, swap error, and phase error, as introduced in Sec.\ \ref{subsec:Error_analysis_ideal_CZ_gate}.
A general propagator $U_{I}$ corresponding to the ideal Hamiltonian
$\mathcal{H}_{I}$ {[}Eq.\ (\ref{eq:full_ideal_H}){]}, when projected
onto the computational basis states $\left\{ \left|0_{B}0_{A}\right\rangle ,\left|0_{B}1_{A}\right\rangle ,\left|1_{B}0_{A}\right\rangle ,\left|1_{B}1_{A}\right\rangle \right\} $,
can be approximated as

\begin{equation}
U_{I,4\times4}\approx\left(\begin{array}{cccc}
a\cdot e^{i\phi_{00}} & 0 & 0 & g\cdot e^{i\phi'_{g}}\\
0 & b\cdot e^{i\phi_{01}} & f\cdot e^{i\phi'_{f}} & 0\\
0 & f\cdot e^{i\phi_{f}} & c\cdot e^{i\phi_{10}} & 0\\
g\cdot e^{i\phi_{g}} & 0 & 0 & d\cdot e^{i\phi_{11}}
\end{array}\right).\label{eq:general_propagator}
\end{equation}
Here, $a$, $b$, $c$, $d$, $f$, and $g$ denote positive real
numbers, while $e^{i\phi_{00}}$, $e^{i\phi_{01}}$, $e^{i\phi_{10}}$,
$e^{i\phi_{11}}$, $e^{i\phi_{f}}$, $e^{i\phi'_{f}}$ ,$e^{i\phi_{g}}$,
and $e^{i\phi'_{g}}$ represent the associated phase factors.

Substituting $U_{I,4\times4}$ from Eq.\ (\ref{eq:general_propagator})
into the definition of $J_{1}$ given in Eq.\ (\ref{eq:J1_definition}),
we obtain

\begin{align}
J_{1} & \approx1-\dfrac{1}{16}\left|a\cdot e^{i\phi_{00}}+b\cdot e^{i\phi_{01}}+c\cdot e^{i\phi_{10}}-d\cdot e^{i\phi_{11}}\right|^{2}\nonumber \\
 & =1-\dfrac{1}{16}\left(a^{2}+b^{2}+c^{2}+d^{2}\right.\nonumber \\
 & \left.+2ab+2ac+2bc+2ad+2bd+2cd\right)\nonumber \\
 & +\dfrac{1}{16}\left\{ 2ab\left[1-\cos\left(\phi_{01}-\phi_{00}\right)\right]+2ac\left[1-\cos\left(\phi_{10}-\phi_{00}\right)\right]\right.\nonumber \\
 & +2bc\left[1-\cos\left(\phi_{01}-\phi_{10}\right)\right]+2ad\left[1-\cos\left(\phi_{11}-\phi_{00}-\pi\right)\right]\nonumber \\
 & +2bd\left[1-\cos\left(\phi_{11}-\phi_{01}-\pi\right)\right]\nonumber \\
 & \left.+2cd\left[1-\cos\left(\phi_{11}-\phi_{10}-\pi\right)\right]\right\} .\label{eq:J1_expansion}
\end{align}
The quantities $a^{2}$, $b^{2}$, $c^{2}$, and $d^{2}$ denote the
residual probabilities in the basis states $\left|0_{B}0_{A}\right\rangle $,
$\left|0_{B}1_{A}\right\rangle $, $\left|1_{B}0_{A}\right\rangle $,
and $\left|1_{B}1_{A}\right\rangle $, respectively, after the application
of $U_{I}$. They can be expressed as $a^{2}=1-P_{\left|0_{B}0_{A}\right\rangle }^{\mathrm{out}}$,
$b^{2}=1-P_{\left|0_{B}1_{A}\right\rangle }^{\mathrm{out}}$, $c^{2}=1-P_{\left|1_{B}0_{A}\right\rangle }^{\mathrm{out}}$,
and $d^{2}=1-P_{\left|1_{B}1_{A}\right\rangle }^{\mathrm{out}}$,
where the $P_{\left|0_{B}0_{A}\right\rangle }^{\mathrm{out}}$, $P_{\left|0_{B}1_{A}\right\rangle }^{\mathrm{out}}$,
$P_{\left|1_{B}0_{A}\right\rangle }^{\mathrm{out}}$, and $P_{\left|1_{B}1_{A}\right\rangle }^{\mathrm{out}}$
represent the total probabilities leaving the corresponding computational
basis states. Specifically, the outflow probability from $\left|0_{B}0_{A}\right\rangle $
is given by $P_{\left|0_{B}0_{A}\right\rangle }^{\mathrm{out}}\approx g^{2}+P_{\left|0_{B}0_{A}\right\rangle }^{\mathrm{leakage}}$,
where $g^{2}$ corresponds to the transition probability to $\left|1_{B}1_{A}\right\rangle $
induced by the $\left|0_{B}0_{A}\right\rangle \leftrightarrow\left|1_{B}1_{A}\right\rangle $
interaction, and $P_{\left|0_{B}0_{A}\right\rangle }^{\mathrm{leakage}}$
accounts for leakage out of the computational subspace. An analogous
expression holds for $\left|1_{B}1_{A}\right\rangle $, namely, $P_{\left|1_{B}1_{A}\right\rangle }^{\mathrm{out}}\approx g^{2}+P_{\left|1_{B}1_{A}\right\rangle }^{\mathrm{leakage}}$.
Similarly, for the states $\left|0_{B}1_{A}\right\rangle $ and $\left|1_{B}0_{A}\right\rangle $,
the outflow probabilities take the form $P_{\left|0_{B}1_{A}\right\rangle }^{\mathrm{out}}\approx f^{2}+P_{\left|0_{B}1_{A}\right\rangle }^{\mathrm{leakage}}$
and $P_{\left|1_{B}0_{A}\right\rangle }^{\mathrm{out}}\approx f^{2}+P_{\left|1_{B}0_{A}\right\rangle }^{\mathrm{leakage}}$,
where $f^{2}$ denotes the transition probability associated with
the $\left|0_{B}1_{A}\right\rangle \leftrightarrow\left|1_{B}0_{A}\right\rangle $
interaction, and $P_{\left|0_{B}1_{A}\right\rangle }^{\mathrm{leakage}}$
and $P_{\left|1_{B}0_{A}\right\rangle }^{\mathrm{leakage}}$ quantify
the corresponding leakage from the computational subspace.

As $U_{I,4\times4}$ approaches the target operation $U_{T}=\mathrm{CZ}$,
the outflow probabilities $P_{\left|0_{B}0_{A}\right\rangle }^{\mathrm{out}}$,
$P_{\left|0_{B}1_{A}\right\rangle }^{\mathrm{out}}$, $P_{\left|1_{B}0_{A}\right\rangle }^{\mathrm{out}}$,
and $P_{\left|1_{B}1_{A}\right\rangle }^{\mathrm{out}}$ become small.
Accordingly, the $U_{I,4\times4}$ diagonal amplitudes and their cross
terms can be expanded to leading order as $a\approx1-P_{\left|0_{B}0_{A}\right\rangle }^{\mathrm{out}}/2$
and $2ab\approx2-P_{\left|0_{B}0_{A}\right\rangle }^{\mathrm{out}}-P_{\left|0_{B}1_{A}\right\rangle }^{\mathrm{out}}$.
Concurrently, the phase differences among the local phases become
small, except for $\phi_{11}$, which retains a $\pi$-phase shift
relative to $\phi_{00}$, $\phi_{01}$, and $\phi_{10}$. Consequently,
the phase-dependent terms can be expanded to leading order as $1-\cos(\phi_{01}-\phi_{00})\approx(\phi_{01}-\phi_{00})^{2}/2$
and $1-\cos(\phi_{11}-\phi_{01}-\pi)\approx(\phi_{11}-\phi_{01}-\pi)^{2}/2$.
Substituting these approximations into $J_{1}$ in Eq.\ (\ref{eq:J1_expansion}),
and neglecting higher-order mixed terms involving products of outflow
probabilities and phase deviations, such as $P_{\left|0_{B}0_{A}\right\rangle }^{\mathrm{out}}\cdot(\phi_{01}-\phi_{00})^{2}/2$,
we obtain

\begin{align}
J_{1} & \approx\dfrac{1}{4}\left(P_{\left|0_{B}0_{A}\right\rangle }^{\mathrm{leakage}}+P_{\left|0_{B}1_{A}\right\rangle }^{\mathrm{leakage}}+P_{\left|1_{B}0_{A}\right\rangle }^{\mathrm{leakage}}+P_{\left|1_{B}1_{A}\right\rangle }^{\mathrm{leakage}}\right)\nonumber \\
 & +\dfrac{1}{2}\left(g^{2}+f^{2}\right)\nonumber \\
 & +\dfrac{1}{16}\left[(\phi_{01}-\phi_{00})^{2}+(\phi_{10}-\phi_{00})^{2}+(\phi_{01}-\phi_{10})^{2}\right]\nonumber \\
 & +\dfrac{1}{16}\left[(\phi_{11}-\phi_{00}-\pi)^{2}+(\phi_{11}-\phi_{01}-\pi)^{2}\right.\nonumber \\
 & \left.+(\phi_{11}-\phi_{10}-\pi)^{2}\right].\label{eq:J1_expansion_approx}
\end{align}

The first term in Eq.\ (\ref{eq:J1_expansion_approx}) represents
the error arising from probability leakage out of the computational
subspace (referred to as the leakage error), which can also be evaluated
using the formula, $1-\left|\mathrm{Tr}(U_{I,4\times4}^{\dagger}U_{I,4\times4})\right|^{2}/4$,
employed in the main text. The second term in Eq.\ (\ref{eq:J1_expansion_approx})
defines the swap error as introduced in the main text, including contributions
of $g^{2}/2$ for the $\left|0_{B}0_{A}\right\rangle \leftrightarrow\left|1_{B}1_{A}\right\rangle $
swap error and $f^{2}/2$ for $\left|0_{B}1_{A}\right\rangle \leftrightarrow\left|1_{B}0_{A}\right\rangle $
swap error. The sum of the remaining terms in Eq.\ (\ref{eq:J1_expansion_approx})
defines the phase error as introduced in the main text, which can
be divided into two contributions. The first contribution, $((\phi_{01}-\phi_{00})^{2}+(\phi_{10}-\phi_{00})^{2}+(\phi_{01}-\phi_{10})^{2})/16$,
represents the relative local phase error arising from the differences
among the local phases $\phi_{00}$, $\phi_{01}$, and $\phi_{10}$.
The second contribution, $((\phi_{11}-\phi_{00}-\pi)^{2}+(\phi_{11}-\phi_{01}-\pi)^{2}+(\phi_{11}-\phi_{10}-\pi)^{2})/16$,
describes the $\phi_{11}$ phase error, which quantifies the difference
between the target coupling phase $\phi_{AB}=\pi$ and the shifted
local phase $\phi_{11}$ relative to the other local phases $\phi_{00}$,
$\phi_{01}$, and $\phi_{10}$. In general, the relative local phase
error is associated with the accuracy of single-qubit phase compensation,
whereas the $\phi_{11}$ phase error characterizes the proximity of
the accumulated coupling phase to the target coupling phase $\phi_{AB}=\pi$.
When the phase shift becomes sufficiently large, the approximation
breaks down, and the omitted higher-order terms and probability-phase
cross terms, such as $P_{\left|0_{B}0_{A}\right\rangle }^{\mathrm{out}}\cdot(\phi_{01}-\phi_{00})^{2}/2$,
may no longer be negligible. Consequently, in this regime, the phase
error defined here is not suitable for precise quantitative analysis.
However, it remains useful for qualitative behavior analysis, for
instance in the case of drift Hamiltonian evolution shown in Figs.\ \ref{fig:Drift_H_evolution}(a)
and \ref{fig:Drift_H_evolution}(b), where the phase error attains
its minimum as $\phi_{11}$ approaches the target value $\pi$ and
increases as $\phi_{11}$ deviates from $\pi$---even exceeding unity
when $\phi_{11}$ is sufficiently far from $\pi$.

\section{Estimated dephasing times}

\label{appendix:estimated_dephasing_times}

In Sec.\ \ref{subsec:Noise_models}, we employ the estimated dephasing
times $T_{\phi,1}$ and $T_{\phi,2}$ to fit the experimental $T_{\phi,1}$
and $T_{\phi,2}$, thereby calibrating the parameters of the noise
models developed herein. Experimentally, $T_{\phi,1}$ and $T_{\phi,2}$
are extracted by fitting the measured Ramsey decay envelope $V(t)$
to the functional form $\exp\left[-t/T_{\phi,1}-(t/T_{\phi,2})^{2}\right]$.
In this Appendix, we start from a Hamiltonian to derive the general
expression for the Ramsey decay envelope, $V(t)\approx\exp\left[-\left\langle \varphi^{2}(t)\right\rangle /2\right]$.
Using this framework, we then apply our noise models to obtain the
corresponding $\left\langle \varphi_{\epsilon}^{2}(t)\right\rangle $
and $\left\langle \varphi_{\phi}^{2}(t)\right\rangle $, as presented
in Eqs.\ (\ref{eq:phi_2_RTCE_noise}) and (\ref{eq:phi_2_flux_noise})
of the main text for the RTCE and flux noise, respectively. This procedure
subsequently provides the corresponding estimated $T_{\phi,1,\epsilon}$
and $T_{\phi,2}$, as reported in Eqs.\ (\ref{eq:T_phi_1_e_est})
and (\ref{eq:T_phi_2_est}) of the main text.

The dephasing dynamics of a qubit can be modeled by the Hamiltonian
\begin{equation}
\mathcal{H}(t)=-\dfrac{1}{2}\hbar\left[\omega_{0}+\beta(t)\right]\sigma_{Z},\label{eq:dephasing_model_Hamiltonian}
\end{equation}
expressed in the computational basis states $\left|0\right\rangle $
and $\left|1\right\rangle $, where $\omega_{0}$ is the qubit frequency
and $\beta(t)$ represents the dephasing noise, both oriented along
$\sigma_{Z}$. Experimentally, the dephasing noise $\beta(t)$ can
be characterized by monitoring the decay of Ramsey fringes. In a Ramsey
experiment, the initial state $\left|0\right\rangle $ is first rotated
onto the $x-y$ plane of the Bloch sphere via a $+\pi/2$ rotation
about the $X$ axis, $R_{X}(+\pi/2)$, yielding the state $R_{X}(+\pi/2)\left|0\right\rangle $.
The state is then allowed to freely evolve for a time $t$ under the
dephasing Hamiltonian in Eq.\ (\ref{eq:dephasing_model_Hamiltonian}),
resulting in $U(t)R_{X}(+\pi/2)\left|0\right\rangle $, where

\begin{equation}
U(t)=\left(\begin{array}{cc}
\exp(+i\dfrac{\omega_{0}t+\varphi(t)}{2}) & 0\\
0 & \exp(-i\dfrac{\omega_{0}t+\varphi(t)}{2})
\end{array}\right)\label{eq:dephasing_model_propagator}
\end{equation}
is the corresponding propagator, with $\varphi(t)\equiv\int_{0}^{t}\beta(t')dt'$.
A second $+\pi/2$ $X$-rotation is subsequently applied to rotate
the state out of the $x-y$ plane, giving the final state $R_{X}(+\pi/2)U(t)R_{X}(+\pi/2)\left|0\right\rangle $.
The probability of measuring the state in $\left|1\right\rangle $
is then

\begin{align}
P_{+X/2}(t) & =\left|\left\langle 1\right|R_{X}(+\pi/2)U(t)R_{X}(+\pi/2)\left|0\right\rangle \right|^{2}\nonumber \\
 & =\dfrac{1}{2}+\dfrac{1}{2}\left\langle \cos[\omega_{0}t+\phi(t)]\right\rangle ,\label{eq:dephasing_model_P_p_x_2}
\end{align}
where the subscript $+X/2$ indicates the $+\pi/2$ $X$-rotation,
$R_{X}(+\pi/2)$, applied immediately prior to measurement. Notably,
$P_{+X/2}(t)$ defined here corresponds directly to the original Ramsey
fringes without invoking any assumptions or approximations. By substituting
the matrix representations of $R_{X}(\pi/2)$ and $U(t)$ from Eq.\ (\ref{eq:dephasing_model_propagator})
into the first line of $P_{+X/2}(t)$ in Eq.\ (\ref{eq:dephasing_model_P_p_x_2}),
one obtains the second line of $P_{+X/2}(t)$, in which the ensemble-averaged
term can be expressed as $\left\langle \cos[\omega_{0}t+\varphi(t)]\right\rangle =\cos(\omega_{0}t)\left\langle \cos[\varphi(t)]\right\rangle -\sin(\omega_{0}t)\left\langle \sin[\varphi(t)]\right\rangle $.
Assuming that the dephasing noise $\beta(t)$ is symmetrically distributed
with zero mean at all times, $\left\langle \sin[\varphi(t)]\right\rangle =0$,
so that the Ramsey fringes reduce to

\begin{equation}
P_{+X/2}(t)=\dfrac{1}{2}+\dfrac{1}{2}\cos(\omega_{0}t)\left\langle \cos[\varphi(t)]\right\rangle .\label{eq:Ramsey_fringes_gauss}
\end{equation}

The Ramsey fringes $P_{+X/2}(t)$ in Eq.\ (\ref{eq:Ramsey_fringes_gauss})
contain the Ramsey decay envelope, $V(t)=\left\langle \cos[\varphi(t)]\right\rangle $,
which characterizes the noise. Experimentally, to isolate the decay
envelope from the oscillatory term $\cos(\omega_{0}t)$, it is necessary
to measure additional Ramsey fringes 
\begin{align}
P_{-X/2}(t) & =\left|\left\langle 1\right|R_{X}(-\pi/2)U(t)R_{X}(\pi/2)\left|0\right\rangle \right|^{2}\nonumber \\
 & =\dfrac{1}{2}-\dfrac{1}{2}\left\langle \cos[\omega_{0}t+\varphi(t)]\right\rangle ,\label{eq:dephasing_model_P_m_x_2}\\
P_{+Y/2}(t) & =\left|\left\langle 1\right|R_{Y}(+\pi/2)U(t)R_{X}(\pi/2)\left|0\right\rangle \right|^{2}\nonumber \\
 & =\dfrac{1}{2}-\dfrac{1}{2}\left\langle \sin[\omega_{0}t+\varphi(t)]\right\rangle ,\label{eq:dephasing_model_P_p_y_2}\\
P_{-Y/2}(t) & =\left|\left\langle 1\right|R_{Y}(-\pi/2)U(t)R_{X}(\pi/2)\left|0\right\rangle \right|^{2}\nonumber \\
 & =\dfrac{1}{2}+\dfrac{1}{2}\left\langle \sin[\omega_{0}t+\varphi(t)]\right\rangle .\label{eq:dephasing_model_P_m_y_2}
\end{align}
These measured values are then combined according to

\begin{align}
 & V(t)=\nonumber \\
 & \sqrt{\left(P_{+X/2}(t)-P_{-X/2}(t)\right)^{2}+\left(P_{+Y/2}(t)-P_{-Y/2}(t)\right)^{2}},\label{eq:Ramsey_Envelope_def}
\end{align}
thereby yielding the general form of the Ramsey decay envelope, $V(t)=\sqrt{\left\langle \cos[\varphi(t)]\right\rangle ^{2}+\left\langle \sin[\varphi(t)]\right\rangle ^{2}}$.
Since $\left\langle \sin[\varphi(t)]\right\rangle =0$ as noted above,
this expression simplifies to $V(t)=\left|\left\langle \cos[\varphi(t)]\right\rangle \right|$,
which is the envelope we aim to isolate.

For the special case of quasi-static and normally distributed dephasing
noise $\beta$, with zero mean and standard deviation $\sigma_{\beta}$,
the decay term becomes $\left\langle \cos[\varphi(t)]\right\rangle =\left\langle \cos(\beta t)\right\rangle =$$\int_{-\infty}^{+\infty}d\beta\cos(\beta t)\mathrm{exp}[-\beta^{2}/(2\sigma_{\beta}^{2})]/\sqrt{2\pi\sigma_{\beta}^{2}}$$=\mathrm{exp}[-(\sigma_{\beta}^{2}t^{2})/2]$.
Consequently, the Ramsey decay envelope can be expressed as

\begin{equation}
V(t)=\exp\left[-\left(t/T_{2}^{\star}\right)^{2}\right],\label{eq:Ramsey_envelope_special}
\end{equation}
where the intrinsic dephasing time is $T_{2}^{\star}=\sqrt{2}/\sigma_{\beta}$.

For the general case of weak, time-dependent dephasing noise, one
can approximate the decay term $\left\langle \cos(\varphi(t))\right\rangle \approx1-\left\langle \varphi^{2}(t)\right\rangle /2$$\approx\mathrm{exp}\left[-\left\langle \varphi^{2}(t)\right\rangle /2\right]$,
yielding

\begin{equation}
V(t)\approx\mathrm{exp}\left[-\dfrac{\left\langle \varphi^{2}(t)\right\rangle }{2}\right],\label{eq:Ramsey_envelope_general}
\end{equation}
where $\left\langle \varphi^{2}(t)\right\rangle =\int_{0}^{t}\int_{0}^{t}dt_{1}dt_{2}\left\langle \beta(t_{1})\beta(t_{2})\right\rangle $.
Substituting the two-time correlation function, $\left\langle \beta(t_{1})\beta(t_{2})\right\rangle =(1/2\pi)\int_{-\infty}^{\infty}S(\omega)e^{-i\omega(t_{1}-t_{2})}d\omega$,
into $\left\langle \varphi^{2}(t)\right\rangle $ via the Wiener--Khinchin
theorem yields

\begin{equation}
\left\langle \varphi^{2}(t)\right\rangle =\dfrac{1}{2\pi}\int_{-\infty}^{\infty}d\omega S(\omega)\dfrac{1}{(\dfrac{\omega}{2})^{2}}\sin^{2}(\dfrac{\omega}{2}t).\label{eq:Ramsey_fringes_phi_square}
\end{equation}
Here $S(\omega)$ denotes the noise spectrum of the dephasing noise
$\beta(t)$.

For the system under consideration, as described in the main text,
the dynamics are influenced by two dephasing mechanisms: the RTCE
noise $\beta_{\epsilon}(t)$ and the flux noise $\beta_{\phi}(t)$,
characterized by the noise Hamiltonians $\mathcal{H}_{\epsilon,A}$
and $\mathcal{H}_{\phi,A}$ in Eqs.\ (\ref{eq:noise_H_RTCE_noise})
and (\ref{eq:noise_H_flux_noise}), respectively. Consequently, the
effective dephasing noise $\beta(t)$ appearing in the model of Eq.\ (\ref{eq:dephasing_model_Hamiltonian})
can be expressed as $\beta(t)=\beta_{\epsilon}(t)\left[\Delta_{A}-\alpha_{A}\right]^{\eta_{\epsilon}/2}+\beta_{\phi}(t)\left[d\Delta_{A}/d(\Phi_{A}/\Phi_{0})\right]\left[\text{\ensuremath{\Phi_{A}}}/\Phi_{0}\right]$.
Assuming $\beta_{\epsilon}(t)$ and $\beta_{\phi}(t)$ are statistically
independent, the two-time correlation function of $\beta(t)$ is $\left\langle \beta(t_{1})\beta(t_{2})\right\rangle =\left[\Delta_{A}-\alpha_{A}\right]^{\eta_{\epsilon}}\left\langle \beta_{\epsilon}(t_{1})\beta_{\epsilon}(t_{2})\right\rangle +\left[d\Delta_{A}/d(\Phi_{A}/\Phi_{0})\right]^{2}\left[\text{\ensuremath{\Phi_{A}}}/\Phi_{0}\right]^{2}\left\langle \beta_{\phi}(t_{1})\beta_{\phi}(t_{2})\right\rangle $.
Substituting the correlation functions of the individual noise sources,
$\left\langle \beta_{\epsilon}(t_{1})\beta_{\epsilon}(t_{2})\right\rangle =(1/2\pi)\int_{-\infty}^{\infty}S_{\epsilon}(\omega)e^{-i\omega(t_{1}-t_{2})}d\omega$
and $\left\langle \beta_{\phi}(t_{1})\beta_{\phi}(t_{2})\right\rangle =(1/2\pi)\int_{-\infty}^{\infty}S_{\phi}(\omega)e^{-i\omega(t_{1}-t_{2})}d\omega$,
where $S_{\epsilon}(\omega)$ and $S_{\phi}(\omega)$ denote the noise
spectra of $\beta_{\epsilon}(t)$ and $\beta_{\phi}(t)$, respectively,
the correlation function of $\beta(t)$ can be written, via the Wiener--Khinchin
theorem, as $\left\langle \beta(t_{1})\beta(t_{2})\right\rangle =(1/2\pi)\int_{-\infty}^{\infty}S(\omega)e^{-i\omega(t_{1}-t_{2})}d\omega$,
with the total noise spectrum given by $S(\omega)=\left[\Delta_{A}-\alpha_{A}\right]^{\eta_{\epsilon}}S_{\epsilon}(\omega)+\left[d\Delta_{A}/d(\Phi_{A}/\Phi_{0})\right]^{2}\left[\text{\ensuremath{\Phi_{A}}}/\Phi_{0}\right]^{2}S_{\phi}(\omega)$.
Substituting $S(\omega)$ into Eq.\ (\ref{eq:Ramsey_fringes_phi_square})
for $\left\langle \varphi^{2}(t)\right\rangle $ then yields a decomposition
into contributions from each noise channel, $\left\langle \varphi^{2}(t)\right\rangle =\left\langle \varphi_{\epsilon}^{2}(t)\right\rangle +\left\langle \varphi_{\phi}^{2}(t)\right\rangle $,
where $\left\langle \varphi_{\epsilon}^{2}(t)\right\rangle $ and
$\left\langle \varphi_{\phi}^{2}(t)\right\rangle $ are given explicitly
in Eqs.\ (\ref{eq:phi_2_RTCE_noise}) and (\ref{eq:phi_2_flux_noise})
of the main text, respectively.

For the RTCE noise, we substitute the white noise spectrum $S_{\epsilon}(\omega)\approx S_{\epsilon,\mathrm{w}}$,
as discussed in Sec.\ \ref{subsec:Noise_models}, into $\left\langle \varphi_{\epsilon}^{2}(t)\right\rangle $
in Eq.\ (\ref{eq:phi_2_RTCE_noise}) of the main text, yielding

\begin{equation}
\left\langle \varphi_{\epsilon}^{2}(t)\right\rangle \approx\dfrac{1}{2\pi}\left[\Delta_{A}-\alpha_{A}\right]^{\eta_{\epsilon}}S_{\epsilon,\mathrm{w}}\int_{-\infty}^{\infty}d\omega\dfrac{1}{(\dfrac{\omega}{2})^{2}}\sin^{2}(\dfrac{\omega}{2}t).\label{eq:RTCE_noise_phi_2_integral_1}
\end{equation}
Employing the integral identity $\int_{-\infty}^{\infty}dx\left[\sin^{2}(xt)/x^{2}\right]=\pi t$
further simplifies this expression to 
\begin{equation}
\left\langle \varphi_{\epsilon}^{2}(t)\right\rangle \approx\left[\Delta_{A}-\alpha_{A}\right]^{\eta_{\epsilon}}S_{\epsilon,\mathrm{w}}t.\label{eq:RTCE_noise_phi_2_integral_2}
\end{equation}
Finally, inserting this reduced $\left\langle \varphi_{\epsilon}^{2}(t)\right\rangle $
into the Ramsey envelope $V(t)$ {[}Eq.\ (\ref{eq:Ramsey_envelope_general}){]}
gives $V(t)\approx\mathrm{exp}[-(t/T_{\phi,1,\epsilon})]$, where
the corresponding dephasing time for the RTCE noise is $T_{\phi,1,\epsilon}=2/[(\Delta_{A}-\alpha_{A})^{\eta_{\epsilon}}\cdot S_{\epsilon,\mathrm{w}}]$,
as shown in Eq.\ (\ref{eq:T_phi_1_e_est}) of the main text.

For the flux noise, substituting the $1/f$ noise spectrum $S_{\phi}(\omega)=S_{1/f}/f$
into $\left\langle \varphi_{\phi}^{2}(t)\right\rangle $ in Eq.\ (\ref{eq:phi_2_flux_noise})
of the main text, and assuming that $S_{\phi}(\omega)$ is symmetric
while contributions from frequencies outside the interval $\left[f_{\mathrm{min}},f_{\mathrm{max}}\right]$
are negligible, yields

\begin{align}
\left\langle \varphi_{\phi}^{2}(t)\right\rangle  & \approx\left[\dfrac{d\Delta_{A}}{d(\Phi_{A}/\Phi_{0})}\dfrac{\Phi_{A}}{\Phi_{0}}\right]^{2}2t^{2}S_{1/f}\cdot\nonumber \\
 & \left\{ \int_{f_{\mathrm{min}}}^{f_{\mathrm{max}}}df\cdot\dfrac{1}{f}\left[\dfrac{1}{(\pi f\cdot t)^{2}}\sin^{2}(\pi f\cdot t)\right]\right\} .\label{eq:flux_noise_phi_2_integral}
\end{align}
In the Ramsey dephasing experiment \citep{BarendsKellyMegrantEtAl2014},
the typical evolution time satisfies $t\lesssim5\,\mu\mathrm{s}$,
and the $1/f$ spectrum has been observed experimentally over an extended
frequency range $f_{\mathrm{min}}\sim10^{-5}\,\mathrm{Hz}$ to $f_{\mathrm{max}}\sim10^{9}\,\mathrm{Hz}$
\citep{Quintana2017}. At the lower bound, $f=f_{\mathrm{min}}\sim10^{-5}\,\mathrm{Hz}$,
the integrand $\left[1/f\right]\left[\sin^{2}(\pi f\cdot t)/(\pi f\cdot t)^{2}\right]$
in Eq.\ (\ref{eq:flux_noise_phi_2_integral}) is dominated by $1/f=1/f_{\mathrm{min}}\sim10^{5}\,\mathrm{Hz^{-1}}$,
since $\left[\sin^{2}(\pi f\cdot t)/(\pi f\cdot t)^{2}\right]\rightarrow1$
for small $f\cdot t=f_{\mathrm{min}}\cdot t\lesssim5\times10^{-11}$.
In contrast, as the frequency approaches $f=1/t$, the squared sinc
function term, $\left[\sin^{2}(\pi f\cdot t)/(\pi f\cdot t)^{2}\right]$,
completes its first oscillation, while $1/f$ decays to $1/(1/t)=t\lesssim5\times10^{-6}\,\mathrm{Hz^{-1}}$.
This indicates that the $1/f$ decay can effectively suppresses the
integrand before the squared sinc term significantly deviates from
unity. Therefore, the integral in Eq.\ (\ref{eq:flux_noise_phi_2_integral})
can be approximated as $\int_{f_{\mathrm{min}}}^{A/t}df\cdot1/f=\mathrm{ln}[A/(f_{\mathrm{min}}t)]$
\citep{Martinis2003}. Here $A\approx0.40048$ is determined by fitting
the approximate expression to the numerical evaluation of the integral.
Next, replacing the integral in Eq.\ (\ref{eq:flux_noise_phi_2_integral})
with $\mathrm{ln}[A/(f_{\mathrm{min}}t)]$ gives

\begin{equation}
\left\langle \varphi_{\phi}^{2}(t)\right\rangle \approx\left[\dfrac{d\Delta_{A}}{d(\Phi_{A}/\Phi_{0})}\dfrac{\Phi_{A}}{\Phi_{0}}\right]^{2}2t^{2}S_{1/f}C_{\mathrm{const}},\label{eq:flux_noise_phi_2_appro}
\end{equation}
where $C_{\mathrm{const}}\equiv\mathrm{ln}[A/(f_{\mathrm{min}}t)]$
varies slowly from $24.4133$ to $22.8039$ as the evolution time
increases from $t=1\,\mu\mathrm{s}$ to $5\,\mu\mathrm{s}$ for $f_{\mathrm{min}}=10^{-5}\,\mathrm{Hz}$
and $A=0.40048$. For the purpose of estimation, we adopt $C_{\mathrm{const}}=22.8039$.
Finally, substituting $\left\langle \varphi_{\phi}^{2}(t)\right\rangle $
from Eq.\ (\ref{eq:flux_noise_phi_2_appro}) into the Ramsey envelope
in Eq.\ (\ref{eq:Ramsey_envelope_general}) yields $V(t)=\mathrm{exp}[-(t^{2}/T_{\phi,2}^{2})]$,
with the estimated slow dephasing time $T_{\phi,2}=1/\left\{ \left[d\Delta_{A}/d(\Phi_{A}/\Phi_{0})\right]\left[\text{\ensuremath{\Phi_{A}}}/\Phi_{0}\right]\sqrt{S_{1/f}\cdot C_{\mathrm{const}}}\right\} $,
as shown in Eq.\ (\ref{eq:T_phi_2_est}) of the main text.

Finally, combining all dephasing contributions---including energy
relaxation characterized by the time $T_{1}$, the RTCE noise $\beta_{\epsilon}(t)$
with a fast dephasing time $T_{\phi,1,\epsilon}$, and the flux noise
$\beta_{\phi}(t)$ with a slow dephasing time $T_{\phi,2}$---yields
an overall Ramsey decay envelope $V(t)=\mathrm{exp}\left[-t/(2T_{1})-t/T_{\phi,1,\epsilon}-t^{2}/T_{\phi,2}^{2}\right]$.
By unifying the terms linear in $t$, the decay envelope can be expressed
as $V(t)=\mathrm{exp}\left[-t/T_{\phi,1}-t^{2}/T_{\phi,2}^{2}\right]$,
where the effective fast dephasing time is $T_{\phi,1}=2T_{1}\cdot T_{\phi,1,\epsilon}/(2T_{1}+T_{\phi,1,\epsilon})$,
as introduced in the main text.

\section{Noise generation by the Ornstein-Uhlenbeck (OU) process}

\label{appendix:noise_generation_OU}

In Sec.\ \ref{subsec:Noise_characterization_dephasing} of the main
text, we employ the single-$\Gamma_{OU}$ OU process and the multi-$\Gamma_{OU}$
OU process to generate realizations of the RTCE noise $\beta_{\epsilon}(t)$
and the flux noise $\beta_{\phi}(t)$, respectively, for simulating
Ramsey dephasing experiments. In this Appendix, we first introduce
the OU process and subsequently describe the implementation of the
single-$\Gamma_{OU}$ OU process for generating white noise, followed
by the implementation of the multi-$\Gamma_{OU}$ OU process for producing
noise with a $1/f^{\alpha}$ spectrum ($\alpha\leq2$) \citep{Milotti1995,Gleeson2005,Huang2019}.

The zero-mean OU process $\beta_{OU}(t)$ can be generated iteratively
via

\begin{equation}
\beta_{OU}(t+dt)=\left(1-2\pi\Gamma_{OU}dt\right)\beta_{OU}(t)+\sigma_{OU}\sqrt{4\pi\Gamma_{OU}}dW(t),\label{eq:OU_process}
\end{equation}
where $dW(t)$ denotes the Wiener process, a zero-mean normal random
variable with standard deviation $\sqrt{dt}$. When the initial condition
$\beta_{OU}(t=0)$ is drawn from a zero-mean normal distribution with
standard deviation $\sigma_{OU}$, the correlation function of $\beta_{OU}(t)$,
defined as $C_{OU}(\tau)=\left\langle \beta_{OU}(t_{1})\beta_{OU}(t_{2})\right\rangle $
with $\tau=\left|t_{1}-t_{2}\right|$, is given by

\begin{equation}
C_{OU}(\tau)=\sigma_{OU}^{2}\exp\left(-2\pi\Gamma_{OU}\tau\right).\label{eq:OU_correlation_function}
\end{equation}
The relation $C_{OU}(\tau=0)=\left\langle \beta_{OU}^{2}(t)\right\rangle =\sigma_{OU}^{2}$
ensures that the standard deviation of $\beta_{OU}(t)$ remains $\sigma_{OU}$
at all times. Here $\Gamma_{OU}$ characterizes the correlation decay
rate. The corresponding spectrum $S_{OU}(f)$, obtained via the Wiener--Khinchin
theorem $S_{OU}(f)=\int_{-\infty}^{\infty}C_{OU}(\tau)e^{i2\pi f\tau}d\tau$,
is a Lorentzian,

\begin{equation}
S_{OU}(f)=\dfrac{\sigma_{OU}^{2}}{\pi}\frac{\Gamma_{OU}}{(\Gamma_{OU}^{2}+f^{2})}.\label{eq:OU_noise_spectrum}
\end{equation}

An OU process $\beta_{OU}(t)$, generated according to Eq.\ (\ref{eq:OU_process}),
is fully characterized by two parameters: $\Gamma_{OU}$ and $\sigma_{OU}$.
Its spectrum $S_{OU}(f)$, given in Eq.\ (\ref{eq:OU_noise_spectrum}),
approaches a constant for $f\ll\Gamma_{OU}$ and exhibits a $1/f^{2}$
decay for $f\gg\Gamma_{OU}$. This property allows a single OU process---hereafter
referred to as the single-$\Gamma_{OU}$ OU process---to model either
white noise or Brown noise by appropriately adjusting $\Gamma_{OU}$:
increasing $\Gamma_{OU}$ extends the flat region of $S_{OU}(f)$
to characterize white noise, while decreasing $\Gamma_{OU}$ extends
the $1/f^{2}$ region to characterize Brown noise. To generate white
noise with a constant spectrum $S_{\mathrm{w}}$ using the single-$\Gamma_{OU}$
OU process, $\Gamma_{OU}$ is chosen to match the upper-bound cutoff
frequency of the desired white noise, and $\sigma_{OU}$ is determined
from $\sigma_{OU}=\sqrt{\pi\Gamma_{OU}S_{\mathrm{w}}}$ by equating
$S_{\mathrm{w}}=S_{OU}(f\ll\Gamma_{OU})\approx\sigma_{OU}^{2}/(\pi\Gamma_{OU})$.
For Brown noise with a $1/f^{2}$ spectrum, $\Gamma_{OU}$ is set
by the lower-bound cutoff frequency of the desired $1/f^{2}$ region,
and $\sigma_{OU}$ is obtained from $\sigma_{OU}=\sqrt{\pi S_{OU}(f_{0})\cdot f_{0}^{2}/\Gamma_{OU}}$,
where $f_{0}\gg\Gamma_{OU}$, ensuring that $S_{OU}(f_{0})\approx(\sigma_{OU}^{2}\Gamma_{OU})/(\pi f_{0}^{2})$.
In the main text, we employ the single-$\Gamma_{OU}$ OU process to
generate the RTCE noise $\beta_{\epsilon}(t)$ with a white noise
spectrum $S_{\epsilon,\mathrm{w}}$. In this context, we set $\sigma_{OU}=\sigma_{\epsilon}$
and $\Gamma_{OU}=\Gamma_{\epsilon}$ in the single-$\Gamma_{OU}$
OU process, yielding $\sigma_{\epsilon}=\sqrt{\pi\Gamma_{\epsilon}S_{\epsilon,\mathrm{w}}}$
from $\sigma_{OU}=\sqrt{\pi\Gamma_{OU}S_{\mathrm{w}}}$ and the corresponding
spectrum $S_{\epsilon}(f)=(\sigma_{\epsilon}^{2}/\pi)\cdot\Gamma_{\epsilon}/(\Gamma_{\epsilon}^{2}+f^{2})$
from $S_{OU}(f)$ in Eq.\ (\ref{eq:OU_noise_spectrum}). For the
characterization of the RTCE noise, we consider two choices of the
upper-bound cutoff, $\Gamma_{\epsilon}=10^{6.5}\,\mathrm{Hz}$ and
$\Gamma_{\epsilon}=10^{10.5}\,\mathrm{Hz}$, for the constant spectrum
$S_{\epsilon,\mathrm{w}}$, and show the resulting RTCE noise spectra
$S_{E}(f)=\left[\Delta_{A}-\alpha_{A}\right]^{\eta_{\epsilon}}S_{\epsilon}(f)$
at $\Delta_{A}/2\pi=5\,\mathrm{GHz}$ in Fig.\ \ref{fig:noise_characterication}(c)
of the main text.

Next, we construct the noise $\beta(t)$ as a composition of independent
OU processes,

\begin{align}
\beta(t) & =\overset{j_{\mathrm{max}}}{\underset{j=1}{\sum}}\beta_{j}(t),\label{eq:composite_noise}
\end{align}
where each OU process $\beta_{j}(t)$ is characterized by its own
standard deviation $\sigma_{j}$ and correlation decay rate $\Gamma_{j}$,
and $j_{\mathrm{max}}$ denotes the total number of constituent OU
processes. We refer to this composite noise as a multi-$\Gamma_{OU}$
OU process. The correlation function of $\beta(t)$, $C(\tau)=\left\langle \beta(t_{1})\beta(t_{2})\right\rangle $,
can be expressed as a sum of the individual correlation functions
$C_{j}(\tau)=\left\langle \beta_{j}(t_{1})\beta_{j}(t_{2})\right\rangle $,

\begin{align}
C(\tau) & =\overset{j_{\mathrm{max}}}{\underset{j=1}{\sum}}C_{j}(\tau)=\overset{j_{\mathrm{max}}}{\underset{j=1}{\sum}}\sigma_{j}^{2}\exp\left(-2\pi\Gamma_{j}\tau\right),\label{eq:composite_CF}
\end{align}
since $\left\langle \beta_{j}(t_{1})\beta_{k}(t_{2})\right\rangle =0$
for $j\neq k$ due to the statistical independence of the constituent
OU processes. Accordingly, the corresponding spectrum, obtained via
the Wiener--Khinchin theorem, $S(f)=\int_{-\infty}^{\infty}C(\tau)e^{i2\pi f\tau}d\tau$,
also decomposes as a sum of the spectra of the individual OU processes,

\begin{equation}
S(f)=\overset{j_{\mathrm{max}}}{\underset{j=1}{\sum}}S_{j}(f)=\overset{j_{\mathrm{max}}}{\underset{j=1}{\sum}}\dfrac{\sigma_{j}^{2}}{\pi}\frac{\Gamma_{j}}{(\Gamma_{j}^{2}+f^{2})},\label{eq:composite_S}
\end{equation}
where $S_{j}(f)$ denotes the spectrum of $\beta_{j}(t)$. The composite
spectrum in Eq.\ (\ref{eq:composite_S}) indicates that noise with
a specific spectrum could be generated using a multi-$\Gamma_{OU}$
OU process by appropriately tuning the parameter set $\left\{ \Gamma_{j},\sigma_{j}\right\} $.
We find that noise exhibiting a $1/f^{\alpha}$ spectrum ($\alpha\leq2$)
within the frequency range $\left[\Gamma_{j,\mathrm{min}},\Gamma_{j,\mathrm{max}}\right]$
can be generated using the multi-$\Gamma_{OU}$ OU process. Specifically,
by selecting the set $\left\{ \Gamma_{j}\right\} $ such that the
$\Gamma_{j}$ are logarithmically evenly spaced between $\Gamma_{j,\mathrm{min}}$
and $\Gamma_{j,\mathrm{max}}$, and defining the corresponding set
$\left\{ \sigma_{j}\right\} $ according to

\begin{equation}
\sigma_{j}=\sigma_{OU}^{0}\cdot\kappa_{\alpha}^{[\log_{10}\Gamma_{j}]/2},\label{eq:composite_kappa}
\end{equation}
the resulting noise spectrum $S(f)$ exhibits $1/f^{\alpha}$ behavior
within the range $\left[\Gamma_{j,\mathrm{min}},\Gamma_{j,\mathrm{max}}\right]$
. In addition, $S(f)$ approaches a constant value $\sum_{j=1}^{j_{\mathrm{max}}}\sigma_{j}^{2}/(\pi\Gamma_{j})$
for $f\ll\Gamma_{j,\mathrm{min}}$, and decays as $1/f^{2}$ for $f\gg\Gamma_{j,\mathrm{max}}$.
The parameter $\kappa_{\alpha}$ in Eq.\ (\ref{eq:composite_kappa})
characterizes the spectrum shape, which is determined by the exponent
$\alpha$ of the $1/f^{\alpha}$ spectrum. Representative values of
$\kappa_{\alpha}$ for specific $\alpha$ are numerically evaluated
and listed in Table.\ \ref{Table:kappa_value}. The parameter $\sigma_{OU}^{0}$
in Eq.\ (\ref{eq:composite_kappa}) quantifies the spectrum amplitude.
It is determined by substituting $\sigma_{j}$ in Eq.\ (\ref{eq:composite_kappa})
into $S(f)$ in Eq.\ (\ref{eq:composite_S}) at a reference frequency
$f=f_{0}$, yielding 
\begin{equation}
\sigma_{OU}^{0}=\sqrt{\pi\cdot S(f_{0})/\left[\overset{j_{\mathrm{max}}}{\underset{j=1}{\sum}}\frac{\kappa_{\alpha}^{\log_{10}\Gamma_{j}}\cdot\Gamma_{j}}{(\Gamma_{j}^{2}+f_{0}^{2})}\right]},\label{eq:composite_noise_sigma_OU_0}
\end{equation}
where $f_{0}$ within $\left[\Gamma_{j,\mathrm{min}},\Gamma_{j,\mathrm{max}}\right]$.
Furthermore, the effective standard deviation of the multi-$\Gamma_{OU}$
OU process is given by 
\begin{equation}
\sigma_{\mathrm{eff}}=\sigma_{OU}^{0}\sqrt{\overset{j_{\mathrm{max}}}{\underset{j=1}{\sum}}\kappa_{\alpha}^{\log_{10}\Gamma_{j}}},\label{eq:composite_noise_strength}
\end{equation}
obtained as the square root of $C(\tau=0)$ in Eq.\ (\ref{eq:composite_CF}).

In the main text, we implement the multi-$\Gamma_{OU}$ OU process
to generate the flux noise $\beta_{\phi}(t)$ with spectrum $S_{\phi}(f)=S_{1/f}/f$
over the frequency range $f=10^{-5}\,\mathrm{Hz}$ to $10^{9}\,\mathrm{Hz}$.
For this case, we first define the set $\left\{ \Gamma_{j}\right\} $
from $\Gamma_{j,\mathrm{min}}/1\,\mathrm{Hz}=10^{-5}$ to $\Gamma_{j,\mathrm{max}}/1\,\mathrm{Hz}=10^{9}$
with a logarithmic interval $0.25$. The corresponding set $\left\{ \sigma_{j}\right\} $
is then determined using Eq.\ (\ref{eq:composite_kappa}), where
$\kappa_{\alpha}=1.0$ for $\alpha=1.0$ from Table.\ \ref{Table:kappa_value},
and $\sigma_{OU}^{0}=\sqrt{\pi S_{1/f}/[\Sigma_{j=1}^{j_{\mathrm{max}}}(\kappa_{\alpha}^{\log_{10}\Gamma_{j}}\cdot\Gamma_{j})/(\Gamma_{j}^{2}+1)]}$
via Eq.\ (\ref{eq:composite_noise_sigma_OU_0}), using the spectrum
at the reference frequency $f_{0}=1\,\mathrm{Hz}$, i.e., $S(f_{0})=S_{\phi}(1\,\mathrm{Hz})=S_{1/f}$.
The generated spectrum $S_{\phi}(f)$ is subsequently obtained from
the composite spectrum $S(f)$ in Eq.\ (\ref{eq:composite_S}) with
the specified parameter set $\left\{ \Gamma_{j},\sigma_{j}\right\} $,
and the corresponding flux noise spectrum $S_{\Phi}(f)=(\Phi_{A}/\Phi_{0})^{2}S_{\phi}(f)$
at $\Delta_{A}/2\pi=5\,\mathrm{GHz}$ is shown in Fig.\ \ref{fig:noise_characterication}(e)
of the main text. In addition, the effective standard deviation $\sigma_{\phi,\mathrm{eff}}$
of the generated multi-$\Gamma_{OU}$ OU process $\beta_{\phi}(t)$
is obtained by substituting the specified $\kappa_{\alpha}$, $\sigma_{OU}^{0}$,
and set $\left\{ \Gamma_{j}\right\} $ into $\sigma_{\mathrm{eff}}$
in Eq.\ (\ref{eq:composite_noise_strength}).

\begin{table}
\caption{$\kappa_{\alpha}$ values for selected exponents $\alpha$ of the
$1/f^{\alpha}$ spectrum.}


\begin{tabular}{c| c| c| c| c| c| c| c} 
\toprule 
$\alpha$  & $0.5$  & $0.8$  & $0.9$  & $1.0$  & $1.01$  & $1.1$  & $1.5$\tabularnewline
\midrule 
$\kappa_{\alpha}$  & $3.162$  & $1.585$  & $1.259$  & $1.0$  & $0.977$  & $0.795$  & $0.316$\tabularnewline
\bottomrule
\end{tabular}

\label{Table:kappa_value} 
\end{table}

\section{Ensemble-averaged infidelity from Lindblad-form master equation}

\label{appendix:LindBlad_master_eq_infid}

In this Appendix, we introduce the ensemble-averaged infidelity $\left\langle \mathcal{I}\right\rangle $,
as defined in Eq.\ (\ref{eq:ensemble_infid_master_eq}) of the main
text, which is expressed in terms of the propagator $G(t)$ corresponding
to the vectorized density matrix evolved under the Lindblad-form master
equation. Starting from this definition of $\left\langle \mathcal{I}\right\rangle $,
we then derive the approximate expression for $\left\langle \mathcal{I}\right\rangle $
presented in Eq.\ (\ref{eq:ensemble_infid_schrodinger_eq}) of the
main text, which is defined in terms of the propagator $U(t)$ generated
by the Schrödinger equation.

We begin by writing the two-qubit Lindblad-form master equation, as
introduced in Sec.\ \ref{subsec:J2_noises} of the main text:

\begin{align}
\dot{\rho} & =-\dfrac{i}{\hbar}\left[\mathcal{H}(t),\rho\right]\nonumber \\
 & +\dfrac{1}{T_{1}}\left\{ a_{A}\rho a_{A}^{\dagger}-\frac{1}{2}a_{A}^{\dagger}a_{A}\rho-\frac{1}{2}\rho a_{A}^{\dagger}a_{A}\right.\nonumber \\
 & \left.+a_{B}\rho a_{B}^{\dagger}-\frac{1}{2}a_{B}^{\dagger}a_{B}\rho-\frac{1}{2}\rho a_{B}^{\dagger}a_{B}\right\} .\label{eq:master_eq_combined}
\end{align}
The two-qubit Hamiltonian is given by $\mathcal{H}(t)=\mathcal{H}_{A}(t)+\mathcal{H}_{B}(t)+\mathcal{H}_{I,AB}(t)$,
where $\mathcal{H}_{A}(t)$ is the realistic Hamiltonian of qubit
$Q_{A}$ introduced in Sec.\ \ref{subsec:Noise_characterization_dephasing},
$\mathcal{H}_{B}(t)$ is the realistic Hamiltonian of qubit $Q_{B}$,
and $\mathcal{H}_{I,AB}(t)$ is the ideal coupling Hamiltonian between
$Q_{A}$ and $Q_{B}$, as described in Sec.\ \ref{subsec:J2_noises}.
Next, we vectorize the master equation using the identity $\mathrm{vec}(ABC)=(C^{T}\otimes A)\mathrm{vec}(B)$,
where $A$, $B$, and $C$ are matrices of dimensions $l\times m$,
$m\times n$, and $n\times p$, respectively. The operation $\mathrm{vec}$
transforms a matrix into a column vector by sequentially stacking
its elements from top to bottom within each column, proceeding from
the leftmost to the rightmost column. The superscript $T$ denotes
matrix transposition. The vectorized form of the master equation,
following Eq.\ (\ref{eq:master_eq_combined}), can then be expressed
as

\begin{equation}
\dot{G}(t)=-iK(t)G(t),\label{eq:vec_master_eq}
\end{equation}
where $G(t)$ is the propagator of the initial vectorized density
matrix $\mathrm{vec}\rho(0)$, such that

\begin{equation}
\left[\mathrm{vec}\rho(t)\right]=G(t)\left[\mathrm{vec}\rho(0)\right].\label{eq:vec_density_matrix}
\end{equation}
Here, $K(t)$ is defined as

\begin{align}
K(t) & =K_{0}(t)+K_{1},\label{eq:K_eq_K0_plus_K1}\\
K_{0}(t) & =\left(I\otimes\dfrac{\mathcal{H}(t)}{\hbar}-\dfrac{\mathcal{H}(t)}{\hbar}\otimes I\right),\label{eq:K0_vec_master_eq}\\
K_{1} & =+i\dfrac{1}{T_{1}}\left(a_{A}\otimes a_{A}-\frac{1}{2}I\otimes a_{A}^{\dagger}a_{A}-\frac{1}{2}a_{A}^{\dagger}a_{A}\otimes I\right.\nonumber \\
 & \left.+a_{B}\otimes a_{B}-\frac{1}{2}I\otimes a_{B}^{\dagger}a_{B}-\frac{1}{2}a_{B}^{\dagger}a_{B}\otimes I\right).\label{eq:K1_vec_master_eq}
\end{align}

To define the two-qubit gate infidelity based on $G(t)$, we first
project the propagator $G(t)$ onto the subspace spanned by the two-qubit
computational basis states, yielding the projected propagator $G_{16\times16}(t)$.
In the case of a closed quantum system, i.e., when $K_{1}=0$, the
projected propagator can be expressed as $G_{16\times16}(t)=U_{4\times4}^{\star}(t)\otimes U_{4\times4}(t)$,
where $U_{4\times4}(t)$ is the propagator obtained by projecting
$U(t)$ onto the two-qubit computational subspace, and $U(t)$ itself
satisfies the Schrödinger equation $\dot{U}(t)=(-i/\hbar)\cdot\mathcal{H}(t)U(t)$.
The target gate for $U_{4\times4}(t)$ is denoted as $U_{T}$; correspondingly,
the target for $G_{16\times16}(t)$ is $G_{T}=U_{T}^{\star}\otimes U_{T}$.
The two-qubit gate infidelity is then defined as

\begin{equation}
\mathcal{I}\equiv1-\dfrac{1}{16}\mathrm{Re}\left\{ \mathrm{Tr}\left[G_{T}^{\dagger}G_{16\times16}(t_{f})\right]\right\} .\label{eq:infid_vec_master_eq}
\end{equation}
For a closed quantum system with $K_{1}=0$, the gate infidelity $\mathcal{I}$
defined above reduces to the familiar form $\mathcal{I}=1-|\mathrm{Tr}[U_{T}^{\dagger}U_{4\times4}(t_{f})]|^{2}/16$,
which coincides with the ideal gate infidelity $J_{1}$ in Eq.\ (\ref{eq:J1_definition})
of the main text, with the replacement of the ideal projected propagator
$U_{I,4\times4}(t_{f})$ by the realistic projected propagator $U_{4\times4}(t_{f})$.
Finally, the ensemble-averaged infidelity $\left\langle \mathcal{I}\right\rangle $,
shown in Eq.\ (\ref{eq:ensemble_infid_master_eq}) of the main text,
is obtained by averaging $\mathcal{I}$, as defined in Eq.\ (\ref{eq:infid_vec_master_eq}),
over an ensemble of realizations.

We next approximate the ensemble-averaged infidelity $\left\langle \mathcal{I}\right\rangle $
obtained above to derive the expression presented in Eq.\ (\ref{eq:ensemble_infid_schrodinger_eq})
of the main text. To this end, we transform the system from the laboratory
frame to the rotating frame associated with $K_{0}(t)$ in Eq.\ (\ref{eq:K0_vec_master_eq}),
yielding the vectorized master equation $\dot{\tilde{G}}(t)=-i\tilde{K_{1}}(t)\tilde{G}(t)$,
where $\tilde{K_{1}}(t)=G_{0}^{\dagger}(t)K_{1}G_{0}(t)$ and $G_{0}(t)$
is generated by $\dot{G_{0}}(t)=-iK_{0}(t)G_{0}(t)$. The propagator
$\tilde{G}(t)$ at $t=t_{f}$ can be expanded using the Dyson series
as $\tilde{G}(t_{f})=I+\overline{\Psi}_{1}+\overline{\Psi}_{2}+\cdots$,
where the first two terms are $\overline{\Psi}_{1}=-i\int_{0}^{t_{f}}\tilde{K_{1}}(t)dt$
and $\overline{\Psi}_{2}=-\int_{0}^{t_{f}}dt_{1}\int_{0}^{t_{1}}dt_{2}\tilde{K_{1}}(t_{1})\tilde{K_{1}}(t_{2})$.
Subsequently, we transform the propagator $\tilde{G}(t_{f})$ back
to the laboratory frame using $G(t_{f})=G_{0}(t_{f})\tilde{G}(t_{f})$,
which gives

\begin{equation}
G(t_{f})=G_{0}(t_{f})\cdot[I+\overline{\Psi}_{1}+\overline{\Psi}_{2}+\cdots].\label{eq:G_Dyson_expansion}
\end{equation}
Next, we project $G(t_{f})$ onto the computational subspace to obtain
$G_{16\times16}(t_{f})=\{G_{0}(t_{f})\cdot[I+\overline{\Psi}_{1}+\overline{\Psi}_{2}+\cdots]\}_{16\times16}$.
When the probability leakage to non-computational states is small,
the off-diagonal blocks of matrix $G_{0}(t)$ that couple the computational
and non-computational subspaces are also small. Consequently, the
projected propagator $G_{16\times16}(t_{f})$ can be approximately
factorized as 
\begin{equation}
G_{16\times16}(t_{f})\approx G_{0,16\times16}(t_{f})\cdot[I+\overline{\Psi}_{1}+\overline{\Psi}_{2}+\cdots]{}_{16\times16},\label{eq:G_16x16_w_approximation}
\end{equation}
where $G_{0,16\times16}(t_{f})$ and $[I+\overline{\Psi}_{1}+\overline{\Psi}_{2}+\cdots]{}_{16\times16}$
denote the projections of $G_{0}(t_{f})$ and $[I+\overline{\Psi}_{1}+\overline{\Psi}_{2}+\cdots]$
onto the computational subspace, respectively. We then substitute
$G_{16\times16}(t_{f})$ from Eq.\ (\ref{eq:G_16x16_w_approximation})
into the infidelity $\mathcal{I}$ defined in Eq.\ (\ref{eq:infid_vec_master_eq}),
and assume that $G_{16\times16}(t_{f})$ is close to the target $G_{T}$,
such that $G_{T}^{\dagger}G_{0,16\times16}(t_{f})\approx I_{16\times16}$.
Under this assumption, the infidelity $\mathcal{I}$ can be approximated
as

\begin{align}
\mathcal{I} & \approx1-\dfrac{1}{16}\mathrm{Re}\left\{ \mathrm{Tr}\left[G_{T}^{\dagger}G_{0,16\times16}(t_{f})\right]\right\} \nonumber \\
 & -\dfrac{1}{16}\mathrm{Re}\left\{ \mathrm{Tr}[\overline{\Psi}_{1}+\overline{\Psi}_{2}+\cdots]{}_{16\times16}\right\} .\label{eq:Infid_approximated_1}
\end{align}
The terms in the first line of the above equation correspond to the
infidelity of a closed quantum system with $K_{1}=0$, which reduces
to $\mathcal{I}=1-|\mathrm{Tr}[U_{T}^{\dagger}U_{4\times4}(t_{f})]|^{2}/16$
as discussed in the preceding paragraph. The term in the second line
represents the contribution to the infidelity from energy relaxation,
where $\overline{\Psi}_{n}\propto(t_{f}/T_{1})^{n}$. When $t_{f}/T_{1}\ll1$,
the lowest-order term $-(1/16)\cdot\mathrm{Re}\{\mathrm{Tr}[\overline{\Psi}_{1}]{}_{16\times16}\}$
dominates. The trace in this dominant term is $\mathrm{Tr}[\overline{\Psi}_{1}]{}_{16\times16}=\mathrm{Tr}[-i\int_{0}^{t_{f}}G_{0}^{\dagger}(t)K_{1}G_{0}(t)dt]{}_{16\times16}=-i\int_{0}^{t_{f}}\mathrm{Tr}[G_{0}^{\dagger}(t)K_{1}G_{0}(t)]_{16\times16}dt$.
Under the assumption of low state probability leakage, the projected
operator satisfies $[G_{0}^{\dagger}(t)K_{1}G_{0}(t)]_{16\times16}\approx$$G_{0,16\times16}^{\dagger}(t)K_{1,16\times16}G_{0,16\times16}(t)$.
Using the cyclic property of the trace, we then have $\mathrm{Tr}[\overline{\Psi}_{1}]{}_{16\times16}\approx-i\int_{0}^{t_{f}}\mathrm{Tr}[K_{1,16\times16}G_{0,16\times16}(t)G_{0,16\times16}^{\dagger}(t)]dt$.
Since $G_{0}(t)$ is unitary, generated by $\dot{G_{0}}(t)=-iK_{0}(t)G_{0}(t)$
with Hermitian $K_{0}(t)$, and thus, when the leakage to non-computational
states is small, the projected propagator $G_{0,16\times16}(t)$ can
be approximated as unitary, leading to $G_{0,16\times16}(t)G_{0,16\times16}^{\dagger}(t)\approx I_{16\times16}$.
Consequently, the trace reduces to $\mathrm{Tr}[\Psi_{1}]{}_{16\times16}\approx-i\int_{0}^{t_{f}}\mathrm{Tr}[K_{1,16\times16}]dt$,
and the dominant contribution to the infidelity becomes $-(1/16)\cdot\mathrm{Re}\{\mathrm{Tr}[\overline{\Psi}_{1}]{}_{16\times16}\}\approx t_{f}/T_{1}$,
after substituting $K_{1}$ as given in Eq.\ (\ref{eq:K1_vec_master_eq}).
Finally, combining all contributions, the infidelity can be written
as

\begin{equation}
\mathcal{I}\approx1-\dfrac{1}{16}\left|\mathrm{Tr}\left[U_{T}^{\dagger}U_{4\times4}(t_{f})\right]\right|^{2}+\dfrac{t_{f}}{T_{1}},\label{eq:Infid_approximated_2}
\end{equation}
and the ensemble-averaged infidelity $\left\langle \mathcal{I}\right\rangle $
is obtained by averaging $\mathcal{I}$ over an ensemble of realizations,
as shown in Eq.\ (\ref{eq:ensemble_infid_schrodinger_eq}) of the
main text.

 \bibliographystyle{apsrev4-2}
\bibliography{citation_robust_gates}

\end{document}